\documentclass{elsart}
\journal{Journal of Functional Analysis}

\usepackage{amssymb}

\usepackage{amsfonts}
\usepackage{amsmath}

\numberwithin{equation}{section}
\newtheorem{theorem}{Theorem}[section]
\newtheorem{proposition}[theorem]{Proposition}
\newtheorem{lemma}[theorem]{Lemma}
\newtheorem{corollary}[theorem]{Corollary}
\newtheorem{definition}[theorem]{Definition}
\newtheorem{remark}[theorem]{Remark}
\newtheorem{assumption}[theorem]{Assumption}
\newenvironment{acknowledgement}{\emph{Acknowledgement.}}

\newenvironment{proof}[1][\proofname]{ \trivlist
  \item[\hskip\labelsep\itshape
    #1]\ignorespaces
}{%
  \qed\endtrivlist
}
\newcommand{\proofname}{Proof.}

\newcommand{\tnorm}[1]{\left|\!\left|\!\left| #1 \right|\!\right|\!\right|}

\renewcommand\H{\mathcal{H}}
\renewcommand\L{\mathrm{L}}
\newcommand\R{\mathbb R}
\newcommand\N{\mathbb N}
\newcommand\C{\mathbb C}
\newcommand\Z{\mathbb Z}
\newcommand\K{\mathcal{K}}
\newcommand\D{\mathcal{D}}
\newcommand\Db{\mathbf{D}}
\newcommand\G{\mathcal{G}}
\newcommand\U{\mathcal{U}}
\newcommand\Ll{\mathcal{L}}
\newcommand\A{\mathbf{A}}

\newcommand\x{\mathbf{x}}
\renewcommand\a{\mathbf{a}}
\newcommand\di{\mathrm d}
\renewcommand\v{\mathbf{v}}
\newcommand{\Hc}{\mathcal{H}_{\mathrm{c}}}
\newcommand{\Cc}{C_{\mathrm{c}}}
\newcommand{\Lc}{\mathrm{L}_{\mathrm{c}}}
\newcommand\El{\mathbf E}
\newcommand\J{\mathbf J}
\newcommand{\T}{\mathcal{T}}
\newcommand\F{\mathbf{F}}
\newcommand\Fc{\mathcal{F}}
\renewcommand\S{\mathcal{S}}
\newcommand\tr{\mathrm{tr}}

\renewcommand\e{\mathrm{e}}
\newcommand{\la}{\langle}
\newcommand{\ra}{\rangle}
\renewcommand\P{\mathbb P}
\newcommand\E{\mathbb E}

\begin{document}

\begin{frontmatter}


\title{Linear response theory for magnetic Schr\"odinger
operators in disordered media}

\author{Jean-Marc Bouclet}
\address{UMR 8524 CNRS, UFR de Math\'ematiques,
Universit\'e de Lille 1,  F-59655 Villeneuve d'Ascq C\'edex, France}
 \ead{Jean-Marc.Bouclet@math.univ-lille1.fr}

\author{Francois Germinet}
\address{ Universit\'e de Cergy-Pontoise,
D\'epartement de Math\'ematiques,
Site de Saint-Martin,
2 avenue Adolphe Chauvin,
95302 Cergy-Pontoise cedex, France}
\ead{germinet@math.u-cergy.fr}

\author{Abel Klein\thanksref{AK}}
\thanks[AK]{A.K. was supported in part   by NSF Grant
DMS-0200710.}
\address{University of California, Irvine,
Department of Mathematics,
Irvine, CA 92697-3875,  USA}
\ead[email]{aklein@uci.edu}

\author{Jeffrey H. Schenker\thanksref{JS}}
\thanks[JS]{J.H.S. was supported in part   by an  NSF postdoctoral
 research fellowship.}
\address{ETH Z\"urich,Theoretische Physik,
CH-8093 Z\"urich,
Switzerland}
\ead[email]{jschenker@itp.phys.ethz.ch}


\begin{abstract}
We justify the linear response theory for an ergodic Schr\"odinger operator
with magnetic field within the non-interacting particle approximation, and
derive a Kubo formula for the electric conductivity tensor. To achieve that, we
construct suitable normed spaces of measurable covariant operators where the
Liouville equation can be solved uniquely. If the Fermi level falls into a
region of localization, we recover the well-known Kubo-St\u{r}eda formula for
the quantum Hall conductivity at zero temperature.
\end{abstract}

\begin{keyword}Kubo formula, Kubo-St\u{r}eda formula,
magnetic Schr\"odinger operator.\\
 {2000 \emph{Mathematics Subject Classification.}
Primary 82B44; Secondary  47B80, 60H25.}
\end{keyword}

\end{frontmatter}

\newpage \tableofcontents

\section{Introduction}

In theoretical works, the electric conductivity tensor is usually expressed in
terms of a ``Kubo formula,"  derived via formal linear response theory. The
importance of this  Kubo formula is enhanced by its links with the quantum Hall
conductivity at zero temperature. During the past two decades a few papers
managed to shed some light on these derivations from the mathematical point of
view, e.g., \cite{Pa,Ku,Be,NB,ASS,BESB,SBB1,SBB2,AG,Na,ES,AES}. While a great
amount of attention has been brought to the derivation of the quantum Hall
conductivity from a Kubo formula, and to the study of this conductivity itself,
not much has been done concerning a controlled derivation of the linear
response and the Kubo formula itself; only the recent papers
\cite{SBB2,Na,ES,AES,CJM} deal with this question.

In this article we consider an ergodic Schr\"odinger operator with magnetic
field, and give a controlled derivation of a Kubo formula for the electric
conductivity tensor, validating the linear response theory within the
noninteracting particle approximation. For an adiabatically switched electric
field, we then recover the expected expression for the quantum Hall
conductivity whenever the Fermi energy lies either in a region of localization
of the reference Hamiltonian or in a gap of the spectrum.

To perform our analysis we develop an appropriate mathematical apparatus for
the linear response theory. We first describe several normed spaces of
measurable covariant operators which are crucial for our analysis. We develop
certain analytic tools on these spaces, in particular the trace per unit volume
and a proper definition of the product of two (potentially unbounded)
operators. (Similar spaces and their relevance were already discussed in
\cite{BESB}.) We then use those tools to compute rigorously the linear response
of the system forced by a time dependent electric field. This is achieved in
two steps. First we set up the Liouville equation which describes the time
evolution of the density matrix under the action of a time-dependent electric
field, in a suitable gauge with the electric field given by a time-dependent
vector potential. In a standard way, this evolution equation can be written as
an integral equation, the so-called Duhammel formula. Second, we compute the
net current per unit volume induced by the electric field and prove that it is
differentiable with respect to the electric field at zero field. This yields
the desired Kubo formula for the electric conductivity tensor. We then push the
analysis further to recover the expected expression for the quantum Hall
conductivity, the Kubo-St\u{r}eda formula.

Our derivation of the Kubo formula is valid for any initial density matrix
${\zeta} = f(H)$ with a smooth profile of energies $f(E)$ that has appropriate decay at
high energies. In particular, the Fermi-Dirac distributions at positive
temperature are allowed. At zero temperature, with the Fermi projection
$P^{(E_F)}$ as the initial profile, our analysis is valid whenever the Fermi
energy $E_F$ lies either in a gap of the spectrum
or in a region of localization of the reference Hamiltonian.
 The latter  is actually one of the main
achievements of this article. There is indeed a crucial difference between
$P^{(E_F)}$ with $E_F$ in a gap (or similarly $f(H)$, with $f$ smooth with
decay at high energies) and $P^{(E_F)}$ with $E_F$ in a region of localization:
in the first case the commutator $[x_k,P^{(E_F)}]$ is a bounded operator while
it is unbounded in the second case. Dealing with the unbounded commutator
$[x_k,P^{(E_F)}]$, which appears naturally in the Kubo-St\u{r}eda formula,
forces us to use the full theory of the normed spaces of measurable covariant
 operators we develop.

We now sketch the main points of our analysis. We consider a system of
non-interacting quantum particles in a disordered background, with the
associated one-particle Hamiltonian described by  an ergodic magnetic
Schr\"odinger operator
\begin{equation}  \label{Homega}
H_\omega=  \left(-i\nabla - \A_\omega\right)^2 + V_\omega  \; \; \;
\mathrm{on} \; \; \;
\mathcal{H} :=\mathrm{L}^2(\mathbb{R}^d),
\end{equation}
where the parameter $\omega$ runs in the probability space $(\Omega, \P)$, and
for $\P$-a.e. $\!\omega$ we assign a  magnetic potential $A_\omega$ and an
electric potential $V_\omega$. The precise requirements are described in
Assumption \ref{RandH} of Section~\ref{sectRandomMedia}. Briefly, $A_\omega$
and $V_\omega$ belong to a very wide class of potentials which ensures that
$H_\omega$ is essentially self-adjoint on $\mathcal{C}_c^\infty(\R^d)$ and
uniformly bounded from below for $\P$-a.e. $\omega$. In particular \emph{no
smoothness} assumption is required on $V_\omega$. The probability space
$(\Omega, \P)$ is equipped with an ergodic group $\{\tau(a); \ a \in \Z^d\}$ of
measure preserving transformations. The crucial property of the ergodic
system is that it satisfies a covariance relation:   there exists a unitary
projective representation   $U(a)$ of $\Z^d$ on $\mathrm{L}^2(\mathbb{R}^d)$,
such that for all  $a ,b \in \Z^d$ and $\P$-a.e. $\omega$ we have
\begin{align} \label{covintro}
U(a) H_\omega U(a)^* &= H_{\tau(a) \omega} \, ,\\[.5ex]
U(a) \chi_b U(a)^* &=\chi_{b+a} \, ,
\end{align}
where $\chi_a$ denotes the multiplication operator by the characteristic
function of a unit cube centered at $a$.
 Operators that satisfy the covariance
relation \eqref{covintro} will be called \emph{covariant operators}.
 (See Subsection~\ref{subsecmblecovop}.) If
$\A_\omega=\A$ is the vector potential of a constant magnetic field, the
operators $U(a)$ are the magnetic translations. Note that the ergodic
magnetic Schr\"odinger operator may be random, quasi-periodic, or even periodic.

At time $t=-\infty$, which we take as reference, the system is in equilibrium
in the state given by a one-particle density matrix ${\zeta}_\omega=f(H_\omega)$
where $f$ is a non-negative function with fast enough decay at infinity. At
zero temperature, we have
${\zeta}_\omega=P_\omega^{(E_F)}=\chi_{(-\infty,E_F]}(H_\omega)$, the Fermi
projection. It is convenient to give the technical statement of the condition
on ${\zeta}_\omega$ in the language of the normed spaces developed in
Section~\ref{sectrandom}. Hence we postpone it to Section~\ref{sec:LinResponse}
where it is stated as Assumption \ref{assumptiona}. We note here, however, that
the {\em key} point in that assumption is that
\begin{equation}  \label{assumpIntro}
\E\left\{ \left\| x_k \,{\zeta}_\omega \chi_0\right \|_2^2\right\} < \infty \, ,
\mbox{ or equivalently } \E\left\{ \left\| [x_k, \, {\zeta}_\omega] \chi_0\right
\|_2^2\right\} < \infty \, ,
\end{equation}
for $k=1,\cdots,d$, where  $\|S\|_2$ denotes the Hilbert-Schmidt norm of
the operator $S$.  (This is essentially the condition identified in
\cite{BESB}.)

Of course, if ${\zeta}_\omega = P^{(E_F)}_\omega$ where $E_F$ falls inside a
gap of the spectrum of $H_\omega$, or ${\zeta}_\omega=f(H_\omega) $ with $f$
smooth and appropriately decaying at high energies, then \eqref{assumpIntro} is
readily fulfilled by general arguments (e.g., \cite{GK2}). The main challenge is
to allow for the Fermi energy $E_F$ to be inside a region of localization, as
described for random operators in \cite{AG,GK1,GK3,AENSS}.  Note that  the
existence of these regions of localization has been proven for random Landau
Hamiltonians with Anderson-type potentials \cite{CH,Wa,GK4}, and that
 assumption \eqref{assumpIntro} holds in these regions of localization
\cite{BGK,GK5}.

Under this assumption,  as expected, the current is proved to be zero at
equilibrium (Lemma~\ref{lemequilib}):
\begin{equation}
\T\left\{ \v_{j,\omega} {\zeta}_\omega \right\} = 0, \;\;\;\mbox{$j=1,2,\dots,d$}
\, ,
\end{equation}
where the velocity operator $\v_{j,\omega}$ is the self-adjoint closure of
$i[H_\omega,x_j]$, initially defined on $C_c^\infty(\R^d)$.  Here $\T$ denotes
the trace per unit volume, and reads, for suitable covariant operators
$Y_\omega$ (applying the Birkhoff ergodic theorem),
\begin{equation} \label{tuvintro}
\T(Y_\omega) := \E \left\{\tr  \left\{\chi_0 Y_\omega \chi_0\right\}\right\} =
\lim_{L \to \infty}\,
\textstyle{ \frac 1 {|\Lambda_L|}} \tr\left\{\chi_{\Lambda_L} Y_\omega
\chi_{\Lambda_L}\right\}  \;\;\;\mbox{for $\P$-a.e. $\omega$} \, ,
\end{equation}
where $\Lambda_L$ denotes the cube of side $L$ centered at $0$.

We then slowly, from time $t= -\infty$ to time $t=0$, switch on a spatially
homogeneous electric field $\El$; i.e., we take (with $t_- = \min \, \{t, 0\}$,
$t_+ = \max \, \{t, 0\}$)
\begin{equation} \label{defEintro}
\mathbf{E}(t)=
\mathrm{e}^{\eta t_-}\mathbf{E}\, .
\end{equation}
In the appropriate gauge, the dynamics are now generated by an ergodic
time-dependent Hamiltonian,
\begin{equation}\label{eq:Homegaintro}
    H_\omega(t) = ({-i\nabla} - \A_\omega - \F(t))^2  + V_\omega(x)  =
G(t) H_\omega G(t)^\ast \, ,
\end{equation}
where
\begin{equation} \label{Ftintro}
\F(t) =  \int_{-\infty}^t \mathbf{E}(s) \di s =
\left(\textstyle{\frac{\mathrm{e}^{\eta t_-}} \eta } + t_+  \right) \El \, ,
\end{equation}
and $G(t)=\mathrm{e}^{i\F(t)\cdot x}$ is a gauge transformation on
$\mathrm{L}^2(\R^d)$. (Note that, if $\psi_t$ is a solution of $i\partial_t
\psi_t = H_\omega(t) \psi(t)$ then, at least formally, $$i\partial_t
G^\ast(t)\psi_t = (H_\omega+ \El(t)\cdot x)G^\ast(t)\psi_t \; ,$$ which
represents $\El(t)$ in a more familiar way via a time dependent scalar
potential.  This fact is made precise for weak solutions in
Subsection~\ref{secgauge}.)

It turns out that for all $t$ the operators  $H_\omega(t)$ are self-adjoint
with the common domain $\D=\D(H_\omega)$, and $H_\omega(t)$ is bounded from
below uniformly in $t$.  Thanks to these facts, a general theory \cite[Theorem
XIV.3.1]{Y} of time evolution for time-dependent operators applies: there is a
unique unitary propagator $U_\omega(t,s)$, i.e.,  a unique  two-parameters
family $U_\omega(t,s)$ of unitary operators,  jointly strongly continuous in
$t$ and $s$, and such that $U_\omega(t,r) U_\omega (r,s)=U_\omega(t,r)$,
$U_\omega(r,r)= I$, $U_\omega(t,s) \D = \D $, and $i \partial_t U_\omega(t,s)
\psi = H(t) U_\omega(t,s) \psi$ for all $\psi \in \D$.

A crucial advantage of our choice of gauge is that $H_\omega(t)$ is a
covariant operator for all $t$, which ensures that the unitary propagator
$U_\omega(t,s)$ is also covariant. This is of great importance in calculating
the linear response \emph{outside} the trace per unit volume, taking advantage
of the centrality of this trace, a key feature of our derivation.

To compute the time evolution of the density matrix $\varrho_\omega(t)$, we
shall have to set up and solve the Liouville equation which formally reads
\begin{equation}\label{Liouvilleeq}
\left\{
\begin{array}{l}i\partial_t \varrho_\omega(t) = [H_\omega(t),\varrho_\omega(t)]
\\
\lim_{t \to  -\infty} \varrho_\omega(t)= {\zeta}_\omega
\end{array}
\right.  \, ,
\end{equation}
where ${\zeta}_\omega$ is the initial density matrix at $t=-\infty$. (Thus
${\zeta}_\omega =P^{(E_F)}_\omega$ at zero temperature.) We shall also give a
meaning to the net current per unit volume (area) in the $j$-th direction,
$j=1,\cdots,d$, induced by the electric field, formally   given by
\begin{equation} J_j(\eta,\El;{\zeta}_\omega) = \T(\v_{j,\omega}(0) \varrho_\omega(0)) -
\T(\v_{j,\omega} {\zeta}_\omega) = \T(\v_{j,\omega}(0) \varrho_\omega(0)),
\end{equation}
with $\v_{j,\omega}(0)$,  the self adjoint closure of $i[H_\omega(0),x_j]$
defined on $C_c^\infty(\R^d)$, being the velocity operator in the
$j$-th direction at time $t$.
Note that $\v_{j,\omega}(0) = G(0)\v_{j,\omega}G(0)^\ast =
\v_{j,\omega}-2\F_j(0)$.

We remark that there is an alternative approach \cite{ES,AES} to a derivation
of the Kubo-St\u{r}eda formula for the quantum Hall current in a two
dimensional sample, based on the calculation of a conductance rather than a
conductivity. Conductance is the linear response coefficient relating a current
to the electric potential difference, whereas conductivity relates a current
density to the electric field strength. In \cite{ES,AES} the effect of a
finite potential drop is analyzed by considering the effect of adding to the
Hamiltonian a term $g(t) \Lambda_1$ with $g(t)$ a time dependent scalar
coupling and $\Lambda_1(x)=\Lambda_1(x_1) \rightarrow \pm 1$ as $x_1
\rightarrow \pm \infty$ a smooth switch function. This term models the effect
of a modulated (in time) potential difference between the left and right edges
of a physical sample, with the edges formally considered to be located at
$x_1=\pm \infty$. With $g(t)$ of the form $g(t) = \phi(t/\tau)$ with $\phi$ a
fixed function, an expression for the net current across the line $x_2=0$  has
been derived, which in the adiabatic ($\tau \rightarrow \infty$) limit gives
the corresponding Kubo-St\u{r}eda formula  for continuum operators
with a gap condition \cite{ES} and  for discrete operators with a localization
assumption \cite{AES}.

Let us now briefly describe the normed spaces of measurable covariant operators
 we construct to
carry out this derivation -- see Section~\ref{sectrandom} for their full
description. We let  $\Hc$ denote the subspace of
functions with compact support, and set $\Ll= \Ll (\Hc,\H)$ to be the vector
space of linear operators on  $\H$ with domain  $\Hc$ (not necessarily
bounded). We introduce the vector space $\K_{mc}$ of measurable  covariant
maps $Y_\omega\colon \Omega \to \Ll$; where we identify maps
that agree $\P$-a.e. We consider the $C^*$-algebra
\begin{equation}
\K_{\infty}= \left\{  {Y_\omega} \in \K_{mc}; \;   \tnorm{{Y_\omega}}_\infty<\infty   \right\} ,
\mbox{ where }
\tnorm{{Y_\omega}}_\infty =  \|\, \|{Y_\omega} \| \, \|_{\mathrm{L}^{\infty}(\Omega, \P)} \, .
\end{equation}
Bounded functions of $H_\omega(t)$ as well as the unitary operators
$U_\omega(t,s)$ belong to this algebra.

However, since we must deal with unbounded operators (think of
$[x_k,P^{(E_F)}_\omega ]$ with $E_F$  in
 a region of localization), we must look outside $\K_\infty$ and consider
subspaces of $\K_{mc}$ which include unbounded operators.   We introduce norms
on $\K_\infty$ given by
\begin{equation}
\tnorm{{Y_\omega}}_1 = \E\, \tr \{\chi_0 |{Y_\omega}| \chi_0\}, \;\;\;
 \tnorm{Y_\omega}_2 = \left\{\E\, \|{Y_\omega}\chi_0\|_2^2\right\}^{\frac12},
\end{equation}
and consider the normed spaces
\begin{equation}
\K_{i}^{(0)} = \{ {Y_\omega}\in\K_{\infty}, \, \tnorm{{Y_\omega}}_i<\infty \}\,
,\;\; i=1,2\, .
\end{equation}
We denote the (abstract) completion of $\K_i^{(0)}$ in the norm
$\tnorm{\cdot}_i$ by $\overline {\K_i}$, $i=1,2$. In principle, elements of the
completion $\overline {\K_i}$ may not be identifiable with elements of
$\K_{mc}$: they may not be {\em covariant operators} defined on the domain
$\Hc$. Since it is important for our analysis that we work with operators, we
set $\K_i = \K_{mc} \cap \overline{ \K_i}$. That is,
\begin{equation}
  \K_i = \{ Y_\omega \in \K_{mc} , \, \tnorm{Y_\omega}_i < \infty \} \; .
\end{equation}
(We are glossing over the technical, but important, detail of defining the
norms $\tnorm{Y_\omega}_i$ on $\K_{mc}$.  In fact, we shall do this only for
\emph{locally bounded}  operators $Y_\omega$ -- see
Definition~\ref{mclbdefn}(iii) -- for which the absolute value $|Y_\omega|$ may be
defined.)

It turns out that $\K_2=\overline {\K_2}$ (Proposition~\ref{propK2}), and the
resulting set is a Hilbert space with inner product  $\la\la
{Y_\omega},Z_\omega\ra\ra=\E \,\tr\{({Y_\omega}\chi_0)^\ast(Z_\omega\chi_0)\}$.
However, $\K_1 \neq \overline {\K_1}$ (Proposition~\ref{notcomplete}), and the
dense subspace $\K_1$ is not complete.  Nonetheless, it represents a natural
space of unbounded  covariant operators on which the trace per unit volume
\eqref{tuvintro} is well defined. The trace per unit volume $\T$ is naturally
 defined on $\K_1$,
where it is bounded by the $\K_1$ norm, and hence it
extends to a continuous linear functional on $\overline {\K_1}$; but
\eqref{tuvintro} is only formal for $Y_\omega \in \overline {\K_1} \setminus
\K_1$.

There is a natural norm preserving conjugation on the spaces $\K_i$, given by
 ${Y_\omega}^\ddagger = ({Y_\omega}^\ast)_{|\H_c}$, which extends to a
conjugation on $\overline {\K_1}$. Moreover, the spaces $\K_i$, $i=1,2$, are left
and right $\K_\infty$-modules, with left and right  multiplication being
explicitly defined for $B_\omega\in\K_\infty$ and ${Y_\omega}\in\K_2$ or $\K_1$
by
\begin{equation} \label{leftrightM}
B_\omega \odot_L {Y_\omega} = B_\omega  {Y_\omega}\,,  \;\;\;
{Y_\omega} \odot_R B_\omega = (B_\omega^\ast \odot_L {Y_\omega}^\ddagger)^\ddagger =
{Y_\omega}^{\ddagger\ast} B_\omega\,  .
\end{equation}
(It is not obvious that the latter equality makes sense!)  The properties of
left and right multiplication, as well as the fact that they commute, can be
read immediately from \eqref{leftrightM}. There is also a bilinear map
$\Diamond: \K_2\times\K_2\to \overline {\K_1}$ with dense range, written
$\Diamond({Y_\omega},Z_\omega) = {Y_\omega} \diamond Z_\omega$,
such that $\T(
{Y_\omega} \diamond Z_\omega)=\la\la {Y_\omega}^\ddagger,Z_\omega\ra\ra$.

Another crucial ingredient is the centrality of the trace per unit volume: if
either  ${Y_\omega},Z_\omega\in \K_2$ or $Y_\omega\in\overline {\K_1}$ and
$Z_\omega \in\K_\infty$, we have either
\begin{equation}
\T ( {Y_\omega} \diamond Z_\omega) = \T ( Z_\omega \diamond {Y_\omega})
\quad \mbox{or} \quad
\T ( Y_\omega \odot_R Z_\omega) = \T ( Z_\omega \odot_L Y_\omega) \, .
\end{equation}

There is a connection with noncommutative integration:  $\K_\infty$ is a
von Neumann algebra, $\T$ is a faithful normal semifinite trace on $\K_\infty$,
$\overline {\K_i}= \L^i (\K_\infty, \T)$ for $i=1,2$ -- see Subsection~\ref{noncom}.
But our explicit construction plays a very important role in our analysis.

The Liouville equation \eqref{Liouvilleeq} will be given a precise meaning and
solved in the spaces $\K_1$ and $\K_2$.  Note that the assumption
\eqref{assumpIntro} is equivalent to  $\left[x_j, {\zeta}_\omega\right]\in
\K_2$ for all $j=1,2,\ldots,d$. (We will also have $\left[x_j,
{\zeta}_\omega\right]\in \K_1$ for all $j=1,2,\ldots,d$. See Remark (i)
following Assumption \ref{assumptiona}, and Proposition~\ref{fHK}.)

If ${Y_\omega}\in\K_i$, $i=1,2,\infty$, is such that
 $\mathrm{Ran}\,{Y_\omega}\subset
\D=\D(H_\omega(t))$ and $H_\omega(t){Y_\omega}\in\K_i$, and similarly
for ${Y_\omega}^\ddagger$, we set
$$ [H_\omega(t),{Y_\omega}]_\ddagger = H_\omega(t){Y_\omega} -
(H_\omega(t){Y_\omega}^\ddagger)^\ddagger \in\K_i \, .
$$
Our first main result is
\begin{theorem} \label{thmrhointro}
Under Assumptions \ref{RandH} and \ref{assumptiona},  the Liouville equation
\begin{equation}
\left\{ \begin{array}{l}i\partial_t \varrho_\omega(t) =
 [H_\omega(t),\varrho_\omega(t)]_\ddagger\\
\lim_{t \to  -\infty} \varrho_\omega(t)= {\zeta}_\omega
\end{array}\right.  \label{dynamicsintro} \,
\end{equation}
 has a solution in $\K_1\cap \K_2$, unique  in both $\K_1$ and $\K_2$,
 given by
\begin{align}
\varrho_\omega(t) &=\lim_{s \to -\infty}{ \U}(t,s)\left( {\zeta}_\omega
\right) = \lim_{s \to -\infty}{ \U}(t,s)\left( {\zeta}_\omega(s) \right)
\\
&= {\zeta}_\omega(t)
- i
 \int_{-\infty}^t \mathrm{d} r \,\mathrm{e}^{\eta {r_{\! -}}}{ \,\U}(t,r) \left([ \mathbf{E}
\cdot \x, {\zeta}_\omega(r) ]\right) \, ,\label{defrho3intro}
\end{align}
where
\begin{align}{ \U}(t,s)(Y_\omega)&=
 U_\omega(t,s)\odot_L Y_\omega \odot_R U_\omega(s,t)
 \;\; \mbox{for
$Y_\omega\in\K_i$, $i=1,2$}\, ,
\\ \label{Ptintro}
{\zeta}_\omega(t)& = G(t) {\zeta}_\omega G(t)^*  = f(H_\omega(t))
\qquad ({\zeta}_\omega= f(H_\omega)) \, .
\end{align}
We  also have
\begin{equation}
\varrho_\omega(t) ={ \U}(t,s) (\varrho_\omega(s))\, ,
\;\;\tnorm{\varrho_\omega(t)}_i=\tnorm{{\zeta}_\omega}_i \, ,
\end{equation}
for all $t,s$ and $i=1,2, \infty$. Furthermore, $\varrho_\omega(t)$ is
non-negative and if ${\zeta}_\omega = P^{E_F}_\omega$ then $\varrho_\omega(t)$
is an orthogonal projection for all $t$.
\end{theorem}

We actually prove a generalization of  Theorem~\ref{thmrhointro}, namely
Theorem~\ref{thmrho}, in which the commutator in \eqref{dynamicsintro} is
replaced by the Liouvillian (defined in Corollary~\ref{Liouvillian}), the
closure of $Y_\omega \mapsto [H_\omega(t),{Y_\omega}]_\ddagger$ as an operator
 on  $\K_i$, $i=1,2$.
As a by-product of the theorem, we prove that
 $\mathrm{Ran} \,\varrho_\omega(t)\in\D$ and $\v_{j,\omega}(t)
\varrho_\omega(t)\in\K_1$, and hence the current $\T(\v_{j,\omega}(t)
\varrho_\omega(t))$ is well-defined for any time $t$. In particular, the net
current per unit volume $J_j(\eta,\El;{\zeta}_\omega)$ is well defined and,  since
 $\varrho_\omega(t)$ is non-negative,  a real number.

Our next main contribution states the validity of the linear response theory,
and provides a Kubo formula.

\begin{theorem}  Let $\eta > 0$. Under
Assumptions \ref{RandH} and \ref{assumptiona},
the map $\El\to \J(\eta,\El;{\zeta}_\omega)$ is differentiable with respect to $\El$ at
$\El=0$ and the derivative $\sigma(\eta;{\zeta}_\omega)$ is given by
\begin{align}\label{sigmajkintro}
\sigma_{jk}(\eta;{\zeta}_\omega) =\textstyle{ \frac{\partial }{\partial
\El_k}}\J_j(\eta,0;{\zeta}_\omega)
= - \T \left\{   \int_{-\infty}^0  \mathrm{d} r\,
\mathrm{e}^{\eta r}
  \v_{j,\omega} \, \U^{(0)}(-r) \left(i [  x_k, {\zeta}_\omega ] \right)  \right\} ,
\end{align}
where $\, \U^{(0)}(r)(Y_\omega)=\mathrm{e}^{-irH_\omega} \odot_L Y_\omega \odot_R
\mathrm{e}^{irH_\omega}$.
\end{theorem}

 Note that we prove a result stronger than the existence of the
partial derivatives of $\J(\eta,\El;{\zeta}_\omega)$ at $\El=0$: we prove
differentiability at $\El=0$.

Next, taking the limit $\eta\to 0$, we recover the expected form for the
quantum Hall conductivity at zero temperature, the
 Kubo-St\u{r}eda formula  (e.g., \cite{St,TKNN,Be,NB,BESB,AG,Na}).

\begin{theorem}\label{thmHallintro}
Under Assumptions \ref{RandH} and \ref{assumptiona}, if ${\zeta}_\omega =
P_\omega^{(E_F)}$,  an orthogonal projection, then for all $j,k=1,2,\ldots,d$, we have
\begin{align}\label{expHallintro}
\sigma_{j,k}^{(E_F)}\! := \lim_{\eta \rightarrow
0}\sigma_{jk}(\eta;P_\omega^{(E_F)}) = -i \T \left\{ P^{(E_F)}_\omega
\odot_L\left[ \left[x_j, P^{(E_F)}_\omega \right]\!, \!\left[x_k,
P^{(E_F)}_\omega
 \right]\right]_\diamond
\right\}\! ,
\end{align}
where $[Z_\omega,Y_\omega]_\diamond= Z_\omega \diamond Y_\omega - Y_\omega
\diamond Z_\omega \in \overline{\K_1}$ if $Z_\omega,Y_\omega\in\K_2$. As a consequence,
the conductivity tensor is antisymmetric; in particular the direct conductivity
is zero in all directions, i.e.,
 $\sigma_{j,j}^{(E_F)} =0$ for $j=1,2,\ldots,d$.
\end{theorem}

If the system is time-reversible the conductivity is zero in the region of localization,
 as expected.

\begin{corollary} Under Assumptions \ref{RandH} and \ref{assumptiona},
if $\A_\omega=0$ (no magnetic field), we have $\sigma_{j,k}^{(E_F)} =0$ for all
$j,k=1,2,\ldots,d$.
\end{corollary}

We remark that under Assumptions  \ref{RandH} and \ref{assumptiona} $\left[
\left[x_j, P^{(E_F)}_\omega \right]\!, \!\left[x_k, P^{(E_F)}_\omega
 \right]\right]_\diamond$ is an element of $\overline{\K_1}$, but may not be in
 $\K_1$. (That is, it may not be representable as a covariant operator with
 domain $\H_c$). In particular, the product $\odot_L$ in \eqref{expHallintro}
 is defined via approximation from $\K_1$ and may not reduce to an ordinary
 operator product.  However, under a stronger localization assumption such as
\begin{equation}\label{strongerassump}
  \E  \left\{\|\chi_x {P^{(E_F)}_\omega} \chi_y \|_2^2 \right\}
 \le \ C \e^{-|x-y|^\alpha}
  \; ,
\end{equation}
which holds throughout the regime in which \eqref{assumpIntro} has been
verified for random Schr\"odinger operators \cite{BGK,GK5}, the products in \eqref{expHallintro} reduce to
ordinary products of (unbounded) operators, and we have
\begin{equation}
    \sigma_{j,k}^{(E_F)}  = -i \T \left\{ P^{(E_F)}_\omega
\left[ \left[x_j, P^{(E_F)}_\omega \right]\!, \!\left[x_k, P^{(E_F)}_\omega
 \right]\right]
\right\} .
\end{equation}

There are several reasons for using  \eqref{assumpIntro} as the key assumption in
 this paper. As discussed in \cite{GK3}, the stronger assumption
\eqref{strongerassump} holds in a region of very strong localization
for random Schr\"odinger operators, analogous
to the region of complete analyticity in classical statistical mechanics.  It is known
that the latter may not hold all the way to the critical point; there are examples where
the single phase region has a transition from complete analyticity at very high
temperatures to another single phase region with fast decay of correlation functions.
The analogy with classical statistical mechanics indicates the possibility
of a weaker localization region, where   \eqref{assumpIntro} may hold, but not
\eqref{strongerassump}.  (In fact \eqref{strongerassump} is equivalent to
being in the region of applicability of the multiscale analysis \cite{GK5}.)
Moreover, the results of this paper apply to ergodic
magnetic Schr\"odinger operators which may be  quasi-periodic or periodic,
 not just random, and for which one may not expect \eqref{strongerassump}.
In addition, note that the use of
 \eqref{strongerassump} as an assumption would not
simplify significantly  the proofs; the normed spaces $\K_1$ and $\K_2$ appear naturally
in linear response theory, and  \eqref{assumpIntro},  which simply states that the relevant
commutators are in $\K_2$, is the natural condition for deriving the linear response
 theory,  as in  \cite{BESB}.


\section{Magnetic and time-dependent electromagnetic Schr\"odinger operators}
\label{sectoperator} \setcounter{equation}{0}
In this section we review some well known facts about
Schr\"odinger operators incorporating a magnetic vector potential
$\A$, and present a basic existence and uniqueness result for
associated propagators in the presence of a time-dependent
electric field.
\subsection{Magnetic Schr\"odinger operators}
\label{subsectMSO}

Let
\begin{equation}  \label{MSO}
H= H(\A,V) = \left(-i\nabla - \A\right)^2 + V  \; \; \; \mathrm{on} \; \; \;
\mathrm{L}^2(\mathbb{R}^d),
\end{equation}
where the magnetic potential $\A$ and the electric potential $V$
 satisfy the Leinfelder-Simader conditions:
\begin{itemize}
\item   $\A(x) \in \mathrm{L}^4_{\mathrm{loc}}(\R^d; \R^d)$  with
$\nabla \cdot \A(x) \in \mathrm{L}^2_{\mathrm{loc}}(\R^d)$.

\item  $V(x)= V_+(x) - V_-(x)$ with
 $V_\pm(x) \in \mathrm{L}^2_{\mathrm{loc}}(\R^d)$, $V_\pm(x) \ge 0$,
 and
$ V_-(x)$  relatively bounded with respect to
$\Delta$ with relative bound $<1$, i.e., there are $0 \le\alpha  < 1$ and $\beta \ge 0$
such that
\begin{equation}\label{relbound}
 \|V_-\psi\| \leq
\alpha \| \Delta \psi \| + \beta \|\psi \| \quad \mbox{for all $\psi \in\D(\Delta)$}.
\end{equation}
\end{itemize}
  Leinfelder and Simader have shown that $ H(\A,V)$
 is essentially self-adjoint on $\Cc^\infty(\R^d)$
 \cite[Theorem 3]{LS} (see also
 \cite[Theorem 1.15]{CFKS},
\cite[Theorem B.13.4]{Si2}), with
\begin{equation}
 H \psi = - \Delta \psi + 2 i  {\A}\cdot\nabla \psi + \left(i \nabla\cdot {\A}  + {\A}^2 +V\right) \psi
\;\; \mbox{for $\psi \in \Cc^\infty(\R^d)$}.
                       \label{expression0}
\end{equation}

Note that \eqref{relbound} implies  that  for all $\alpha^\prime > \alpha$ we
have
 \cite[Proof of Theorem X.18]{RS2}
\begin{equation}\label{relbound3599}
0\le \la \psi, V_-\psi\ra\le
\alpha^\prime \la \psi,-\Delta\psi\ra +
\textstyle{\frac{\alpha^\prime}{\alpha^\prime- \alpha}}\beta \|\psi \|^2 \, .
\end{equation}
A similar bound holds for $ H({\A},V_+)$ \cite[Eq. (4.11)]{LS}:
  for all $\alpha^\prime > \alpha$ we have
\begin{equation}\label{relbound2}
 \|V_-\psi\| \leq
\alpha^\prime \| H({\A},V_+) \psi \| +
\textstyle{\frac{\alpha^\prime}{\alpha^\prime- \alpha}}\beta \|\psi \|
\quad \mbox{for all $\psi \in\D(H({\A},V_+))$} \,,
\end{equation}
from which we immediately get the lower bound \cite[Theorem V.4.11]{Ka}\cite[Theorem X.12]{RS2}
\begin{equation} \label{lowerboundbad}
H({\A},V) \ge
- \min_{ \alpha^\prime \in (\alpha,1)}
 {\frac{\alpha^\prime \beta}{(\alpha^\prime- \alpha)(1 - \alpha^\prime)}}
= -\, \frac \beta {(1 - \sqrt{\alpha})^2} \,  .
\end{equation}

But we can get a better lower bound.  We have the a.e. pointwise
inequality
\cite[Proof of Lemma 2]{LS} \cite{BeG}
\begin{equation}\label{diamnabla}
\left| \nabla (|\psi|) \right| \le \left|  \left(-i\nabla - \A\right) \psi \right|
\quad \mbox{for all $\psi \in\Cc^\infty(\R^d)$} \, .
\end{equation}
Thus it follows for all $\alpha^\prime > \alpha$ that we have
(using \eqref{relbound3599})
\begin{align}\label{relbound35}
\la \psi, V_-\psi \ra  &\le  \la |\psi|, V_-|\psi| \ra \le \alpha^\prime \la
|\psi|,-\Delta |\psi|\ra +
\textstyle{\frac{\alpha^\prime}{\alpha^\prime- \alpha}}\beta ||\psi ||^2\\
\nonumber
&  =
\alpha^\prime  \left\| \nabla |\psi| \right\|^2+
\textstyle{\frac{\alpha^\prime}{\alpha^\prime- \alpha}}\beta ||\psi ||^2
\le
\alpha^\prime  \left\|  \left(-i\nabla - \A\right)\psi \right\|^2+
\textstyle{\frac{\alpha^\prime}{\alpha^\prime- \alpha}}\beta ||\psi ||^2\\
& \nonumber  \le
\alpha^\prime \la \psi,H(\A,V_+)\psi\ra +
\textstyle{\frac{\alpha^\prime}{\alpha^\prime- \alpha}}\beta ||\psi ||^2
\end{align}
for all $\psi \in\Cc^\infty(\R^d)$. We conclude that
\begin{equation} \label{lowerbound}
H({\A},V) \ge
- \min_{ \alpha^\prime \in (\alpha,1)}
 {\frac{\beta}{(\alpha^\prime- \alpha)}}
= -\, \frac \beta {(1 - {\alpha})} \,  .
\end{equation}
For convenience we write
\begin{equation}\label{gamma}
\gamma = \gamma(\alpha, \beta) := {\frac \beta {1 - {\alpha}}} +1 \, ,
\end{equation}
and note that
\begin{equation} \label{gammalowerbound}
H + \gamma \ge 1  \; .
\end{equation}

We  also have the diamagnetic inequality
\begin{equation} \label{diamag}
\left| \e^{-tH({\A},V)}\psi\right|  \le  \e^{-tH({0},V)}|\psi|
\end{equation}
for all $\psi \in \mathrm{L}^2(\R^d)$ and  $t >0$, see \cite[Proof of Theorem
1.13]{CFKS}. Note that the  diamagnetic inequality  and \eqref{lowerbound}
imply (using $\int_0^\infty t^q \e^{-t(x+\lambda)} {\rm d} t = \Gamma(q)
(x+\lambda)^{-q}$)
\begin{equation}  \label{diamag2}
\left| \left(H({\A},V) + \lambda\right)^{-q}\psi\right|\le
\left(H({0},V) + \lambda\right)^{-q}|\psi|
\end{equation}
for all $\psi \in \mathrm{L}^2(\R^d)$,
$\lambda >\textstyle{\frac \beta {(1 - {\alpha})}}$, and $q>0$.

An important consequence of \eqref{diamag2} is that the usual trace estimates
for $-\Delta + V$ are valid for the magnetic Schr\"odinger operator
$H({\A},V)$, with bounds independent of $\A$ and depending on $V$ only through
$\alpha$ and $\beta$. We state them as in \cite[Lemma A.4]{GK3}. (We do not
need the Leinfelder-Simader conditions here, just the conditions for the
diamagnetic inequality:  $\A(x) \in \mathrm{L}^2_{\mathrm{loc}}(\R^d; \R^d)$,
$V_+(x) \in \mathrm{L}^1_{\mathrm{loc}}(\R^d; \R^d)$, and $ V_-(x)$ relatively
form bounded with respect to $\Delta$ with relative bound $<1$. See
\cite[Theorem 1.13]{CFKS} where this is shown for $V_- =0$.  The general case,
with $V_-$ relatively bounded as above, is proved by an approximation argument,
see \cite[Theorems 7.7, 7.9]{F}.)   We use the notation
 $\langle x \rangle= \sqrt{1 + |x|^2}$ throughout this paper.

\begin{proposition}\label{ltrace} Let $\nu > \frac d 4$.   There is a finite
constant $\mathcal{T}_{\nu, d, \alpha,\beta}$,
depending only on the indicated constants, such that
\begin{equation} \label{trSGEE}
\mathrm{tr}\left\{ \left\langle x \right\rangle^{-2\nu}
 \left(  H({\A},V) +\gamma \right)^{-2[[\frac d 4]]}
\left\langle x \right\rangle^{-2\nu}\right\} \le
\mathcal{T}_{\nu, d, {\alpha},\beta}
 \, ,
\end{equation}
where $[[\frac d 4]]$ is the smallest integer bigger than $ \frac d 4$ and $\gamma$ is the
constant defined in \eqref{gamma}. Thus, letting
\begin{equation}\label{PhiE}
\Phi_{d,{\alpha}, \beta} (E) = \chi_{\left[-\frac \beta {1 - {\alpha}},\infty\right)} (E)
 \left(  E + \gamma \right)^{2[[\frac d 4]]} \, ,
\end{equation}
we have
\begin{equation} \label{trSGEE2}
\mathrm{tr}\left( \left\langle x \right\rangle^{-2\nu}
f(H)
\left\langle x\right\rangle^{-2\nu}\right) \le
\mathcal{T}_{\nu, d,{\alpha},\beta}
 \|f \Phi_{d,{\alpha},\beta}\|_\infty < \infty
\end{equation}
 for every Borel measurable function $f\ge 0$ on the real line.
\end{proposition}

\begin{proof}  The proposition follows once the estimate \eqref{diamag2}
is converted into an estimate on traces, because then the well known trace
estimates for $-\Delta +V$, e.g., \cite[Lemma A.4]{GK3}, finish the argument.
Hence \eqref{trSGEE} follows from the following lemma, with
\begin{equation}\begin{split}
A &= \left\langle x \right\rangle^{-2\nu}
 \left( H({\A},V) +
\gamma \right)^{-2[[\frac d 4]]}
\left\langle x \right\rangle^{-2\nu}\, ,\\
B &= \left\langle x \right\rangle^{-2\nu}
 \left( H({0},V) +
\gamma \right)^{-2[[\frac d 4]]}\left\langle x
\right\rangle^{-2\nu} \, ,
\end{split}\end{equation}
using the  fact that the operator $ \left( H({0},V) +
\gamma \right)^{-2[[\frac d 4]]}$ is positivity preserving.
\end{proof}

\begin{lemma}  Let $A$ and $B$ be bounded positive operators on
$\L^2(\R^d)$, with $B$ a positivity preserving operator, such that
\begin{equation} \label{diatr}
\la \psi, A \psi \ra  \le \la |\psi|, B |\psi| \ra
\;\;\; \mbox{for all $\psi \in \L^2(\R^d)$.}
\end{equation}
Then \ $\tr\,  A \le \tr \, B$.
\end{lemma}

\begin{proof}First note that the lemma is obvious if we replace $\L^2(\R^d)$ by $\ell^2(\Z^d)$,
since in this case we have a basis of positive functions
($|\delta_x| = \delta_x $). Note also that we may assume  $\tr \,
B<\infty$ without loss of generality.

For $\L^2(\R^d)$, let $\H_n$ be the sub-Hilbert space with ortho-normal basis
$$\{\tilde{\chi}_{n,x}=2^{\frac{nd}2} \chi_{\Lambda_{2^{-n}} (2^{-n} x)} ; \; x \in \Z^d\},$$
where $\Lambda_L(x)$ denotes the cube centered at $x$ and of length $L$; and
let $P_n$ be the orthogonal projection onto $\H_n$.  Note that
$P_n$ is positivity preserving.   Set
\begin{equation}
A_n = P_n A P_n \quad \text{and} \quad
B_n = P_n B P_n
\end{equation}
It follows from \eqref{diatr} and the fact that both $B$ and $P_n$ are
 positivity preserving  that
 \begin{equation} \label{diatr2}
\la \psi, A_n \psi \ra \le \la |P_n \psi|, B |P_n\psi| \ra \le \la |\psi|, B_n |\psi| \ra
\;\;\; \mbox{for all $\psi \in \H_n$.}
\end{equation}
Since $\H_n$ has a basis of positive functions, we get
\begin{equation}
 \tr A_n \le \tr B_n \le  \tr B\, .
\end{equation}
Thus $\sqrt{A}P_n$ is Hilbert-Schmidt, and it follows that
\begin{equation}
\tr \sqrt{A} P_n\sqrt{A} \le \tr {B}\, .
\end{equation}
Since $P_n \to I$ strongly, we conclude that $\tr\,  A \le \tr \, B$.
\end{proof}

The velocity operator $\mathbf{v} = i[H,\x  ]$, where $\x$ is the operator from
$\mathrm{L}^2(\R^d)$ to $\mathrm{L}^2(\R^d; \C^d)$ of multiplication by the
coordinate vector $x$, plays an important role in the linear response theory.
To give precise meaning to $\v$, we note that on $\Cc^\infty(\R^d)$ we have
\begin{equation}
  i[H,\x  ] = 2(-i\nabla - \A) \, .
\end{equation}
We let $\Db=\Db(\A)$ be the closure of $ (-i\nabla - \A)$ as an
operator from $\mathrm{L}^2(\R^d)$ to $\mathrm{L}^2(\R^d; \C^d)$
with domain $\Cc^\infty(\R^d)$. Each of its components $\Db_j=
\Db_j(\A) =  (-i \frac \partial {\partial x_j} - \A_j)$,
$j=1,\ldots,d$, is essentially self-adjoint on $\Cc^\infty(\R^d)$
since $\A(x) \in \mathrm{L}^2_{\mathrm{loc}}(\R^d; \R^d)$ (see
\cite[Lemma 2.5]{Si1}). We define
\begin{equation}\label{velocity}
\v = \v(\A) = 2 \Db(\A) \ .
\end{equation}

\begin{proposition} \label{propDA}
We have
\begin{description}
\item[(i)] $\D(\sqrt{H +\gamma}) \subset \D(\Db)$. In fact there exists $C_{\alpha,\beta} <
\infty$ such that
\begin{equation} \label{MH}
\left \| {\Db} \left ( H + \gamma \right )^{-\frac 1 2} \right \| \ \le \
C_{\alpha,\beta}.
\end{equation}

\item[(ii)] For all $\chi \in \Cc^\infty(\R^d)$ we have $\chi \D(H) \subset \D(H)$ and
\begin{equation}\label{chiH}
H \chi\psi = \chi H \psi - (\Delta\chi)\psi - 2i (\nabla \chi )\cdot {\Db}\psi
\quad \mbox{for all $\psi \in\D(H)$}.
\end{equation}

\item[(iii)] Let
\begin{equation}\label{PhitE}
\widetilde \Phi_{d,\alpha,\beta}(E)  \ := \ (E+\gamma)^{\frac{1}{2}}
\Phi_{d,\alpha,\beta}(E) \ = \ \chi_{\left[- \frac \beta {1 -
{\alpha}},\infty\right)} (E)
 \left(  E + \gamma \right)^{2[[\frac d 4]] + \frac{1}{2}} \; .
\end{equation}
If  $f$ is Borel measurable function  on the real line with
  $\| f \widetilde \Phi_{d,\alpha,\beta}\|_\infty < \infty$, the bounded operator
 $|\Db f(H)|=\left\{\overline{f}(H) \Db^*\Db f(H)\right\}^{\frac 1 2}$
satisfies
\begin{equation}
\tr \left\{\langle x \rangle^{-2 \nu} \left | \Db f(H) \right | \langle x \rangle^{-2
\nu} \right\} \ \le \
  \widetilde {\mathcal T}_{\nu, d,{\alpha},\beta},
\end{equation}
where $\widetilde {\mathcal T}_{\nu, d,{\alpha},\beta} < \infty$ for $\nu
> d/4$ and depends only on the indicated constants.
\end{description}
\end{proposition}

\begin{proof}
To prove (i), note that $\Db^* \Db =(-i \nabla - \A)^2$ and by
\eqref{relbound35}
\begin{equation}
  \delta \alpha' \Db^* \Db \ \le \ (1+\delta)\alpha'(-i \nabla - \A)^2 - V_-
  + \frac{\alpha'}{\alpha - \alpha'} \beta \ \le \ H
  + \frac{\alpha'}{\alpha - \alpha'} \beta
\end{equation}
for $\alpha' \in (\alpha,1)$ and $\delta$ such that
 $(1+ \delta) \alpha' \le 1$.
Choosing $\alpha'$ and $\delta$ such that
\begin{equation}
  \frac{\alpha'}{\alpha - \alpha'} \beta = \gamma  \quad \text{and }
\quad  (1+\delta) \alpha' = 1
   \; ,
\end{equation}
we have
\begin{equation}\label{Db2<Hg}
  (1-\alpha') \Db^*\Db \le \ H + \gamma
\end{equation}
as quadratic forms. Since $\alpha'=\alpha'(\alpha,\beta)$ is strictly less than
one, it follows that
 $\D(\sqrt{H +\gamma}) \subset \D(\Db)$ and furthermore
\begin{equation}
   \left ( H + \gamma \right )^{-\frac{1}{2}} \Db^* \Db
   \left ( H + \gamma \right )^{-\frac{1}{2}} \ \le \ \frac{1}{1 - \alpha'} \; ,
\end{equation}
which gives \eqref{MH} with $C_{\alpha,\beta} = \sqrt{\frac 1{1-\alpha'}}\,$.

Part (ii) follows from \eqref{MH}, since the identity holds for $\psi \in
C^\infty_c$ by \eqref{expression0}. Part (iii) is a result of combining
Proposition \ref{ltrace}, and the estimate
\begin{equation}
\left | \Db f(H) \right | \ \le \ C_{\alpha,\beta} ( H+ \gamma )^{\frac 1 2}
|f|(H) \; ,
\end{equation}
which follows from \eqref{Db2<Hg} and monotonicity of the square root.
\end{proof}

We shall also need to consider commutators $[x,f(H)]$ with functions of $H$.
For smooth functions, the easiest way to do this is to use the
Helffer-Sj\"ostrand formula  \cite{HS,D}. Specifically, we restrict our
attention to functions which are finite in one of the following norms:
\begin{equation} \label{sdfn}
|\!|\!|f|\!|\!|_m = \sum_{r=0}^m \int_{\mathbb{R}} |f^{(r)}(u)|\langle u
\rangle^{r-1} \mathrm{d}u \;\; , \;\;\; m=1,2,\ldots \, .
\end{equation}
If $|\!|\!|f|\!|\!|_m < \infty$ with $m \ge 2$, then we have \cite{HS,D}
\begin{equation}\label{HS}
 f (H) = \int d\tilde{f}(z) (z-H)^{-1} \, ,
\end{equation}
where the integral converges absolutely in operator norm:
\begin{equation}\label{HSnormbound}
  \|f(H)\| \ \le \ \int |d\tilde{f}(z)| \frac{1}{\mathrm{Im}\,  z} \ \le \ c \
  |\!|\!|f|\!|\!|_m < \infty \; ,
\end{equation}
with $c$ independent of $m \ge 2$. Here $z= x + iy$,  $\tilde{f}(z)$ is an
\emph{almost analytic} extension of $f$ to the complex plane,  and
$d\tilde{f}(z)  = - \frac 1 {2\pi}\partial_{\bar{z}}\tilde{f}(z) \,\mathrm{d}
x\, \mathrm{d} y $, with $\partial_{\bar{z}}= \partial_x + i \partial_y$. For
various purpose it is useful to note that
\begin{equation}\label{HShigherorder}
  \int |d\tilde{f}(z)| \frac{\langle \mathrm{Re}\, z \rangle^{p-1}}
    {|\mathrm{Im}\, z|^p} \ \le \ c_p \
  |\!|\!|f|\!|\!|_m < \infty \; ,
\end{equation}
for $m \ge p+1$. (See \cite[Appendix B]{HuS} for details.) Note that if $f \in
\S(\R)$ we have $|\!|\!|f|\!|\!|_m < \infty$ for all  $ m=1,2,\ldots\,$.
We recall that  $\H_c$  denotes the dense linear subspace of functions with
compact support.

\begin{proposition}  \label{x,f}
Let $ f \in C^\infty(\R)  $ with $|\!|\!|f|\!|\!|_{3} < \infty$ .  Then
\begin{description}
\item[(i)]  $f(H) \H_c\subset  D(H)\cap D(\x) $.

\item[(ii)]  The operator $[\x,f(H)]$ is well defined on $\H_c$ and has a bounded
closure: there exists a constant
 $C_{\alpha, \beta} < \infty$ such that
\begin{equation} \label{HSbound}
\left\| \overline{[\x,f(H)]}\right\| \le C_{\alpha, \beta} |\!|\!|f|\!|\!|_{3}
\, .
\end{equation}
\end{description}
\end{proposition}

\begin{proof}
The Combes-Thomas argument \cite{CT} shows that $R(z) \H_c \subset \D(\mathbf
x)$, with $R(z) = (H-z)^{-1}$, whenever $\mathrm{Im} \,z \neq 0$ . In fact, we
have $R(z) \H_c \subset \D(\e^{\mu(z) |\mathbf x|})$ with the explicit estimate
\begin{equation}
    \left \| \e^{\mu(z) |\mathbf x-y|}R(z) \chi_y \right \| \ \le
\ C_{\alpha, \beta}\
    \frac{1}{|\mathrm{Im}\, z|} \; , \quad \text{ for every unit cube $\chi_y$,}
\end{equation}
where $\mu(z) = C_{\alpha,\beta} \,
 |\mathrm{Im}\, z| / (\langle \mathrm{Re} \, z
\rangle + |\mathrm{Im} z|)$.
 (See \cite[Theorem 1]{GK2} for details in this
context.  We denote by the same $C_{\alpha,\beta}$
possibly different constants
depending only on the parameters  $\alpha$ and $\beta$
given in \eqref{relbound}.) We conclude that
\begin{equation}
    \left \| \mathbf{x} R(z) \chi_y \right \| \ \le \ C_{\alpha,\beta,y}\frac{1}{\mu(z) |
    \mathrm{Im}\, z|} \ \le \ C_{\alpha,\beta,y} \
    \begin{cases}
    \frac{\langle \mathrm{Re} \,z \rangle}{|\mathrm{Im}\,z|^2} \; , & |\mathrm{Im}\, z| \le \langle
        \mathrm{Re}\, z \rangle \; , \\
    \frac{1}{ |\mathrm{Im}\, z|} \; , & |\mathrm{Im}\, z| \ge \langle
            \mathrm{Re}\,  z \rangle \; ,
    \end{cases}
\end{equation}
which gives (i) in light of \eqref{HShigherorder}.

Furthermore, we see that $[\mathbf x, R(z)]$ is well defined on $\H_c$. In
particular, for $\psi \in \H_c \cap \D$ we have
\begin{equation}
    \left [ \mathbf{x} , R(z) \right ] (H- z) \psi \ = \ \mathbf{x} \psi
    - R(z) \mathbf{x} (H-z) \psi \; ,
\end{equation}
where $(H-z)\psi \in \H_c$, since $H$ is local. As $\psi$ is compactly
supported, the components of $\mathbf{x}\psi$ are in $\D$ by Prop.
\ref{propDA}ii. Thus
\begin{equation}
     (H-z) \left [ \mathbf{x} , R(z) \right ] (H- z) \psi \ = \
    (H-z) \mathbf{x} \psi
    - \mathbf{x} (H-z) \psi \ = \ 2i\Db(\A) \psi \; ,
\end{equation}
where to obtain the last equality it is useful to consider $\psi \in
C_c^\infty$ initially and pass to $\psi \in \H_c \cap \D$ by a limiting
argument. Thus
\begin{equation}
    \left [ \mathbf{x} , R(z) \right ] (H- z) \psi \ = \ 2i R(z) \Db(\A) R(z)
    (H- z) \psi  \; ,
\end{equation}
whenever $\psi \in \H_c \cap \D$, which is a domain of essential
self-adjointness for $H$.  Thus $(H-z) \H_c \cap \D$ is dense, and we conclude
that $[\mathbf x, R(z)]$ is a bounded operator with
\begin{equation}\label{x,R=RDR}
    \left [ \mathbf{x} , R(z) \right ] \ = \ 2 iR(z) \Db(\A) R(z) \; .
\end{equation}
Specifically we have
\begin{equation}
    \left \| \left [ \mathbf{x} , R(z) \right ] \right \| \ \le \ 2
    \left \| R(z) \sqrt{H+\gamma} \right \| \cdot \left \|
    \frac{1}{\sqrt{H+\gamma}}\Db(\A) \right \| \cdot \| R(z) \| \; ,
\end{equation}
with the middle factor bounded by Proposition~\ref{propDA}(iii), and the first and
last factors bounded by $\sqrt{|z+\gamma|}/|\mathrm{Im} z|$ and $1/|\mathrm{Im
z}|$ respectively. Plugging these bounds into the Helffer-Sj\"ostrand formula
\eqref{HS}, and using \eqref{HShigherorder},  we find
\begin{equation}
    \left \| \left [ \mathbf{x} , f(H) \right ] \right \| \ \le C_{\alpha,\beta}\
    \int |d\tilde{f}(z)| \frac{\sqrt{|z| + \gamma}}{|\mathrm{Im} z|^2} \
  \le C_{\alpha,\beta} \ |\!|\!|f|\!|\!|_3 < \infty \, .
\end{equation}
\end{proof}


\subsection{Time-dependent electric fields}\label{secgauge}

Consider a quantum particle in the presence of a background
potential $V(x)$, a magnetic vector potential $\A(x)$, and a time
dependent spatially uniform electric field $ \El(t)$. We will
refer to the time-dependent self-adjoint generator of the unitary
evolution as the Hamiltonian.

One's initial impulse might be to add the electric potential
$\El(t) \cdot x$ to the magnetic Schr\"odinger operator $H(A,V)$
and consider the Hamiltonian:
\begin{equation}\label{eq:initial}
\widetilde H(t) =  H(\A,V) + \El(t) \cdot x= (-i \nabla -\A(x))^2 + V(x) + \El(t) \cdot x
 \, .
\end{equation}
However, {\em this choice is not dictated by the physics under
consideration.}  In fact, there is an infinite family of choices
for the Hamiltonian, related to one another by time-dependent
gauge transformations, all equally valid from the standpoint of
the underlying physics.

The operators defined by \eqref{eq:initial} suffer from the fact
that they are unbounded from below, and for general $\A,V$ it is
not obvious if there is a unitary propagator $\widetilde U(t,s)$
obeying
\begin{equation}\label{eq:Prop}
\left \{ \begin{array}{lll} i \partial_t \widetilde U(t,s) & =&  \widetilde
H(t) \widetilde U(t,s)
    \\   \widetilde  U(s,s) & =& \ I \end{array} \right . \; .
\end{equation}
However, there is a physically equivalent choice of Hamiltonian:
\begin{equation}\label{eq:Htilde}
    H(t) = ({-i\nabla} - \A - \F(t))^2  + V(x)=  H(\A + \F(t),V)  \, ,
\end{equation}
with $\F(t) = \int_{t_0}^t \El(s)\di s$ (with perhaps $t_0 =
-\infty$),  for which the propagator can be shown to exist for
quite general $\A,V$. It turns out that there is a general theory
of propagators with a time dependent generator \cite[Theorem
XIV.3.1]{Y} which applies to $H(t)$ but does not obviously apply
to $\widetilde H(t)$. Note that $H= H(t_0)$.

What is the justification for taking the Hamiltonian
\eqref{eq:Htilde}? In classical electrodynamics (Maxwell's
equations), one expresses the electric and magnetic field
$\El(x,t)$ and $\mathbf{B}(x,t)$ in terms of a ``scalar
potential'' $\phi(x,t)$ and a ``vector potential'' $\A(x,t)$:
\begin{equation}
    \begin{array}{lll}
    \El(x,t) & =&  -\partial_t\A(x,t) - \nabla \phi(x,t) \, ,\\
    \mathbf{B}(x,t) &=&  \nabla \times \A(x,t) \; .
    \end{array}
\end{equation}
The key observation is that $\El$ and $\mathbf{B}$ are not changed if $\A$ and $\phi$ are
perturbed by a ``gauge transformation'':
\begin{equation}
    \begin{array}{lll}\label{eq:EMgauge}
\A(x,t) &\mapsto& \A(x,t) + \nabla \alpha(x,t) \; ,\\
\phi(x,t) &\mapsto & \phi(x,t) - \partial_t \alpha(x,t) \; .
    \end{array}
\end{equation}

In particular, $\A$ and $\phi$ are not uniquely determined by the
``observable" fields $\El$ and $\mathbf{B}$.  Note that a
spatially uniform electric field $\El(t)$ may be obtained from the
time dependent vector potential $\F(t)$.

This non-uniqueness carries over to one particle quantum
mechanics. Consider a Hamiltonian associated to an electron in the
presence of the electromagnetic field described by $\A(x,t)$ and
$\phi(x,t)$:
\begin{equation}
   H(\A(x,t),\phi(x,t)) = (-i\nabla - \A(x,t))^2 + \phi(x,t) \; ,
\end{equation}
acting on $\L^2(\R^d)$ (in units with the electric charge equal to one).  To implement the gauge transformation
\eqref{eq:EMgauge}, we must also transform the wave function
$\psi(x,t)$ by
\begin{equation}\label{eq:psigauge}
    \psi(x,t) \mapsto \e^{i\alpha(x,t)} \psi(x,t) \; .
\end{equation}
Indeed, if $\psi(x,t)$ obeys the Schr\"odinger equation
\begin{equation} i
\partial_t \psi(x,t) = H(\A(x,t),\phi(x,t)) \psi(x,t)
\end{equation}
then it is easy to check that, {\em formally},
\begin{align} \nonumber
& i \partial_t \e^{i\alpha(x,t)} \psi(x,t) =-
      (\partial_t \alpha(x,t))\e^{i \alpha(x,t)} \psi(x,t) \ + \
       i \e^{i \alpha(x,t)} \partial_t\psi(x,t) \\
   &\quad = \left [\e^{i \alpha(x,t)}H(\A(x,t),\phi(x,t))\e^{-i
       \alpha(x,t)} - \partial_t \alpha(x,t) \right ] \e^{i \alpha(x,t)}
       \psi(x,t) \nonumber\\
      &\quad  = H(\A(x,t)+\nabla \alpha(x,t),\phi(x,t)- \partial_t
       \alpha(x,t))\e^{i \alpha(x,t)}
       \psi(x,t) \; . \label{eq:formalgt}
\end{align}
Effectively the gauge transformation \eqref{eq:psigauge}
implements a ``moving frame'' in $\L^2(\R^d)$, and we must
transform the Hamiltonian accordingly to account for the shift in
the time derivative in Schr\"odinger's equation.

 The possibility always exists to ``choose a gauge'' with
$\phi\equiv0$ and work only with $\A$: take $\partial_t
\alpha(x,t) = \phi(x,t)$, effectively replacing $\phi$ by zero and
$\A$ by $\A + \int_{t_o}^t
 \nabla \phi(x,s)\di s$. Generally, this gauge transformation is not
used in time independent quantum mechanics, since it replaces a
{\em time-independent} scalar potential with a {\em
time-dependent} vector potential, introducing an extra level
complexity. However, our Hamiltonian is {\em intrinsically}
time-dependent, and there is not really any greater complexity to
be found working with $\A(x,t)$ in place of $\phi(x,t)$.

For the problem at hand, we do not want to take the extreme step
of setting the scalar potential identically to zero. Instead it is
convenient to fix a time independent scalar potential $\phi(x,t) =
V(x)$ and a time dependent vector potential $\A(x,t) = \A(x) +
\F(t)$ with $\F(t) = \int_{t_0}^t \El(s) \di s$.  This leads to
the Hamiltonian $H(t)$ presented in \eqref{eq:Htilde}. Note that
on $\Cc^\infty(\R^d)$ we have
\begin{equation}\label{gaugeC}
    H(t) \ = \ G(t) \left [ (-i\nabla - \A)^2 + V \right ]
    G(t)^*\; ,
\end{equation}
where $G(t)$ denotes the gauge transformation
\begin{equation}\label{gaugedef}
[G(t)\psi](x) = \e^{i \F(t) \cdot x} \psi(x) \; .
\end{equation}
Repeating the formal calculation leading to \eqref{eq:formalgt},
we find that if $\psi(t)$ obeys Schr\"odinger equation
\begin{equation}\label{schreq1}
i \partial_t \psi(t) = H(t)
\psi(t),
\end{equation}
 then, \emph{formally},
\begin{equation}
 i \partial_t G(t)^\ast \psi(t) = \left [ (-i \nabla - \A)^2 + V + \El(t)\cdot x \right ]
    G(t)^\ast \psi(t) = \widetilde H(t) G(t)^\ast \psi(t)\; ,
\end{equation}
although this begs the question of whether $G(t)^\ast \psi(t)$ is in the
domain of either  $\El(t) \cdot x$ or $\widetilde H(t)$.

While there is no physical reason to work with one particular
gauge, it is comforting to know that the choice truly does not
affect the results.  One difficulty
is that we do not know (in general) if strong solutions to the
Schr\"odinger equation
\begin{equation}\label{schreq2}
    i \partial_t \psi_t \ = \ \widetilde H(t) \psi_t
\end{equation}
exist with $\widetilde H(t)$ given by \eqref{eq:initial}.
Thus we must consider weak solutions.
 Given a time dependent Hamiltonian $K(t)$ with
$C^\infty_c(\R^d) \subset \D(K(t))$ for all $t \in \R$,  a {\em weak
solution} to the Schr\"odinger equation
$
   i \partial \psi_t  =  K(t) \psi_t
$
is a map $t \mapsto \psi_t \in L^2(\R^d)$ such that
\begin{equation}
i \partial_t \langle \phi,\psi_t \rangle \ = \ \langle K(t)
    \phi, \psi_t \rangle \quad \text{for all }\quad  \phi \in C^\infty_c(\R^d)\; .
\end{equation}
It is easy to see that the weak solutions of
the Schr\"odinger equations \eqref{schreq1} and \eqref{schreq2}
are related by the gauge transformation $G(t)$: $\psi_t$ is a weak solution of
\eqref{schreq1} if and only if the gauge transformed $G(t)^* \psi_t$
is a weak solution of
\eqref{schreq2}.


\subsection{Time-dependent Hamiltonians and their propagators}

\emph{We assume throughout that  $\A(x)$ and  $V(x)$ satisfy the
Leinfelder-Simader conditions and $\El(t) \in C(\R; \R^d)$}.  (If
in addition $\El(t) \in \mathrm{L}^1((-\infty,0] ; \R^d)$ we take
$t_0 = -\infty$.)

\begin{proposition}\label{propH(t)}
$H(t)$, given in \eqref{eq:Htilde},  is essentially self-adjoint on
 $\Cc^\infty(\R^d)$ with
\begin{align}\label{HtH1}
H(t) &=H - 2 \F(t)\cdot (-i\nabla - \A) + \F(t)^2 \;\;
\mbox{on  $\Cc^\infty(\R^d)$}\, ,\\
&= H - 2 \F(t)\cdot{\Db}(\A) + \F(t)^2 \;\;\mbox{on  $\D(H)$}\, .\label{HtH}
\end{align}
Hence
\begin{equation}  \label{DDH}
 \D :=  \D(H)=  \D(H(t)) \quad \mbox{for all $t \in \R$},
\end{equation}
and on  $\D$ we have that for all $t$ and $s$,
\begin{equation}
H(t) =H(s)  - 2 (\F(t)- \F(s))\cdot{\Db}(\A) + (\F(t)^2- \F(s)^2)\, .  \label{HtH2}
\end{equation}
In addition, all $H(t)$  satisfy the lower bound given in
\eqref{lowerbound}:
\begin{equation}\label{lowerboundt0}
H(t) \ge  -\,{\frac \beta {1 - {\alpha}}}  \;\;\mbox{for all $t \in \R$}.
\end{equation}
\end{proposition}

\begin{proof}  Clearly
$\A(x) + \F(t)$ and $V(x)$ satisfy  the Leinfelder-Simader
conditions with the parameters $\alpha, \beta$ independent of $t$,
hence $H(t)$ is essentially self-adjoint on
 $\Cc^\infty(\R^d)$, \eqref{HtH1} follows from \eqref{expression0},
and we have \eqref{lowerboundt0}.  The equality
 \eqref{HtH} follows from  \eqref{HtH1} and Proposition~\ref{propDA}(i), and implies
\eqref{DDH}.
\end{proof}

\begin{lemma}\label{lemDA} Let $G(t)$ be as  in \eqref{gaugedef}.  Then
\begin{align}
G(t)\D &=\D \, ,\\
\label{gaugeC1}
    H(t) &=  G(t) H G(t)^*\, ,\\
\Db(\A + \F(t)) &=  \Db(\A) - \F(t)= G(t) \Db(\A) G(t)^*\, .
\end{align}
Moreover, $i[x_j,H(t)]=2 \Db(\A + \F(t))$ as quadratic forms on
$\D\cap\D(x_j)$, $j=1,2, \dots,d$.
\end{lemma}

\begin{proof}
The lemma follows from \eqref{gaugeC} and Propositions~\ref{propH(t)} and \ref{propDA}.
\end{proof}

We now discuss the existence of a propagator $U(t,s)$
satisfying
\begin{equation}
    i \partial_t U(t,s) = H(t)U(t,s) , \quad U(s,s)=I.
\end{equation}
We note that
\begin{equation} \label{lowerboundt}
H(t) + \gamma \ge 1  \;\;\mbox{for all $t \in \R$},
\end{equation}
where $\gamma$ is given in \eqref{gamma}.
We also set
\begin{align}\label{C(t,s)}
C(t,s) &=
 \left (H(t) - H(s)\right )
   \left( H(s) + \gamma \right)^{-1} \\
& =(\F(t)- \F(s))\cdot\left\{  - 2{\Db}(\A) + (\F(t) + \F(s))  \right \}
 \left( H(s) +\gamma\right)^{-1} \nonumber  . \nonumber
\end{align}
By Proposition~\ref{propDA}(i), we have
\begin{equation}\label{DbH}
\left\|{\Db}(\A) \left( H(s) + \gamma \right)^{-1}\right\|  \le
\left \| {\Db}(\A) \left( H + \gamma \right)^{-1}\right\| + |\F(s)|  \le
\  C_{\alpha,\beta}  + | \F(s)| \, ,
\end{equation}
with $C_{\alpha,\beta}$ a finite constant. Since $F(t) \in C^1(\R; \R^d)$, we
conclude that both $C(t,s)$ and $\textstyle{\frac 1 {t-s}} C(t,s)$ (with  $t
\not=s$) are uniformly continuous and uniformly bounded in operator norm for
$t,s$ restricted to a compact interval. Moreover,
\begin{align} \label{C(t)}
C(t) & = \lim_{s \to t} \textstyle{\frac 1 {t-s}} C(t,s) = 2 \El(t)\cdot (
{\Db}(\A) - \F(t))
 \left( H(t) +\gamma \right)^{-1} \\
 \nonumber & = 2 \El(t) \cdot G(t) \Db(\A) \left ( H + \gamma \right)^{-1} G(t)^*
\end{align}
exists,   is continuous in operator norm, and satisfies
\begin{equation}\label{HUHest45}
\|C(t)\| \le 2 C_{\alpha,\beta} |\El(t)| \, .
\end{equation}

\begin{theorem} \label{tpropagator}  The time-dependent  Hamiltonian $H(t)$
has a unique unitary propagator $U(t,s)$, i.e., there is a unique
two-parameter family $U(t,s)$ of unitary operators,  jointly
strongly continuous in $t$ and $s$, such that
\begin{align}  \label{U1}
U(t,r) U(r,s)&= U(t,s) \\
U(t,t)&=  I\\
\label{Dinv}
U(t,s) \D &= \D \, ,\\
i \partial_t U(t,s) \psi &=  H(t) U(t,s) \psi \;\; \mbox{for all $\psi \in \D$}\, ,  \label{leftdiff}\\
\label{reverseddif}
 i \partial_s U(t,s) \psi &= -\, U(t,s) H(s) \psi \;\; \mbox{for all $\psi \in \D$}\, .
\end{align}
In addition,
$W(t,s)=  \left( H(t) +\gamma \right)
U(t,s) \left( H(s) + \gamma \right)^{-1} $ is a bounded operator,
jointly strongly continuous  in $t$ and $s$, with
\begin{equation} \label{HUHest}
 \left\| W(t,s) \right\|
\le \e^{\int_{\min \{s,t\}}^{\max \{s,t\}}  \|C(r)\| \,\mathrm{d} r} \, ,
\end{equation}
 the operators $U(t,s)\left( H(s) + \gamma \right)^{-1} $ and
$\left( H(t) + \gamma \right)^{-1}U(t,s) $ are jointly
continuous in $t$ and $s$ in operator norm, and
\begin{align}\label{dirdiff}
i \partial_t\left\{ U(t,s)\left( H(s) + \gamma \right)^{-2} \right\}&=
 H(t)U(t,s)\left( H(s) + \gamma \right)^{-2} \, ,\\
 \label{reverseddif2}
 i \partial_s \left\{\left( H(t) + \gamma \right)^{-2}U(t,s)  \right\}&=
 -\,\left( H(t) + \gamma \right)^{-2}U(t,s) H(s) \, ,
\end{align}
in operator norm.

Furthermore, if we define the unitary operators $U_k(t,s)$,
$k=1,2,\ldots$,
 by
\begin{equation}
 U_k(t,s)  =  \e^{ -i(t-s) H \left(m + \textstyle{\frac{i-1}{k}} \right) }  \;\; \mbox{if}\;\;
 m+ \textstyle{\frac{i-1}{k}} \le  s ,t < m  + \textstyle{\frac{i}{k}} \, ,
\end{equation}
where $m\in \Z $, $i=1,2,\ldots,k$, and
\begin{equation}
 U_k(t,r) = U_k (t,s) U_k(s,r)\;\;  \mbox{for all}\;\;  t,s,r \, ,
\end{equation}
then
\begin{equation}\label{normlimit}
U(t,s)\left( H(s) + \gamma \right)^{-1} =
\lim_{k \to \infty}U_k(t,s) \left( H(s) + \gamma \right)^{-1}
\end{equation}
in operator norm, uniformly for $t,s$ restricted to a compact
interval.
\end{theorem}

\begin{proof} The uniqueness
and unitarity of the propagator $U(t,s)$ follows from existence
and the fact that  $i \partial_t \phi_t = H(t) \phi_t$ with $H(t)$
self-adjoint implies $\partial_t \|\phi_t \|^2=0$.

To prove the existence of the propagator we  apply \cite[Theorem
XIV.3.1]{Y} (see also \cite[Theorem X.70]{RS2}) with
\begin{equation}
A(t) = - i(H(t) +\gamma)\, .
\end{equation}
Note that
\begin{equation}
{C(t,s)=  A(t)A(s)^{-1} - I  = (A(t) -A(s))A(s)^{-1}}\, .
\end{equation}
The hypotheses of \cite[Theorem XIV.3.1]{Y} (and \cite[Theorem
X.70]{RS2}) require that (a) $0 \notin \sigma(A(t))$, (b) $A(t)$
have a common domain, and (c) $C(t,s)$ and $C(t) = \lim_{t
\rightarrow s} (t-s)^{-1} C(t,s)$ are uniformly bounded and
strongly continuous for $t,s$ restricted to a compact interval.
Clearly $\D(A(t)) = \D(H(t))= \D$ for all $t$, and it follows from
\eqref{lowerboundt} that $0 \notin \sigma(A(t))$ for all $t$.
Boundedness and continuity of $C(t,s)$ and $C(t)$ were discussed
before the statement of the theorem.

Thus the hypotheses of \cite[Theorem XIV.3.1]{Y} are satisfied. If
we set
\begin{equation}
U(t,s) = \e^{i(t-s) \gamma}\,\widehat{U}(t,s) \, ,
\end{equation}
where $\widehat{U}(t,s)$ is the propagator for the $A(t)$ given in
\cite[Theorem XIV.3.1]{Y} (and \cite[Theorem X.70]{RS2}) if $s\le t$,  and
 $\widehat{U}(t,s) = \widehat{U}(s,t)^*$ if  $s\ge t$, we  obtain unitary operators
$U(t,s)$, strongly continuous in $t$ and $s$, satisfying
\eqref{U1}-\eqref{leftdiff}.
To prove \eqref{reverseddif}, we use the chain rule:  Since $U(t,s)U(s,t)=I$, it follows
 from \eqref{Dinv} and \eqref{leftdiff} that for $\varphi \in \D$ we have,
with $\psi = U(s,t)\varphi$,
\begin{align}
0&=  \partial_s U(t,s)U(s,t) \varphi =
 \partial_s  U(t,s) \psi + U(t,s) \partial_s U(s,t) \varphi\\
&=  \partial_s  U(t,s) \psi - i  U(t,s) H(s) U(s,t) \varphi=
 \partial_s  U(t,s) \psi - i  U(t,s) H(s)\psi  \nonumber \, ,
\end{align}
since $\D = U(s,t) \D$.

The estimate  \eqref{HUHest} is given in \cite[Theorem XIV.3.1]{Y}.
A careful
 reading of the proof of  \cite[Theorem XIV.3.1]{Y}, using our stronger
hypotheses on $C(t,s)$,
 shows that  the operators $U(t,s)\left( H(s) + \gamma \right)^{-1} $ and
$\left( H(t) + \gamma \right)^{-1}U(t,s) $ are jointly
continuous in $t$ and $s$ in operator norm, and we have
\eqref{dirdiff}.
Since
the adjoint operation is an isometry in operator norm,   \eqref{reverseddif2} follows from
 \eqref{dirdiff}.
\end{proof}

To compute the linear response, we shall make use of the following
``Duhamel formula".
\begin{lemma} \label{propUU0}
 Let $ U^{(0)}(t) = \mathrm{e}^{-itH} $. For all  $ \psi \in \D $
and  $ t,s \in \R $ we have
\begin{equation}\label{DuhamelU}
U(t,s) \psi = U^{(0)}(t-s)\psi  + i \int_{s}^t U^{(0)}(t-r) (2  \F(r)\cdot
\Db(\A)-\F(r)^2)U(r,s) \psi \ \mathrm{d}r  \, .
\end{equation}
Moreover,
\begin{equation}\label{DuhamelU2}
 \lim_{\El \rightarrow 0} U(t,s) = U^{(0)}(t-s) \;\;\mbox{strongly}\, .
\end{equation}
\end{lemma}

\begin{proof}
Eq.  \eqref{DuhamelU}  follows simply
by calculating $ \partial_t U^{(0)}(s-t)U(t,s)\psi $ with
$ \psi  \in \D $, using \eqref{Dinv}, \eqref{leftdiff}, and  \eqref{HtH}.     The strong limit in \eqref{DuhamelU2}
follows from \eqref{DuhamelU} for vectors in $\D$, and hence everywhere since all
the operators are unitary.
\end{proof}


\section{Covariant operators and the  trace per unit volume}
\label{sectrandom}
\setcounter{equation}{0}

\subsection{Measurable covariant  operators}\label{subsecmblecovop}

We fix the notation
$\mathcal{H}=\mathrm{L}^2(\mathbb{R}^d)$
and let  $\Hc=\Lc^2(\R^d)$,  the dense linear subspace of functions with
compact support. We set $\Ll= \Ll (\Hc,\H)$  to be the vector space of
linear operators on  $\H$ with domain $\Hc$. Elements of $\Ll$ need
not be bounded.

We also fix ``magnetic translations'': for each $a\in \Z^d$ we define a unitary
operator
\begin{equation}
U(a) = \e^{ia\cdot Sx}T(a), \;\;\;\mbox{with}\;\; \left(T(a) \psi\right)(x)= \psi(x-a) \, ,
\end{equation}
where $S$ is a given $d \times d$ real matrix. Note that $a
\mapsto U(a)$ is a projective representation of the translation
group $\Z^d$ since
\begin{equation}
    U(a) U(b) = \e^{-i b\cdot Sa} U(a+b),
\end{equation}
and that $U(a)$ leaves $\Hc$ invariant, in fact
\begin{equation} \label{translation}
U(a) \chi_b U(a)^* = \chi_{b+ a} \, .
\end{equation}

Let $(\Omega, \P)$ be a probability space equipped with an ergodic
group $\{\tau(a); \ a \in \Z^d\}$ of measure preserving
transformations.  We study  operator--valued maps  $A \colon \Omega \to
\Ll$, which we will  simply call  operators $A_\omega $.
 We identify maps that agree $\P$-a.e., and all properties stated are supposed to hold for
$\P$-a.e. $\omega$.

\begin{definition}\label{mclbdefn} Let  $A= A_\omega \colon  \Omega \to \Ll$.  Then
\begin{description}
\item[(i)] $A_\omega$ is measurable if $\langle \varphi, A_\omega \psi\rangle$ is a
measurable function
for all $\varphi, \psi \in \Hc$. (Or, equivalently, if $A_\omega$
 is strongly measurable on $\Hc$, i.e.,
$\ A_\omega \psi$ is a measurable $\H$-valued
function
for all $ \psi \in \Hc$.)

\item[(ii)]  $A_\omega$ is covariant if
\begin{equation} \label{cov}
U(a) A_\omega U(a)^* = A_{\tau(a) \omega}\; \;\;\mbox{for all $a \in \Z^d$.}
\end{equation}

\item[(iii)] $A_\omega$ is locally bounded if
\begin{equation}
    \|A_\omega \chi_x\|< \infty \mbox{ and }  \|\chi_x A_\omega \| <\infty \quad
    \mbox{ for all } x \in \Z^d.
\end{equation}
\end{description}
\end{definition}

We  let $\K_{mc}$ denote the vector space of measurable
covariant operators $A_\omega$, with
 $\K_{mc,lb}$ being the   subspace of  locally bounded operators.
 We define
 the Banach space
\begin{equation}
\K_{\infty}= \left\{  A_\omega \in \K_{mc}; \;
\tnorm{A_\omega}_\infty<\infty   \right\} \subset
\K_{mc,lb}\, ,
\end{equation}
where
\begin{equation} \label{norminfty}
\tnorm{A_\omega}_\infty =  \|\, \|A_\omega \| \,
\|_{\mathrm{L}^{\infty}(\Omega, \P)} \, .
\end{equation}
If $A_\omega \in \K_{\infty}$, we identify $A_\omega$ with its
extension to $\H$ (i.e., with its closure $\overline{A_\omega}$).  If we
define multiplication in $\K_{\infty}$ by
$A_\omega B_\omega :=\overline{A_\omega}B_\omega$, and the adjoint
by $(A_\omega)^*:= A_\omega^* $, then
$\K_{\infty}$ becomes a $C^*$-algebra.

Whenever $A_\omega \in \K_{mc,lb}$, we have $\D(A_\omega^*) \supset \Hc$, since
$\chi_x A_\omega$ is bounded for all $x$. We define $A_\omega^\ddagger$ to be
the restriction of $A_\omega^*$ to $\H_c$.   It follows that $A_\omega^\ddagger
\in \K_{mc,lb}\,$, and the map $A_\omega \to A_\omega^\ddagger$ is a
conjugation in $ \K_{mc,lb}$. (Note that $A_\omega \in \K_{mc,lb}$ if and only
if there exist symmetric operators  $B_\omega, C_\omega \in \K_{mc} $
 such that $  \|B_\omega \chi_x\| + \|C_\omega \chi_x\|< \infty $
for all $x \in \Z^d$ and
$A_\omega = B_\omega + i  C_\omega$.  In this case
$A_\omega^\ddagger = B_\omega - i  C_\omega$.)

Thus, given $A_\omega \in \K_{mc,lb}$, we have that $A_\omega^*$ is densely
defined and therefore $A_\omega$ is closable. The closure of $A_\omega$,
denoted $\overline{A_\omega}$, has a polar decomposition and $\Hc$ is a core
for the self-adjoint operator $| \overline{A_\omega}|$. \emph{We will abuse
 notation and denote the restriction of $| \overline{A_\omega}|$ to $\Hc$
 by $|A_\omega|$}. It is not hard to see that $| {A_\omega}|$ is
covariant, i.e., it satisfies \eqref{cov}.  Similarly, local
boundedness of $|A_\omega|$ is a simple consequence of the
identities
\begin{equation}
    \lVert\, \lvert A_\omega\rvert \chi_x \rVert = \lVert A_\omega \chi_x\rVert
 \mbox{ and }
   \lVert\chi_x \lvert A_\omega \rvert \,\rVert =
\lVert\, \lvert A_\omega\rvert \chi_x \rVert .
\end{equation}
It is also true that $|A_\omega|$ is measurable, so $|A_\omega| \in
\K_{mc,lb}$, but this requires a little more work.

\begin{lemma} \label{lem|A|}   Let $A_\omega \in \K_{mc,lb}\,$, and consider
the polar decomposition $\overline{A_\omega} = U_\omega |\overline{A_\omega}|$.
Then $|A_\omega| \in \K_{mc,lb}$ and $U_\omega \in \K_\infty$. We also have
$f(|\overline{A_\omega}|) \in \K_{\infty}$ for any bounded Borel function $f$
on the real line.
\end{lemma}

\begin{proof}
Let $A_\omega \in\K_{mc,lb}$. We start by proving that
$(|\overline{A_\omega}|^2 +1)^{-1}$ is strongly measurable on $\H$, from which
it follows that $g(|\overline{A_\omega}|^2)$ is also strongly measurable for
any bounded Borel function $g$ on the real line. It then follows that
$f(|\overline{A_\omega}|) \in \K_{\infty}$ for any bounded Borel function $f$
on the real line (covariance is easy to see). Picking $f_n(t)=  t
\chi_{[-n,n]}(t)$, it is clear that $f_n(|\overline{A_\omega}|) \to |A_\omega|$
strongly on $\Hc$, and hence $|A_\omega|$ is  strongly measurable. We conclude
that $|A_\omega|\in \K_{mc,lb}$.

To prove measurability of $(|\overline{A_\omega}|^2 +1)^{-1}$, we pick an
ortho-normal basis $\{\varphi_n\}_{n \in \mathbb{N}}$ for the subspace
$\mathcal{H}_0= \chi_0\H \cong \mathrm{L}^2(\mathbb{R}^d,\chi_0(x)
{\mathrm{d}}x) $
 of $\mathcal{H}$, and set  $\varphi_n^{(a)} = T(a)\varphi_n$
for $a \in \mathbb{Z}^d$.  Then
  $\{\varphi_n^{(a)}\}_{n \in \mathbb{N}, a \in \mathbb{Z}^d}$
 is a an ortho-normal basis for $\mathcal{H}$, which we relabel as
$\{\phi_n\}_{n \in \mathbb{N}}$, and let $\widehat{\mathcal{H}_{c}}$ be the
subspace of finite linear combinations of the ${\phi_n}$'s.  Note that
$\widehat{\mathcal{H}_{c}}$ is a dense subspace of ${\mathcal{H}_{c}}$ and
hence is a core for $\overline{A_\omega}$, since $A_\omega$ is locally bounded.

Let $P_n$ be the orthogonal projection onto the finite dimensional
 subspace spanned by
$\phi_1,\phi_2,\ldots,\phi_n$.  We set
\begin{equation}
M_\omega^{(n)}=  (A_\omega P_n)^* A_\omega P_n\, ,
\end{equation}
 a bounded operator since $A_\omega$ is locally
bounded. Since we have  $ \langle \varphi,M_\omega^{(n)} \psi\rangle  = \langle
A_\omega P_n\varphi, A_\omega P_n\psi\rangle $ for $\varphi, \psi \in \H$, we
conclude that $M_\omega^{(n)}$ is weakly, and hence strongly, measurable on
$\H$. Proceeding as in \cite[Proof of Lemma 2.8]{PF}, we see that
$(M_\omega^{(n)} +1)^{-1}$ is measurable  on $\H$ (basically, because a matrix
element of the inverse may be expressed as a ratio of determinants, which are
measurable functions). We now show that $(M_\omega^{(n)} +1)^{-1} \to (
|\overline{A_\omega}|^2 +1)^{-1}$ weakly as $n \to \infty$, and hence  $(
|\overline{A_\omega}|^2 +1)^{-1}$ is measurable on $\H$.

For this purpose, let $\varphi, \psi \in
\widehat{\mathcal{H}_{c}}$. For sufficiently large $n$ we have
\begin{align}\nonumber
\langle A_\omega \varphi, A_\omega  (M_\omega^{(n)} +1)^{-1} \psi\rangle &=
\langle A_\omega P_n\varphi, A_\omega  P_n (M_\omega^{(n)} +1)^{-1} \psi\rangle\\
 &= \langle\varphi, M_\omega^{(n)}  (M_\omega^{(n)} +1)^{-1} \psi\rangle \, ,
\end{align}
and hence
\begin{equation} \label{pf}
\langle A_\omega \varphi, A_\omega  (M_\omega^{(n)} +1)^{-1} \psi\rangle  +
\langle \varphi,   (M_\omega^{(n)} +1)^{-1} \psi\rangle =
\langle \varphi,   \psi\rangle \, .
\end{equation}
Now let $\phi \in \mathcal{D}(\overline{A_\omega})$. Given $\varepsilon
>0$ we pick $\varphi \in \widehat{\mathcal{H}_{c}}$ such that
\begin{equation}
 \| (\phi - \varphi)\| + \| \overline{A_\omega} (\phi - \varphi)\|  < \varepsilon \ .
\end{equation}
Since \begin{equation}
\|A_\omega P_n  (M_\omega^{(n)} +1)^{-1} \|^2 =
 \|(M_\omega^{(n)} +1)^{-1} M_\omega^{(n)} (M_\omega^{(n)} +1)^{-1} \|
 \le    \frac 1 4 \, ,
\end{equation}
we have
\begin{align}\nonumber
&\left| \langle\overline{A_\omega} (\phi -\varphi), A_\omega
(M_\omega^{(n)} +1)^{-1} \psi\rangle  + \langle  \phi -\varphi,
(M_\omega^{(n)} +1)^{-1} \psi\rangle -  \langle  \phi -\varphi, \psi\rangle
\right | \\
  &\qquad  \qquad
 \le 3 \varepsilon \| \psi\| \, ,  \label{pf2}
\end{align}
whenever $ \psi \in \widehat{\mathcal{H}_{c}}$ and $n$ is correspondingly
large. Therefore, it follows from \eqref{pf} that for all $\phi \in
\mathcal{D}(\overline{A_\omega})$ we have
\begin{equation} \label{pf3}
\lim_{n \to \infty}\langle\overline{A_\omega}\phi, A_\omega  (M_\omega^{(n)}
+1)^{-1} \psi\rangle  + \langle \phi,  (M_\omega^{(n)} +1)^{-1} \psi\rangle =
\langle \phi,  \psi\rangle
\end{equation}
for all  $ \psi \in \widehat{\mathcal{H}_{c}}$.

Taking $\phi \in \mathcal{D}(A_\omega^* \overline{A_\omega})
 \subset \mathcal{D}(\overline{A_\omega})$, we get
\begin{equation} \label{pf4}
\lim_{n \to \infty}\langle (A_\omega^*\overline{A_\omega} +1) \phi,
(M_\omega^{(n)} +1)^{-1} \psi\rangle
 =  \langle \phi,   \psi\rangle
\end{equation}
for all $ \psi \in \widehat{\mathcal{H}_{c}}$, and hence for all $\psi \in \H$.
Writing $\eta =  (|\overline{A_\omega}|^2 +1) \phi$, we get
\begin{equation} \label{pf45}
\lim_{n \to \infty}\langle \eta,   (M_\omega^{(n)} +1)^{-1} \psi\rangle
 =  \langle  (|\overline{A_\omega}|^2 +1)^{-1}\eta,   \psi\rangle
\end{equation}
for all $\eta, \psi \in \H$. We conclude that $  (M_\omega^{(n)} +1)^{-1}  \to
(|\overline{A_\omega}|^2+1)^{-1}$ weakly.

We now turn to the partial isometry $U_\omega$.  We recall that
\begin{equation}
U_\omega = \lim_{\varepsilon \to 0}  \overline{A_\omega}(| \overline{A_\omega}|
+ \varepsilon)^{-1}\;\;\; \mbox{strongly on $\H$} \, .
\end{equation}
Thus $U_\omega$ is clearly covariant and measurable, so $U_\omega \in
\K_\infty$.
\end{proof}

\begin{lemma}  \label{lemAn}
Let $A_\omega \in \K_{mc,lb}$. Then, for each $n$,
\begin{equation}  \label{An}
A_\omega^{(n)} = \left(\textstyle{\frac 1 n}|\overline{A_\omega^\ddagger}|^2 +1
\right)^{-\frac12} A_\omega \in \K_{\infty} \, ,
\end{equation}
with $\|A_\omega^{(n)}\| \le n$, and
$A_\omega^{(n)} \to A_\omega$ strongly on $\Hc$.
\end{lemma}

\begin{proof} We clearly have  $A_\omega^{(n)} \in \K_{mc}$ since
$\left(\textstyle{\frac 1 n}|\overline{A_\omega^\ddagger}|^2 +1
\right)^{-\frac12} \in \K_{\infty}$ by Lemma~\ref{lem|A|}. As
 $\left(\textstyle{\frac 1 n}|\overline{A_\omega^\ddagger}|^2 +1 \right)^{-\frac12} \to I$ strongly,
we conclude that $A_\omega^{(n)} \to A_\omega$ strongly on $\Hc$.

Thus we only need to show that $\|A_\omega^{(n)}\| \le n$.       To do so, let
\begin{equation}
\widetilde{A_\omega^{(n)}} =
\left(\textstyle{\frac 1 n}|A_\omega^*|^2 +1 \right)^{-\frac12} A_\omega  \, ,
\end{equation}
and recall $\|\widetilde{A_\omega^{(n)}}\| \le n$.  Since $A^\ddagger$ is the
restriction of $A^*$ to $\mathcal{H}_c$, we have $|A_\omega^*|^2 \le
|\overline{A_\omega^\ddagger}|^2$ as quadratic forms (see \cite[p. 375]{RS1})
and hence
\begin{equation}
\left(\textstyle{\frac 1 n}|\overline{A_\omega^\ddagger}|^2 +1 \right)^{-1} \le
\left(\textstyle{\frac 1 n}|A_\omega^*|^2 +1 \right)^{-1}
\end{equation}
by  \cite[Theorem S.17]{RS1}.  We conclude that
\begin{equation}
 \|A_\omega^{(n)}\| \le \|\widetilde{A_\omega^{(n)}}\| \le n \, .
\end{equation}
\end{proof}

\begin{lemma} \label{lemRMgeneral}
If  $A_\omega\in  \K_{mc,lb}$, $B_\omega \in \K_{\infty}$, and $B_\omega
A_\omega\in \K_{mc,lb}$, we have that
$\mathcal{D}(A_\omega^*)\supset B_\omega^*\H_c$
and
\begin{equation}\label{adj277}
(B_\omega A_\omega)^\ddagger\varphi  =  A_\omega^*B_\omega ^*\varphi \;\;
\mbox{for all $\varphi \in \H_c$}\, .
\end{equation}
\end{lemma}
\begin{remark}
  Note that $B_\omega A_\omega$ is not necessarily in $\K_{mc,lb}$, since we have
  no control on $\|\chi_x B_\omega A_\omega\|$ for $x \in \Z^d$.
\end{remark}
\begin{proof}
For any $\varphi,\psi  \in \H_c$ we have
\begin{equation}
\langle \varphi,  B_\omega A_\omega \psi\rangle = \langle  (B_\omega
A_\omega)^\ddagger \varphi,  \psi\rangle \, .
\end{equation}
On the other hand,
\begin{equation}
\langle \varphi,  B_\omega A_\omega \psi\rangle =
\langle  B_\omega ^*\varphi, A_\omega \psi\rangle \, .
\end{equation}
It follows that
\begin{equation}\label{adj}
B_\omega ^*\varphi \in \D(A_\omega^*) \;\, \mbox{for all $\varphi \in \H_c$}\,
 \end{equation}
and \eqref{adj277} holds.
\end{proof}

Let us define
\begin{equation}
  \K_{\odot}\ = \ \left\{  A_\omega \in \K_{mc,lb}; \; B_\omega
A_\omega, B_\omega A_\omega^\ddagger  \in \K_{mc,lb}\, \mbox{if $B_\omega \in
\K_\infty$}   \right\} .
\end{equation}
Note that  $\K_{\odot} \subset \K_{mc,lb}$ is a vector space, and in
$\K_{\odot}$ we can define left and, using Lemma~\ref{lemRMgeneral}, right
multiplication by an element of $\K_\infty$:
\begin{align}  \label{multleft}
B_\omega \odot_L A_\omega& = B_\omega A_\omega\, ,  \\
A_\omega \odot_R B_\omega& =   A_\omega^{\ddagger *} B_\omega |_{\H_c}\, ,
\end{align}
where $A_\omega \in \K_{\odot}$ and $B_\omega \in \K_\infty$. Note that for
$B_\omega \in \K_\infty$ we have $B_\omega^{\ddagger *}=B_\omega$ since we
identify $B_\omega$ with its closure, so \eqref{multleft} could also have been
 written as
\begin{equation}
B_\omega \odot_L A_\omega= B_\omega^{\ddagger *} A_\omega \, .
\end{equation}

\begin{proposition}\label{propK2588}
Let  $A_\omega \in \K_{\odot}$ and $B_\omega, C_\omega  \in \K_{\infty}$. We
then have $B_\omega \odot_L A_\omega, A_\omega \odot_R B_\omega \in \K_{
\odot}$. Moreover,
\begin{equation}\label{rightM}
 A_\omega \odot_R B_\omega=
\left(B_\omega^* \odot_L A_\omega^\ddagger\right)^\ddagger \, ,
\end{equation}
\begin{equation} B_\omega \odot_L A_\omega \odot_R C_\omega  :=
\left(B_\omega \odot_L A_\omega\right) \odot_R C_\omega =
B_\omega \odot_L \left( A_\omega \odot_R C_\omega\right)
\, ,\label{BACassoc}
\end{equation}
\begin{equation}\label{BACdag}
\left(B_\omega \odot_L A_\omega \odot_R C_\omega\right)^\ddagger
=
C_\omega^\ast \odot_L A_\omega^\ddagger \odot_R B_\omega^\ast \, ,
\end{equation}
\begin{equation}
\{B_\omega \odot_L A_\omega \odot_R C_\omega\} \varphi =
 B_\omega A_\omega^{\ddagger *} C_\omega \varphi  \;\;\mbox{for all $\varphi \in \H_c$}\, .
\end{equation}
\end{proposition}

\begin{proof} The proof is a simple exercise.
\end{proof}


 \subsection{The Hilbert space $\K_2$}
\label{subsectK2}

Let
\begin{align}
\K_{2}&= \left\{  A_\omega \in \K_{mc}; \;
    \tnorm{A_\omega}_2<\infty   \right\}\, ,\\
\K_{2}^{(0)}& = \K_{2}\cap \K_{\infty}\, ,
\end{align}
where
\begin{equation}
\tnorm{A_\omega}_2 =
 \left\{\E\left( \|A_\omega \chi_0\|_2^2\right)\right\}^{\frac12} .
\end{equation}

\begin{proposition}
\label{propK2}
\emph{\textbf{(i)}} $\K_{2}$ is a Hilbert space with the inner product
\begin{equation}\label{<<}
\langle\langle A_\omega, B_\omega \rangle\rangle = \E \left\{
\tr \, \left\{(A_\omega \chi_0)^* B_\omega \chi_0\right\}\right\} \, ,
\end{equation}
and $\tnorm{ \ }_2$ is the corresponding norm, i.e.,
\begin{equation}
\tnorm{ A_\omega }_2^2= \langle\langle A_\omega, A_\omega \rangle\rangle \, .
\end{equation}
\noindent{\emph{\textbf{(ii)}}}  $\K_{2} \subset \K_{mc,lb}$ and the
conjugation $A_\omega \to A_\omega^\ddagger$ is antiunitary in $\K_{2}$, i.e.,
\begin{equation}\label{<<2}
\langle\langle A_\omega, B_\omega \rangle\rangle
=\langle\langle B_\omega^\ddagger, A_\omega^\ddagger \rangle\rangle \, .
\end{equation}

\noindent{\emph{\textbf{(iii)}}}  For all $A_\omega \in \K_2$ we have
\begin{equation} \label{takeadj}
(A_\omega \chi_0)^* =\overline{\chi_0 A_\omega ^*}=
 \overline{\chi_0 A_\omega ^\ddagger} \, ,
\end{equation}
and hence
\begin{align}\label{<<3}
\langle\langle A_\omega, B_\omega \rangle\rangle& = \E \left\{
\tr \, \left\{\overline{\chi_0 A_\omega ^\ddagger}  B_\omega \chi_0\right\}\right\} \, ,\\
\tnorm{ A_\omega }_2 &=
\left\{\E\left(\|\chi_0  A_\omega^\ddagger \|_2^2\right)\right\}^{\frac12}
=\left\{\E\left(\|\chi_0  A_\omega \|_2^2\right)\right\}^{\frac12}\, .
\label{<<37}
\end{align}

\noindent{\emph{\textbf{(iv)}}}  $\K_{2}^{(0)}$ is dense in   $\K_{2}$.
\end{proposition}

\begin{proof}
 We first note that $\K_{2}$ is
a vector space, since
\begin{equation}
\tnorm{A_\omega+B_\omega}_2^2 \le
 \E\left\{ \left(\|A_\omega \chi_0\|_2 +\|B_\omega \chi_0\|_2\right)^2\right\}
\le 2\left( \tnorm{A_\omega}_2^2 +\tnorm{B_\omega}_2^2\right)  \, .
\end{equation}
Since  the right hand side of \eqref{<<} is well defined for $A_\omega,
B_\omega \in \K_2$, it clearly  defines an inner product.

To show that $\K_{2}$ is complete it suffices to show that every summable series
in $\K_{2}$ converges. So consider the series
\begin{equation}
\sum_{n=1}^\infty \tnorm{A_{n,\omega}}_2 < \infty\, ,\;\;\;
A_{n,\omega} \in \K_{2}\, .
\end{equation}
It follows that
\begin{equation} \label{a1}
\E\left( \sum_{n=1}^\infty \|A_{n,\omega} \chi_0\|_2\right) =
\sum_{n=1}^\infty\E\left(  \|A_{n,\omega} \chi_0\|_2\right) \le
\sum_{n=1}^\infty \tnorm{A_{n,\omega}}_2 < \infty \, ,
 \end{equation}
and hence
\begin{equation}
 \sum_{n=1}^\infty \|A_{n,\omega} \chi_0\|_2 < \infty \, .
\end{equation}
Using the completeness of $\H$ and the covariance property we conclude that
$\sum_{n=1}^\infty A_{n,\omega}$ converges strongly in $\H_c$ to an
operator $A_\omega \in \K_{mc}$.  Since the Hilbert-Schmidt
operators on $\H$ are also complete, we also conclude that
$A_\omega\chi_0 =\sum_{n=1}^\infty A_{n,\omega}\chi_0 $ with convergence in
Hilbert-Schmidt norm.  Thus, using Fatou's lemma,
\begin{align} \nonumber
\tnorm{A_\omega}_2^2 &=
 \E\left( \lim_{N \to \infty}\left\|\sum_{n=1}^N A_{n,\omega}\chi_0\right \|_2^2\right)
\le  \liminf_{N \to \infty}
 \E\left(\left\|\sum_{n=1}^N A_{n,\omega}\chi_0\right \|_2^2\right)\\
&\le  \left( \lim_{N \to \infty}  \sum_{n=1}^N\tnorm{A_{n,\omega}}_2\right)^2
=   \left(  \sum_{n=1}^\infty\tnorm{A_{n,\omega}}_2\right)^2 < \infty \, ,
\end{align}
and hence $A_\omega \in \K_2$.  Since
$A_\omega -\sum_{n=1}^N A_{n,\omega} =\sum_{n=N+1}^\infty A_{n,\omega}$,
the same  argument gives
\begin{equation}  \label{a2}
\tnorm{A_\omega - \sum_{n=1}^N A_{n,\omega} }_2^2
\le   \left(  \sum_{n=N+1}^\infty\tnorm{A_{n,\omega}}_2\right)^2
  \to 0 \;\; \mbox{as $N \to \infty$} \, ,
\end{equation}
and hence $\K_2$ is complete.

To show  $\K_{2} \subset \K_{mc,lb}$ it suffices to show $ A_\omega^*\chi_0$ is
well defined and almost surely bounded, since $A_\omega \chi_0$ is almost
surely Hilbert-Schmidt and thus bounded. Given $A_\omega \in \K_{2}$, we set
$A_{\omega,x,y}= \chi_x A_\omega\chi_y$ for
 $x,y \in\mathbb{Z}^2$, a Hilbert-Schmidt operator.
Then note that $ (A_{\omega,x,y})^*= \chi_y  (A_{\omega,x,y})^* \chi_x$ and
\begin{align}
&\sum_{y \in \mathbb{Z}^2}  \mathbb{E}\left\{ \tr
\left(  A_{\omega,x,y} (A_{\omega,x,y})^*  \right)\right\} =
\sum_{y \in \mathbb{Z}^2}  \mathbb{E}\left\{ \tr
\left(\chi_x  A_{\omega,x,y}\chi_y (A_{\omega,x,y})^*  \chi_x\right)\right\}
\nonumber\\
& \qquad  \qquad =
 \sum_{y \in \mathbb{Z}^2}  \mathbb{E}\left\{ \tr
\left(\chi_{x-y}  A_{\tau(y)\omega, x-y,0}
\chi_0 A_{\tau(y)\omega, x-y,0}^*  \chi_{x-y}\right)\right\} \label{a3}\\
&\qquad   \qquad=
\sum_{y \in \mathbb{Z}^2}  \mathbb{E}\left\{ \tr
\left(\chi_{0}  A_{\omega, x-y,0}^*
\chi_{x-y} A_{\omega, x-y,0}  \chi_{0}\right)\right\}
=\tnorm{A_\omega}_2^2 \label{ad2}  \, ; \nonumber
\end{align}
we used \eqref{cov}, the invariance of the expectation under  the
transformations $\{\tau(a); \ a \in \Z^d\}$, and cyclicity of the trace, plus
the fact that, as all terms in the expressions are positive, we can interchange
the sum with the trace and the expectation. Proceeding as in
\eqref{a1}-\eqref{a2} we conclude that
 the operator $B_\omega=\sum_{x,y \in \mathbb{Z}^2} (A_{y,x})^*$
is in $\mathcal{K}_2$.  (Note that  covariance only holds
for the sum over all ${x,y \in \mathbb{Z}^2}$).  It is easy to see that
 $B_\omega \subset A_\omega^*$, so $\mathcal{D}(A_\omega^*) \supset \H_c$
and $B_\omega = A_\omega^\ddagger$.  Thus
\begin{equation}
\tnorm{A_\omega^\ddagger}_2^2= \sum_{y \in \mathbb{Z}^2}  \mathbb{E}\left\{ \tr
\left(  A_{\omega,0,y} (A_{\omega,0,y})^*  \right)\right\} =
\tnorm{A_\omega}_2^2
\end{equation}
by \eqref{a3}, and
\eqref{<<2} follows
using the polarization identity.

The equality \eqref{takeadj} is an easy consequence of $\D(A^*) \supset \H_c$;
\eqref{<<3} and \eqref{<<37} then follow from \eqref{<<} and \eqref{<<2}.

It remains to show that  $\K_{2}^{(0)}$ is dense in   $\K_{2}$. Let $A_\omega
\in \K_2$, then $A_\omega, A_\omega^\ddagger  \in \K_{mc,lb}$, and
$A_\omega^{(n)}$, defined in \eqref{An}, is clearly in  $\K_{2}^{(0)}$, and
$\tnorm{A_\omega -A_\omega^{(n)}}_2 \to 0$ by a dominated convergence argument.
\end{proof}

Left and right multiplication by elements of $\K_\infty$ leave $\K_2$ invariant.

\begin{proposition}\label{propK25}  $\K_2 \subset  \K_{\odot}$.
Moreover, if $A_\omega \in \K_2$ and $B_\omega \in K_\infty$ we have
$B_\omega
\odot_L A_\omega, A_\omega \odot_R B_\omega \in \K_2$ with
\begin{align} \label{LK2}
\tnorm{ B_\omega \odot_L A_\omega}_2 &\le   \tnorm{B_\omega}_\infty
\tnorm{A_\omega}_2 \,,\\
\tnorm{ A_\omega \odot_R B_\omega }_2 &\le   \tnorm{B_\omega}_\infty
\tnorm{A_\omega}_2 \, . \label{RK2}
\end{align}
\end{proposition}

\begin{proof}  Since we clearly have $B_\omega  \odot_LA_\omega \in \K_2$
with \eqref{LK2},
 Proposition~\ref{propK2}(ii) gives
  $\K_2 \subset  \K_{mc\odot}$.
 The estimate
 \eqref{RK2} follows from \eqref{rightM}, \eqref{LK2}, and \eqref{<<2}.
\end{proof}

The following lemma will be very useful.

\begin{lemma}\label{stK2}  Let $B_{n,\omega}$ be a bounded sequence in $\K_\infty$
such that $B_{n,\omega}\to B_\omega$ strongly.   Then for all $A_\omega \in \K_2$
we have
$ B_{n,\omega} \odot_L A_\omega\to B_{\omega} \odot_L A_\omega$
and $A_\omega \odot_R  B_{n,\omega} \to A_\omega \odot_R  B_{\omega}$ in $\K_2$.
\end{lemma}

\begin{proof}  It suffices to prove the result for left multiplication in view of \eqref{rightM}.
By considering the sequence $B_{n,\omega}-B_{\omega}$ we may assume $B_{\omega}=0$.
  We have, with $A_\omega \in \K_2^{(0)}$,
\begin{equation}
\tnorm{ B_{n,\omega} \odot_L A_\omega}_2^2 =
 \E \, \tr \{\chi_0 A_\omega^*B_{n,\omega}^*B_{n,\omega}A_\omega \chi_0  \}  \to 0
\end{equation}
by dominated convergence.  Since  $B_{n,\omega}$ is  bounded and $\K_2^{(0)}$
is dense in $\K_2$, this extends to general $A_\omega \in \K_2$.
\end{proof}


\subsection{The normed space $\K_1$.}
\label{subsectK1}

Let
\begin{align}
\K_{1}&= \left\{  A_\omega \in \K_{mc,lb}; \;
 \tnorm{A_\omega}_1<\infty   \right\}\, ,\\
\K_{1}^{(0)}& = \K_{1}\cap \K_{\infty} ,
\end{align}
where
\begin{equation}
\tnorm{A_\omega}_1 =
 \E \left\{\tr\left\{\chi_0 |A_\omega| \chi_0\right\}\right\} .
\end{equation}
Note that $\tnorm{A_\omega}_1$ is well defined (possibly infinite) for
 $A_\omega \in \K_{mc,lb}$ by Lemma~\ref{lem|A|}.

\begin{lemma}\label{lemtrace} Let $A_\omega \in \K_{1}$.  Then
\begin{equation} \label{trace}
\E\left\{ \tr \left |\chi_0 A_\omega \chi_0\right |\right\}
 \le \tnorm{A_\omega}_1 < \infty\, ,
\end{equation}
and hence
$\E\left\{ \tr\left\{\chi_0 A_\omega \chi_0\right\} \right\}$ is well defined.
\end{lemma}

\begin{proof}
Let $\overline{A_\omega}=  U_\omega |\overline{A_\omega}|$ be the polar
decomposition of $\overline{A_\omega}$.  We have
\begin{equation}\label{trace2}
\chi_0 A_\omega \chi_0 = \chi_0  U_\omega |\overline{A_\omega}|^{\frac12}
|\overline{A_\omega}|^{\frac12}\chi_0  \, .
\end{equation}
Since  $A_\omega \in \K_{1}$,  $|\overline{A_\omega}|^{\frac12} \in \K_2$ and,
by Lemma~\ref{lem|A|},  $U_\omega \in \K_\infty$.  (More precisely,
 the restriction $|A_\omega|^{\frac12}$ of $|\overline{A_\omega}|^{\frac12}$ to $\H_c$
 is in $\K_2$.  Note that $\H_c$ is
a core for $|\overline{A_\omega}|^{\frac12}$.) Thus
 $ U_\omega |A_\omega|^{\frac12}  \in \K_2$,
and $\overline{\chi_0  U_\omega |\overline{A_\omega}|^{\frac12}}$  is a
Hilbert-Schmidt operator by \eqref{takeadj}.   Hence  it follows from
\eqref{trace2} that  $\chi_0 A_\omega \chi_0$ is trace class.   The inequality
\eqref{trace} now follows from \eqref{trace2},   H\"older's inequality, and
\eqref{<<37}.
\end{proof}

\begin{lemma} \label{lemtrace2}  Let $A_\omega \in \K_1$ and  $B_\omega \in \K_\infty$.  Then
$B_\omega A_\omega \in \K_1$ and
\begin{equation}\label{K1bound}
\tnorm{B_\omega A_\omega}_1 \le
\tnorm{B_\omega}_\infty  \tnorm{A_\omega}_1 \, .
\end{equation}
\end{lemma}

\begin{proof}   We  have
\begin{equation}
|B_\omega A_\omega|= W_\omega^* B_\omega A_\omega =
 W_\omega^* B_\omega U_\omega |A_\omega| =
 W_\omega^* B_\omega U_\omega |\overline{A_\omega}|^{\frac12}   |A_\omega|^{\frac12}\, ,
\end{equation}
where $W_\omega$ and $U_\omega$ are  partial isometries coming from the polar
decompositions of $B_\omega A_\omega$ and $A_\omega$ respectively.  Since
$|A_\omega|^{\frac12} \in \K_2$ and $B_\omega U_\omega |{A}_\omega|^{\frac12}
\in\K_2$, we may proceed as in Lemma~\ref{lemtrace} to conclude that $B_\omega
A_\omega \in \K_1$ and \eqref{K1bound} holds.
\end{proof}

\begin{proposition} \label{propK1}
 \noindent{\emph{\textbf{(i)}}} $\K_{1}$
is a normed vector space with the norm   $\tnorm{ \ }_1$.
.

\noindent{\emph{\textbf{(ii)}}} The conjugation
$A_\omega \to A_\omega^\ddagger$ is an isometry on $\K_{1}$, i.e.,
\begin{equation} \label{K1iso}
\tnorm{A_\omega^\ddagger}_1 =\tnorm{A_\omega}_1 \, .
\end{equation}

\noindent{\emph{\textbf{(iii)}}} $\K_{1}^{(0)}$ is dense in $\K_{1}$
\end{proposition}

\begin{proof}  We first prove the triangle inequality for  $\tnorm{ \ }_1$.
So let $A_\omega, B_\omega \in \K_1$.  We have
\begin{equation}
 |A_\omega + B_\omega| =W_\omega^*  (A_\omega + B_\omega)=
W_\omega^*A_\omega + W_\omega^* B_\omega \, ,
\end{equation}
with $W_\omega$ a partial isometry. It follows from Lemmas~\ref{lemtrace} and
\ref{lemtrace2} that $A_\omega + B_\omega \in K_1$ and $\tnorm{A_\omega +
B_\omega}_1 \le \tnorm{A_\omega  }_1 + \tnorm{B_\omega}_1$.  We conclude that
$\K_1$ is a normed space.

Given $A_\omega \in \K_1$, we have
\begin{align}\nonumber
\chi_0 |A_\omega^\ddagger| \chi_0 &=\chi_0 V_\omega^* A_\omega^\ddagger \chi_0
=\chi_0 V_\omega^* A_\omega^* \chi_0 =
 \chi_0 V_\omega^* |A_\omega |U_\omega^* \chi_0\\ & =
 \left(\overline{\chi_0 V_\omega^*|A_\omega |^{\frac12}}\right)
\left(|A_\omega |^{\frac12}U_\omega^* \chi_0\right)
\, ,
\end{align}
where $\overline{A_\omega}=  U_\omega |\overline{A_\omega}|$ and
$\overline{A_\omega^\ddagger}=  V_\omega |\overline{A_\omega^\ddagger}|$, and
the operators in parentheses are Hilbert-Schmidt by Propositions~\ref {propK2}
and \ref{propK25}.  It also follows that
\begin{equation}
\tnorm{A_\omega^\ddagger  }_1 \le \tnorm{A_\omega  }_1 \, .
\end{equation}
Since $A= A^{\ddagger \ddagger}$, the reverse inequality follows, yielding  \eqref{K1iso}.

Finally, we prove that $\K_{1}^{(0)}$ is dense in $\K_{1}$.
Given  $A_\omega \in \K_1$, let
 $A_\omega^{(n)} \in \K_\infty$ be as  in \eqref{An}.   Since
\begin{equation}
\mathrm{Ran}
 \,\left(\textstyle{\frac 1 n}|\overline{A_\omega^\ddagger}|^2 +1 \right)^{-\frac12}  =
\D(|\overline{A_\omega^\ddagger}|)= \D(\overline{A_\omega^\ddagger}) \subset
\D(A_\omega^*) \, ,
\end{equation}
we have
\begin{equation}
{A_\omega^{(n)}}^* = A_\omega^* \left(\textstyle{\frac 1
n}|\overline{A_\omega^\ddagger}|^2 +1 \right)^{-\frac12}
\end{equation}
and
\begin{equation}
|A_\omega^{(n)}|^2 = A_\omega^* \left(\textstyle{\frac 1
n}|\overline{A_\omega^\ddagger}|^2 +1 \right)^{-1} A_\omega \le |A_\omega|^2 \,
,
\end{equation}
and hence $|A_\omega^{(n)}| \le |A_\omega|$.  It follows that
 $A_\omega^{(n)} \in \K_1^{(0)}$.  To prove
that we have $\tnorm{A_\omega -A_\omega^{(n)}}_1\to 0$, we first remark that by a similar argument we have
\begin{equation}
| A_\omega -A_\omega^{(n)}| \le  |A_\omega|\, .
\end{equation}
So let  $\{\varphi_k\}_{k \in \mathbb{N}}$ be an ortho-normal basis for the
subspace $\chi_0 \H$, we have
\begin{equation}
\tnorm{A_\omega -A_\omega^{(n)}}_1 =\E \left\{   \sum_{k \in \mathbb{N}}
\langle \varphi_k, |A_\omega -A_\omega^{(n)}|\varphi_k\rangle\right \}
\le \,  \tnorm{A_\omega}_1 <\infty,
\end{equation}
since $A_\omega \in \K_1$ and
\begin{equation}
\langle \varphi_k, |A_\omega -A_\omega^{(n)}|\varphi_k\rangle \le
\langle \varphi_k, |A_\omega|\varphi_k\rangle \, .
\end{equation}
On the other hand, using Jensen's inequality we get
\begin{align}
\langle \varphi_k, |A_\omega -A_\omega^{(n)}|\varphi_k\rangle &\le
\langle \varphi_k, |A_\omega -A_\omega^{(n)}|^2\varphi_k\rangle^{\frac12}\\
&=\| (A_\omega -A_\omega^{(n)}) \varphi_k\| \to 0 \;\; \mbox{as $k \to \infty$}\, .
\nonumber
\end{align}
Thus   $\tnorm{A_\omega -A_\omega^{(n)}}_1\to 0$
 by the Dominated Convergence Theorem.
\end{proof}

We will denote  the (abstract) completion of $\K_1$ by $\overline{\K_1}$.

\begin{proposition}\label{notcomplete}
 The normed space $\K_1$ is not complete, i.e.,
 $\K_1\not=\overline{\K_1}$.
\end{proposition}

\begin{proof}
Let us denote by $\K_{mc,lb}^{\mathrm{(cst)}}$ and $K_1^{\mathrm{(cst)}}$ the
subset of \emph{constant} operators in $\K_{mc,lb}$ and $\K_1$,
 respectively.
In view of \eqref{cov}, $A \in \K_{mc,lb}^{\mathrm{(cst)}}$ can always be
written in the form
\begin{equation} \label{per}
 A = \sum_{x,y \in \Z^d} \chi_x  U(x) S_{x-y} U(-y) \chi_y \, ,
\end{equation}
where $S=\{S_x\}_{x \in \Z^d}$ is a family of bounded operators in $\chi_0 \H$
such that the series $\sum_{x\in \Z^d} \chi_x  U(x) S_{x}  \chi_0 $ converges
strongly to a bounded operator.  A sufficient condition for the latter is
\begin{equation} \label{S2}
 \sum_{x\in \Z^d} \|S_x\|^2  <  \infty \, .
\end{equation}

Operators $A$ as in \eqref{per} can be partially diagonalized by a Floquet
transform given by
\begin{equation}
\Fc =
 (2\pi)^{-\frac d 2} \sum_{x\in \Z^d}   \mathrm{e}^{i k \cdot x} U(-x) \chi_{x}\, ,
\end{equation}
a unitary map from $\H=\mathrm{L}^2(\mathbb{R}^d, {\mathrm{d}}x)$ to
 $\mathrm{L}^2(\mathbb{T}^d, {\mathrm{d}}k; \chi_0 \H)$, where
$\mathbb{T}^d = [-\frac \pi 2, \frac \pi 2)^{d}$ is the $d$-dimensional torus. Its inverse,
$\Fc^*$, is given by
\begin{equation}
\Fc^* =  (2\pi)^{-\frac d 2} \sum_{x\in \Z^d} \chi_{x}  U(x)
\langle  \mathrm{e}^{i k \cdot x}, \cdot \rangle_{\mathrm{L}^2(\mathbb{T}^d, {\mathrm{d}}k)}
\end{equation}

For $A$ as in \eqref{per} with  $\sum_{x\in \Z^d} \|S_x\|^2 <\infty $ we have
\begin{equation}
({\Fc} A {\Fc}^*  \Phi)(k)=   \hat{A}(k)  \Phi(k)   \;\; \mbox{ for all $\Phi \in {\Fc} \H_c$ } \, ,
\end{equation}
where
\begin{equation} \label{hatAk}
 \hat{A}(k) =   (2\pi)^{-\frac d 2} \sum_{x\in \Z^d}   \mathrm{e}^{i k \cdot x} S_x \, .
\end{equation}
Since ${\Fc}$ is unitary, in this case we also  have
\begin{equation}
({\Fc}| A| {\Fc}^*  \Phi)(k)=  | \hat{A}(k)|  \Phi(k)   \;\; \mbox{ for all $\Phi \in {\Fc} \H_c$} \, ,
\end{equation}
and
\begin{equation}\label{Ak}
\tnorm{A}_1 =
\tr\, \chi_0 |A| \chi_0 =   (2\pi)^{- d } \int_{\mathbb{T}^d} \tr \,  | \hat{A}(k)|  \, \mathrm{d} k \, .
\end{equation}
It follows that the completion  $\overline{\K_1^{\mathrm{(cst)}}}$ of
 $\K_1^{\mathrm{(cst)}}$
is isomorphic to the Banach space
 $$\mathrm{L}^1(\mathbb{T}^d, (2\pi)^{- d }{\mathrm{d}}k; \T_1(\chi_0 \H))\; ,$$ where
$\T_1(\chi_0 \H))$ denotes the Banach space of trace class operators on $\chi_0 \H$.

To see that there are elements in
 $\mathrm{L}^1(\mathbb{T}^d, (2\pi)^{- d }{\mathrm{d}}k; \T_1(\chi_0 \H))$
that do not correspond to operators in
  $\K_1^{\mathrm{(cst)}}$, let
us consider  $A$ as in \eqref{per} with
$S_x = s_x  Y$ for all $x \in \Z^d$,  where $Y \in \T_1(\chi_0 \H))$
and the scalars $ \{s_x\}_{x \in \Z^d}$ are chosen such
$\hat{s}(k) \in \mathrm{L}^1(\mathbb{T}^d,{\mathrm{d}}k)$ but
$\hat{s}(k) \notin \mathrm{L}^2(\mathbb{T}^d,{\mathrm{d}}k)$,
where $\hat{s}(k)$ is defined as in \eqref{hatAk}.  (This can always be done.)
We clearly have
$\hat{A}(k) \in
\mathrm{L}^1(\mathbb{T}^d, (2\pi)^{- d }{\mathrm{d}}k; \T_1(\chi_0 \H))$,
but for each $\varphi \in \chi_0\H$ we have
\begin{equation}
\| A \varphi \|^2 = \left(\sum_{x\in \Z^d} |s_x|^2\right) \|Y \varphi\|^2 =
\|\hat{s}(k)\|_ {\mathrm{L}^2(\mathbb{T}^d,{\mathrm{d}}k)}^2 \|Y \varphi\|^2 = \infty \,
\end{equation}
unless $Y \varphi = 0$.  Thus $A \notin \K_1^{\mathrm{(cst)}}$ as it does not
contain $\H_c$ in its domain.  (In fact,  $A \notin
\K_{mc,lb}^{\mathrm{(cst)}}\,$.)

Note that we proved  that for any  $\varphi \in \chi_0\H$ we can find
$A \in \overline{K_1^{\mathrm{(cst)}}}$ which cannot be represented
 by an operator with $\varphi$ in its domain. In fact, we proved more:
for appropriate $Y$ the constructed $A$ has the property that
 its domain is disjoint from $\H_c$.
\end{proof}

\begin{remark}  More generally, it follows from
\eqref{cov} that $A_\omega \in \K_{mc,lb}$ can always be written in the form
\begin{equation} \label{per6}
 A = \sum_{x,y \in \Z^d} \chi_x  U(x) S_{\tau(-y)\omega, x-y} U(-y) \chi_y \, ,
\end{equation}
where $S_\omega=\{S_{\omega,x}\}_{x \in \Z^d}$ is a family of  bounded
operators on $\chi_0 \H$  such
that the
series $\sum_{x\in \Z^d} \chi_x  U(x)S_{\omega,x}  \chi_0 $
 converges strongly to a bounded operator.  As in \eqref{S2}, we have
\begin{equation} \label{S27}
\|A_\omega\chi_x\|^2  \le
 \sum_{y\in \Z^d} \| S_{\tau(-x)\omega, y}\|^2  , \;\; \text{and also}
\quad  \|A_\omega\chi_x\|^2_2 =
 \sum_{y\in \Z^d} \| S_{\tau(-x)\omega, y}\|^2_2 \, .
\end{equation}
In particular,
\begin{equation} \label{S278}
\tnorm{A_\omega}_2^2 =
 \sum_{y\in \Z^d} \E \left( \|S_{\omega, y}\|^2_2\right)  .
\end{equation}

In the constant case we could write  $\tnorm{A}_1$ as in \eqref {Ak},
 but we do not have a similarly simple expression for
 $\tnorm{A_\omega}_1$.
\end{remark}

Although $\K_1$ is not complete, it is closed in the following sense:

\begin{proposition} \label{K1Fatou}  Let $A_\omega \in \K_{mc,lb}$ and
suppose there exists a Cauchy sequence $A_{n,\omega}$ in $\K_1$ such that
$A_{n,\omega}\chi_0 \to A_{\omega}\chi_0$ weakly.  Then
 $A_\omega \in \K_1$ and  $A_{n,\omega}\to A_{\omega}$ in $\K_1$.
\end{proposition}

\begin{proof}
Let   $\overline{A_\omega}=  U_\omega |\overline{A_\omega}|$ be the polar
decomposition. It follows that
 \begin{equation}
U_\omega^* A_{n,\omega}\chi_0 \to |A_{\omega}|\chi_0 \;\;\;\;\mbox{weakly}.
\end{equation}
Thus, if   $\{\varphi_j\}_{j\in \mathbb{N}}$ is an ortho-normal basis for the
subspace $ \chi_0\H$, we have, using Fatou's Lemma,
\begin{align}
\tnorm{A_\omega}_1& = \E  \sum_{j\in \N}  \la\varphi_j,|A_\omega|\varphi_j \ra =
\E  \sum_{j\in \N} \lim_{n \to \infty}
 \left|\la\varphi_j,U_\omega^* A_{n,\omega}\varphi_j \ra\right|\\
&  \le   \liminf_{n \to \infty}\E  \sum_{j\in \N}
 \left|\la\varphi_j,U_\omega^* A_{n,\omega}\varphi_j \ra\right|
 \le \liminf_{n \to \infty}  \tnorm{A_{n,\omega}}_1 < \infty \, , \nonumber
\end{align}
and hence $A_\omega \in \K_1$.

For fixed $m$ we have that $A_{n,\omega} -A_{m,\omega}$ is a Cauchy sequence in $\K_1$,
and that $(A_{n,\omega} -A_{m,\omega})\chi_0 \to (A_{\omega} -A_{m,\omega})\chi_0$
weakly as $n \to \infty$.  Thus the above argument gives
\begin{equation}
\tnorm{A_{\omega} -A_{m,\omega}}_1 \le
\liminf_{n \to \infty}  \tnorm{A_{n,\omega} -A_{m,\omega}}_1 \to 0
\;\;\mbox{as $m \to \infty$}.
\end{equation}
\end{proof}

\begin{corollary} Let $\K_{1,2} =\K_1 \cap \K_2 $ with the norm
 $\tnorm{ \ }_{1,2}=\tnorm{ \ }_1 + \tnorm{ \ }_2$.  Then $\K_{1,2}$ is a
Banach space.
\end{corollary}

The corollary is an immediate consequence of Propositions~\ref{propK2}(i) and
\ref{K1Fatou}. Its value is that given a sequence $A_{n,\omega} \in \K_{mc,lb}$
which converges in  $\overline{\K_1}$, if it also converges in $\K_2$ then its
limit in $\overline{\K_1}$ is actually in $\K_1$.

Left and right multiplication by elements of $\K_\infty$ leave $\K_1$ invariant.

\begin{proposition}\label{lemRM3}  $\K_1 \subset  \K_{\odot}$.
Moreover, if $A_\omega \in \K_1$ and $B_\omega \in K_\infty$
we have  $B_\omega \odot_L A_\omega,
A_\omega \odot_R B_\omega \in \K_1$ with
\begin{align} \label{LK1}
\tnorm{ B_\omega \odot_L A_\omega}_1 &\le   \tnorm{B_\omega}_\infty
\tnorm{A_\omega}_1 \, ,\\
\tnorm{ A_\omega \odot_R B_\omega }_1 & \le   \tnorm{B_\omega}_\infty
\tnorm{A_\omega}_1 \, . \label{RK1}
\end{align}
\end{proposition}

\begin{proof}  We have $B_\omega  \odot_LA_\omega \in \K_2$
and \eqref{LK2} from  Lemma~\ref{lemtrace2}, so it follows from
 Proposition~\ref{propK1}(ii) that
  $\K_1 \subset  \K_{\odot}$.
 The estimate
 \eqref{RK1} follows from \eqref{rightM}, \eqref{LK1}, and \eqref{K1iso}.
\end{proof}

We consider one other sort of multiplication, namely the bilinear map $
\Diamond\colon \, \K_2^{(0)} \times \K_2^{(0)} \to \K_1$ given by
\begin{equation}
A_\omega\diamond B_\omega := \Diamond (A_\omega,B_\omega)
=A_\omega B_\omega \, .
\end{equation}

\begin{proposition} \label{propD}
We have
\begin{equation}\label{diamond}
\tnorm{A_\omega\diamond B_\omega}_1 \le \tnorm{A_\omega}_2 \tnorm{B_\omega}_2
\;\; \;\;\mbox{for all $A_\omega, B_\omega \in \K_2^{(0)}$}.
\end{equation}
Thus $\Diamond$ extends by continuity to a bilinear map
 (we do not change notation)
 $\Diamond\colon  \,  \K_2  \times  \K_2 \to\overline{ \K_1}$,
which satisfies \eqref{diamond} and has dense range.  In fact,
\begin{equation}\label{K1D0}
\K_1^{(0)}= \Diamond\left(\K_2^{(0)}\times \K_2^{(0)}\right)
\end{equation}
and
\begin{equation}
\K_1 \varsubsetneq\mathrm{Ran}\, \Diamond \, .  \label{K1D}
\end{equation}
Moreover,
given  $A_\omega, B_\omega \in \K_2$, we have
\begin{align}  \label{diamond1}
A_\omega\diamond B_\omega &=
A_\omega\odot_L B_\omega \;\;\mbox{if}\;\; A_\omega \in \K_2^{(0)} \, ,
\\ \label{diamond2}
A_\omega\diamond B_\omega &=
A_\omega\odot_R B_\omega \;\;\mbox{if}\;\; B_\omega \in \K_2^{(0)} \, \\
\left(A_\omega\diamond B_\omega \right)^\ddagger &=
B_\omega^\ddagger\diamond A_\omega^\ddagger \, . \label{diamond3}
\end{align}
\end{proposition}

\begin{proof} To prove \eqref{diamond} we proceed as in the proof of
Lemma~\ref{lemtrace2}.  The  inclusion in \eqref{K1D} was exhibited in the
proof of Lemma~\ref{lemtrace}; note that it also gives \eqref{K1D0}.
\eqref{diamond1} is proven by an approximation argument. \eqref{diamond3}
follows from the special case when $A_\omega, B_\omega \in \K_2^{(0)}$ and
\eqref{K1iso}. \eqref{diamond2} follows from \eqref{diamond1}, \eqref{diamond3}
and \eqref{rightM}.

To show that we do not have equality in \eqref{K1D} we proceed as
 in the proof of
Proposition~\ref{notcomplete}. Let   $A$ be as in \eqref{per} with
$S_x = s_x  Z$ for all $x \in \Z^d$,  where $Z \in \T_2(\chi_0 \H))$
and
$\hat{s}(k) \in \mathrm{L}^2(\mathbb{T}^d,{\mathrm{d}}k)$ but
$\hat{s}(k) \notin \mathrm{L}^4(\mathbb{T}^d,{\mathrm{d}}k)$.
  (This can always be done.) Then $A \in \K_2$ but $A \diamond A \notin \K_1$
since $\hat{s}(k)^2 \notin \mathrm{L}^2(\mathbb{T}^d,{\mathrm{d}}k)$.
\end{proof}

\begin{lemma}\label{stK1}  Let $B_{n,\omega}$ be a bounded sequence in
$\K_\infty$ such that $B_{n,\omega}\to B_\omega$ strongly.  Then for all
$A_\omega \in \K_1$ we have $ B_{n,\omega} \odot_L A_\omega\to B_{\omega}
\odot_L A_\omega$ and $A_\omega \odot_R  B_{n,\omega} \to A_\omega \odot_R
B_{\omega}$ in $\K_1$.
\end{lemma}

\begin{proof}  Again it suffices to prove the result for left multiplication
in view of \eqref{rightM}.  Since the sequence $B_{n,\omega}$ is bounded and
$\K_1^{(0)}$ is dense in $\K_1$ it suffices to prove the result for $A_\omega
\in  \K_1^{(0)}$. But then we can write $A_\omega = C_\omega D_\omega =
C_\omega \diamond D_\omega$, with  $C_\omega , D_\omega \in  \K_2^{(0)}$. Since
\begin{equation}
B_{n,\omega} \odot_L A_\omega = B_{n,\omega}  C_\omega D_\omega =
( B_{n,\omega}  C_\omega) D_\omega =
( B_{n,\omega}  \odot_L C_\omega)\diamond D_\omega \, ,
\end{equation}
the desired conclusion follows from Lemma~\ref{stK2} and Proposition~\ref {propD}.
\end{proof}

 \subsection{The trace per unit volume}
Given $A= A_\omega \in \K_1$ we define
\begin{equation}
\T(A) = \E\left\{ \tr\left\{ \chi_0 A_\omega \chi_0\right\}\right\}.
\end{equation}
Lemma~\ref{lemtrace} says that $\T$ is a well defined linear functional on
$\K_1$ such that
\begin{equation}\label{Testimate}
|\T(A)| \le \tnorm{A}_1 \, .
\end{equation}
In fact, $\T$ is the \emph{trace per unit volume}.

\begin{proposition} Given $A= A_\omega \in \K_1$ we have
\begin{equation} \label{tuv}
\T(A) = \lim_{L \to \infty}\, \textstyle{ \frac 1 {|\Lambda_L|}}
\tr\left\{\chi_{\Lambda_L} A_\omega \chi_{\Lambda_L}\right\}  \;\;\;\mbox{for $\P$-a.e. $\omega$} \, ,
\end{equation}
where $\Lambda_L$ denotes the cube of side $L =1,3,5,\ldots$ centered at $0$.
\end{proposition}

\begin{proof}  We have
\begin{align}
\tr\left\{\chi_{\Lambda_L} A_\omega \chi_{\Lambda_L}\right\}= \sum_{x \in \Z^d
\cap \Lambda_L} \tr\left\{\chi_x A_\omega \chi_x \right\} =
 \sum_{x \in \Z^d \cap \Lambda_L}
\tr\left\{\chi_0 A_{\tau(x)\omega} \chi_0 \right\}\, .
\end{align}
Thus \eqref{tuv} follows from \eqref{trace} and the ergodic theorem.
\end{proof}

\begin{lemma}\label{centralTK2}  Let $A_\omega, B_\omega \in \K_2$.  Then
\begin{equation}\label{centralK2}
\T(A_\omega\diamond B_\omega) =
\langle\langle A_\omega^\ddagger, B_\omega \rangle\rangle  \, .
\end{equation}
In particular we have centrality for the trace per unit volume:
\begin{equation}  \label{central}
\T(A_\omega\diamond B_\omega) =
\T(B_\omega\diamond A_\omega)   \, .
\end{equation}
Moreover, given $C_\omega \in \K_\infty$, we have
\begin{equation} \label{central2}
\T( (C_\omega\odot_L A_\omega)\diamond B_\omega) =
\T(A_\omega \diamond (B_\omega \odot_R C_\omega))   \, .
\end{equation}
\end{lemma}

Note that if $A_\omega, B_\omega \in \K_2^{(0)}$ equation  \eqref{central} reads
\begin{equation}  \label{central3}
\T(A_\omega B_\omega) =
\T(B_\omega A_\omega)   \, ,
\end{equation}
and  equation  \eqref{central2} reads
\begin{equation} \label{central5}
\T(C_\omega A_\omega B_\omega)=
\T(A_\omega  B_\omega C_\omega)   \, .
\end{equation}

\begin{proof}  It suffices to prove the Lemma for
 $A_\omega, B_\omega \in \K_2^{(0)}$, in which case it follows from
Propositions~\ref{propK2} and \ref{propK25}
\end{proof}

We also have a ``$\K_\infty$, $\K_1$'' version of centrality for the trace per
unit volume:

\begin{lemma} \label{lemmacentral} Let $A_\omega \in \K_1$ and
 $C_\omega \in \K_\infty$, then
\begin{equation}
\T(C_\omega\odot_L  A_\omega) =\T(A_\omega \odot_R C_\omega) \, .
\end{equation}
\end{lemma}

\begin{proof}  Just use $A_\omega =
(U_\omega |A_\omega|^{\frac12})\diamond |A_\omega|^{\frac12}$, with $U_\omega
|\overline{A_\omega}|$ the polar decomposition of $\overline{A_\omega}$, and
\eqref{central2}.
\end{proof}

We will also use the following lemmas.

\begin{lemma} \label{lemmaduality}  Let $A_\omega \in \K_1$ be such that
$\T(C_\omega\odot_L  A_\omega)=0$ for all  $C_\omega \in \K_\infty$.
Then $A_\omega = 0$.
\end{lemma}

\begin{proof}
Let
 $U_\omega
|\overline{A_\omega}|$  be the polar decomposition of $\overline{A_\omega}$.
Then $U_\omega\in \K_\infty$ and
$\tnorm{A_\omega}_1=  \T (U_\omega^* A_\omega)=0 $.
\end{proof}

\begin{lemma} \label{lemmaduality2}  Let $B_{n,\omega}$ be a bounded sequence in
$\K_\infty$ such that $B_{n,\omega}\to B_\omega$ weakly.  Then for all
$A_\omega \in \K_1$ we have $ \T (B_{n,\omega} \odot_L A_\omega) \to
\T(B_{\omega}\odot_L A_\omega)$ and
 $ \T ( A_\omega\odot_R B_{n,\omega}  ) \to
\T(A_\omega\odot_R B_{\omega})$.
\end{lemma}

\begin{proof}  It suffices to consider the case $B_\omega=0$.
If $U_\omega
|\overline{A_\omega}|$ is the polar decomposition,
\begin{equation}
\T (B_{n,\omega} \odot_L A_\omega)=
\T ( |A_\omega|^{\frac12} \diamond
 \{B_{n,\omega} \odot_L (U_\omega |A_\omega|^{\frac12})\}) \to 0
\end{equation}
by dominated convergence.  The other limit then follows
 from Lemma~\ref{lemmacentral}.
\end{proof}

\subsection{The connection with noncommutative integration} \label{noncom}

There is a connection with noncommutative integration:  $\K_\infty$ is a
von Neumann algebra, $\T$ is a faithful normal semifinite trace on $\K_\infty$, and
$\overline {\K_i}= \L^i (\K_\infty, \T)$ for $i=1,2$. (We assume that $\K_1^{(0)}$
is not trivial, which is guaranteed by  Assumption \ref{RandH} in view of
 Proposition~\ref{fHK}.)
But our explicit construction plays a very important role in our analysis.

That $\K_\infty$ is a von Neumann algebra can be seen a follows. Let
$$\widetilde{\H}:= \L^2 ((\Omega,\P); \H) = \int_{\Omega}^\oplus \H\, \di \P$$
(see \cite[Section XIII.16]{RS4} for the notation). Then the collection
$\widetilde{\K}_\infty$ of strongly measurable maps $A=A_\omega: \Omega \to
\mathcal{B}(\H)$ with $\tnorm{A_\omega}_\infty < \infty$, where
$\tnorm{A_\omega}_\infty$ is as in \eqref{norminfty},
 form the von Neumann algebra
 of decomposable operators on $\widetilde{\H}$  \cite[Theorems XIII.83 and XIII.84]{RS4}.
If we define unitary operators $\widetilde{U}(a)$ on $\widetilde{\H}$ for $a
\in \Z^d$ by $(\widetilde{U}(a) \Phi)(\omega)=U(a)  \Phi(\tau(-a) \omega)$ for
$\Phi \in \tilde{\H}$, it follows  that $\K_\infty = \{ A_\omega \in
\widetilde{\K}_\infty; \ [\widetilde{U}(a), A_\omega] = 0 \ \text{for all $a
\in \Z^d$}\}$, and hence $\K_\infty$ is a von Neumann algebra.

$\T$ is a faithful normal semifinite trace (e.g., \cite[Definition~2.1]{T}) on $\K_\infty$.
That $\T$ is faithful is clear; to see that $\T$ is
 normal note that the condition given in
\cite[Theorem~2.7.11(i)]{BR} can be verified using
 properties of the usual trace and  the monotone convergence theorem. To show that
$\T$ is semifinite, pick a self-adjoint $0\neq B_\omega \in \K_1^{(0)}$, note
that we have the orthogonal projections $Q_{n,\omega} := \chi_{[-n,n]}(B_\omega) \in\K_1^{(0)} $ by
Lemma~\ref{lem|A|}, and hence   we conclude that $\T$ is semifinite since
$Q_{n,\omega} \nearrow I$ strongly.

Note that if   $A_\omega \in \K_{mc,lb}$ then its closure
$ \overline{A_\omega}$ is affiliated with $\K_\infty$ by Lemma~\ref{lem|A|}.
The converse cannot be true in view of Proposition~\ref{notcomplete}.


\section{Ergodic magnetic media}\label{sectRandomMedia}
 \setcounter{equation}{0}
\subsection{The ergodic Hamiltonian}
\label{subsectMSOrandom}

We now state the technical assumptions on our ergodic Hamiltonian $H_\omega$.
\begin{assumption}\label{RandH}
  The ergodic Hamiltonian $\omega \mapsto H_\omega$ is a measurable map
 from the
  probability space $(\Omega, \P)$ to the self-adjoint operators on $\H$ such
  that
  \begin{equation}
    H_\omega= H(\A_\omega,V_\omega) = \left(-i\nabla - \A_\omega\right)^2 +
    V_\omega \; ,
  \end{equation}
  almost surely, where $\A_\omega$ ($V_\omega$) are vector (scalar) potential
  valued random variables which satisfy the Leinfelder-Simader conditions
(see Subsection~\ref{subsectMSO})
  almost surely. It is furthermore assumed that $H_\omega$ is covariant:
  \begin{equation}
    U(a) H_\omega U(a)^* = H_{\tau(a) \omega} \text{ for all } a \in \Z^d \; .
  \end{equation}
\end{assumption}

Measurable in this context means that $\langle \psi,H_\omega \phi \rangle$ is a
Borel measurable function for every $\psi, \phi \in C^\infty_c(\R^d)$. As a
consequence $f(H_\omega) \in \K_\infty$ for every bounded Borel function $f$ on
the real line.  (The only subtle point here is measurability, but that is well
known. See \cite{PF}.)

Note that it follows from ergodicity that
 ${V_\omega}_-$ satisfies \eqref{relbound} almost surely
 with \emph{the same
constants} $\alpha,\beta$.

We remark that much more detailed knowledge of $H_\omega$ is required to verify
Assumption \ref{assumptiona} below, at least for ${\zeta}_\omega =
P^{(E_F)}_\omega$.  In particular, one might require $V_\omega$ to be of the
form $V_\omega(x)  = \sum_{a\in \Z^d} \eta_a u(x-a)$, where $\eta_a$ are
independent, identically, distributed random variables and $u$ is a function of
compact support.  However, the only fact we need here regarding localization
for ergodic Schr\"odinger operators  is \eqref{assumption} below for suitable
functions $h$.  Thus we prefer to take the general Assumption~\ref{RandH} and
note that Assumption~\ref{assumptiona} for $\zeta_\omega= P^{(E_F)}_\omega$
follows, for suitable $\A_\omega,V_\omega$ and $E_F$, by the methods of, for
example, \cite{GK1,BGK,AENSS,GK5}.

It is absolutely crucial to our analysis that the parameters $\alpha,\beta$ in
the Leinfelder-Simader conditions may be chosen independently of $\omega$. In
particular, this allows us to prove:

\begin{proposition}
\label{fHK} Let $f$ be a  Borel measurable function on the real line such that
$ \|f \Phi_{d,{\alpha},\beta}\|_\infty < \infty$, where $
\Phi_{d,{\alpha},\beta}$ is given in \eqref{PhiE}. Then
\begin{description}
\item[(i)] We have $f(H_\omega) \in \K_1^{(0)}$, and if \
$\|f^2 \Phi_{d,{\alpha},\beta}\|_\infty < \infty$ then $f(H_\omega) \in
\K_2^{(0)}$.

\item[(ii)]  If  $f(H_\omega)= g(H_\omega)$ for some $g \in \S(\R) $, we have
$\left[x_j, f(H_\omega)\right] \in \K_1^{(0)} \cap \K_2^{(0)} $, $j=1,2,\dots,d$.

\item[(iii)] If $f(H_\omega)= g (H_\omega)h(H_\omega)$ with $g \in \S(\R) $
and $h$  a  Borel measurable function with $ \|h^2\Phi_{d,{\alpha},\beta}\|_\infty < \infty$,
and  for some $j \in \{1,2,\ldots,d\}$ we have
 $\left[x_j, h(H_\omega)\right] \in \K_2$, then we also have
$\left[x_j, f(H_\omega)\right] \in \K_1\cap \K_2$.

\item[(iv)] We have $P^{(E)}_\omega \in
\K_1^{(0)} \cap \K_2^{(0)}$, where $P^{(E)}_\omega
=\chi_{(-\infty,E]}(H_\omega)$, i.e., $P^{(E)}_\omega =f(H_\omega)$ with   $f=
\chi_{(-\infty, E]}$. If in addition we have
 $\left[x_j, P^{(E)}_\omega\right] \in \K_2$ for some $j \in \{1,2,\ldots,d\}$, then we also have
 $\left[x_j, P^{(E)}_\omega\right] \in \K_1$.

\item[(v)]  If $f$  is as in either (ii), (iii), or (iv), we also have
\begin{equation}\label{Tcom=0}
\T\left\{ \left[x_j, f(H_\omega)\right]\right\} = 0 \, .
\end{equation}

\end{description}

\end{proposition}
\begin{proof}   (i) is an immediate consequence of \eqref{trSGEE2}.
To prove (ii), first note that $\left[x_j, f(H_\omega)\right]$ is in
$\K_\infty$ by Proposition~\ref{x,f}(ii).  We recall that \cite[Eq.
(3.8)]{GK3}
\begin{equation}
\| \chi_x
 {f({H_\omega})} \chi_0\|_2^2
  \le {C}_{d,\alpha,\beta,\nu,k}\,
 \left\|  f
\Phi_{d,{\alpha},\beta}\right\|_\infty    |\!|\!|g|\!|\!|_{k+2} \,
 \left\langle x \right\rangle^{- k+2\nu }
 \label{E23}
\end{equation}
for $\mathbb{P}$-a.e.~$\omega$ and all $k=1,2,\ldots$ and   $\nu > \frac d 4$,
and set  $\a$ to be a step function approximation to
the operator  $\x$; i.e., $\a$ is the operator given by multiplication by the discretized
coordinates $a \in \Z^d$:
$
\a= \sum_{a \in \Z^d} a \chi_a
$.
Note that multiplication by $ x_j - a_j$ is a bounded operator for each  $j \in \{1,2,\ldots,d\}$;
 in fact, we have
$
\| x_j - a_j\|\le  \frac 1 2
$.
Since
\begin{equation}
[x_j,f({H_\omega})] = [a_jf({H_\omega})] + [x_j - a_j,f({H_\omega})]\, ,
\end{equation}
to prove  $\left[x_j,
f(H_\omega)\right] \in \K_2 $ it suffices to prove  $\left[a_j,
f(H_\omega)\right] \in \K_2 $.
This follows from \eqref {E23} with sufficiently large $k$:
\begin{align}\notag
& \left\| [a_j,f({H_\omega})] \chi_0\right\|_2^2 =
\Biggl\|  \sum_{a \in Z^d} \chi_a [a_j,f({H_\omega})] \chi_0\Biggr\|_2^2\\
& \qquad =  \sum_{a \in Z^d}  \left\| \chi_a [a_j,f({H_\omega})] \chi_0\right\|_2^2
= \sum_{a \in Z^d} |a_j|^2 \left\| \chi_a f({H_\omega}) \chi_0\right\|_2^2\\& \qquad
\le  {C}_{d,\alpha,\beta,\nu,k}\,
 \left\|  f
\Phi_{d,{\alpha},\beta}\right\|_\infty    |\!|\!|g|\!|\!|_{k+2}  \sum_{a \in Z^d} |a_j|^2
 \left\langle a \right\rangle^{- k+2\nu }.
\nonumber
\end{align}

 That $\left[x_j, f(H_\omega)\right]$ it is also in $\K_1$ follows from (iii),
since we can write $g(t) =( \la t\ra^{n} g(t))\la t\ra^{-n}$ with $n \in \N$,
$( \la t\ra^{n} g(t))\in \S(\R)$ and $h(t)= \la t\ra^{-n}$ is as in (iii) for
$n$ large.

To prove (iii), we note that   $\left[x_j,  g (H_\omega)\right] \in \K_\infty$
by \eqref{HSbound} and, since  $\left[x_j, h(H_\omega)\right] \in \K_2$, $x_j
h(H_\omega) \chi_0$ is a bounded operator. Hence
\begin{align}
\left[x_j, f(H_\omega)\right] \chi_0&=
\left[x_j,  g (H_\omega)h(H_\omega)\right]\chi_0\\
&=\left[x_j,  g (H_\omega)\right] h(H_\omega) \chi_0 + g (H_\omega)\left[x_j,
h(H_\omega)\right]\chi_0 \, . \nonumber
\end{align}
Noting that $g(H_\omega),h(H_\omega)\in  \K_2$ by (i), we conclude that
\begin{equation}
\left[x_j, f(H_\omega)\right]=
\left[x_j,  g (H_\omega)\right]\odot_R h(H_\omega)
+ g (H_\omega)\odot_L \left[x_j,  h(H_\omega)\right]  \in \K_2 \, ,
\end{equation}
and, as $\left[x_j,  g (H_\omega)\right] \in \K_2$ by (ii),
\begin{equation}\label{xfK1}
\left[x_j, f(H_\omega)\right]=
 \left[x_j,  g (H_\omega)\right]\diamond h(H_\omega)
+ g (H_\omega)\diamond \left[x_j,  h(H_\omega)\right]  \in \K_1 \, .
\end{equation}

Item (iv) is an immediate consequence of  (i) and (iii).   To see (v), note
$x_j \chi_0 = \chi_0 x_j \chi_0 $ is bounded and $\chi_0 f(H_\omega) x_j \chi_0
= (\chi_0f(H_\omega) \chi_0) (x_j \chi_0)$ is trace class. Since $\left[x_j,
f(H_\omega)\right] \in \K_1$, we conclude that $\chi_0 x_j f(H_\omega) \chi_0$
is also trace class, and
\begin{equation}
\T\left\{ \left[x_j, f(H_\omega)\right]\right\} =
\E\, \tr \left( \chi_0 x_j f(H_\omega)\chi_0 \right) -
\E\, \tr \left( \chi_0 f(H_\omega) x_j\chi_0 \right)=0
\end{equation}
using centrality of the ordinary trace $\tr$.
\end{proof}


\subsection{Commutators  of measurable covariant operators}

In this subsection, $H_\omega$ stands either for the time independent
$H_\omega$ or for  $H_\omega(t)$ incorporating a time-dependent electric
 field. By
$H_\omega A_\omega \in \K_i$ we mean $A_\omega \H_c \subset \D$ and the
operator $H_\omega A_\omega$ with domain $\H_c$ is in $\K_i$.

\begin{definition}\label{defcommut}
We define the following (generalized)  commutators:
\begin{description}
\item[(i)]
If $A_\omega\in\K_{\odot} $ and $B_\omega\in\K_\infty$, then
\begin{align}\label{defcombd1}
[B_\omega,A_\omega]_\odot & =
B_\omega \odot_L A_\omega - A_\omega \odot_R  B_\omega \ \in\K_{\odot} \, ,\\
\label{defcombd2}
[A_\omega,B_\omega]_\odot &=
A_\omega \odot_R B_\omega - B_\omega \odot_L  A_\omega
= \left([B_\omega^\ast,A_\omega^\ddagger]_\odot\right)^\ddagger \in\K_{\odot}  .
\end{align}
\item[(ii)]
If $A_\omega,B_\omega\in\K_2$, then
\begin{equation}\label{defcomK2}
[B_\omega,A_\omega]_\diamond  =
 B_\omega \diamond A_\omega -  A_\omega \diamond B_\omega \;
\in\overline{\K_1}\, .
\end{equation}
\item[(iii)]
If $A_\omega\in\K_\odot$ is such that $H_\omega A_\omega$ and $H_\omega
A_\omega^\ddagger$ are in $\K_\odot$, then
\begin{equation}\label{defcomHA}
[H_\omega,A_\omega]_\ddagger = H_\omega A_\omega -
 (H_\omega A_\omega^\ddagger)^\ddagger \in\K_\odot \, .
\end{equation}
\end{description}
\end{definition}

\begin{remark}  These commutators agree when any two of them
make sense.  More precisely:
\begin{description}
\item[(a)]  If $A_\omega,B_\omega\in\K_\infty$ then
 $[B_\omega,A_\omega]_\odot = [B_\omega,A_\omega]
=B_\omega  A_\omega - A_\omega  B_\omega $, the usual commutator.
\item[(b)]\eqref{defcomK2} agrees with either \eqref{defcombd1} or \eqref{defcombd2}
if either $B_\omega$ or $A_\omega$ are in $\K_\infty$.

\item[(c)] \eqref{defcomHA} should be interpreted as an extension of
\eqref{defcombd1} to unbounded $B_\omega$. Note that \eqref{defcombd1} can be
rewritten as $[B_\omega,A_\omega]_\odot = B_\omega A_\omega - (B_\omega^\ast
A_\omega^\ddagger)^\ddagger$, and the right hand side makes sense as long as
$B_\omega A_\omega$ and $B_\omega^\ast A_\omega^\ddagger$ are in $\K_{mc,lb}$.
In addition, \eqref{defcomHA} reduces to the usual commutator on $\H_c\cap \D$,
as shown in the following lemma.

\end{description}
\end{remark}

\begin{lemma}\label{lemcomHA}
Let $A_\omega\in K_\odot$ be such that $H_\omega A_\omega\in K_\odot$.
Then
\begin{equation}
(H_\omega A_\omega)^\ddagger\psi = A_\omega^\ddagger H_\omega\psi
\;\;\;\mbox{for all  $\psi\in \H_c\cap \D$}\, . \label{adjHAi}
\end{equation}
In addition, we have $\D((H_\omega A_\omega)^*) \cap \D
=\D(A_\omega ^\ast H_\omega )  $
and
\begin{equation}
(H_\omega A_\omega)^*\psi = A_\omega^*H_\omega\psi
\;\;\;\mbox{for all  $\psi\in D((H_\omega A_\omega)^*) \cap \D $}\, . \label{adjHAi8}
\end{equation}
As a consequence, if $H_\omega A_\omega$ and $H_\omega A_\omega^\ddagger$ are
in $\K_\odot$, then
\begin{equation}
 [H_\omega,A_\omega]_\ddagger \psi= H_\omega A_\omega\psi - A_\omega H_\omega\psi
\;\;\;\mbox{for all  $\psi\in \H_c\cap \D$}\, .
\end{equation}

\end{lemma}

\begin{proof}
If $H_\omega A_\omega\in K_\odot$, for all
   $\psi\in \H_c\cap \D$ and  $\xi\in \H_c$ we have
\begin{equation}
\la (H_\omega A_\omega)^\ddagger \psi, \xi \ra = \la \psi, H_\omega A_\omega\xi \ra
= \la H_\omega\psi,  A_\omega\xi \ra =
\la A_\omega^\ddagger H_\omega\psi,  \xi \ra  \, ,
\end{equation}
where we used the fact that $H_\omega \psi \in \H_c$ since $H_\omega$ is a local
operator.  Thus \eqref{adjHAi} follows. A similar argument proves \eqref{adjHAi8}.
\end{proof}

The following lemma will also be useful.

\begin{lemma}\label{formulaCommutator}
  Let $A_\omega, B_\omega \in \K_2$, $C_\omega \in \K_\infty$.  Then
\begin{equation}
\T \left\{  [C_\omega, A_\omega]_\odot \diamond B_\omega\right\}=
\T \left\{ C_\omega \odot_L [A_\omega, B_\omega]_\diamond\right\}\, .
\end{equation}
\end{lemma}

\begin{proof}
It follows from \eqref{defcombd1}, \eqref{defcomK2}, and Lemma~\ref{centralTK2}.
\end{proof}


\subsection{Time evolution on spaces of covariant  operators}

For $\P$-a.e. $\omega$ let $U_\omega(t,s)$ be the  unitary propagator given by
Theorem~\ref{tpropagator}.   Note that $U_\omega(t,s) \in
\K_\infty$. (Since we apply Theorem~\ref{tpropagator} independently for
each $\omega$, there is the subtle question of measurability for
$U_\omega(t,s)$. However, measurability follows from the construction
\eqref{normlimit}, since the propagator $U_\omega(t,s)$ is expressed as a limit
of ``Riemann products,'' i.e., multiplicative Riemann sums, each of which is
manifestly measurable since it is a product of finitely many propagators
$\e^{-i \Delta t H_\omega(t_k)}$)

It will be important at times to keep track of the dependence of   $U_\omega(t,s)$
on the electric field $\El $, in which case we will write $U_\omega(\El, t,s)$.  Note that
\begin{equation} \label{U0}
U_\omega(\El= 0, t,s)  = U^{(0)}_\omega(t-s) := \e^{-i(t-s) H_\omega} \, .
\end{equation}
We omit $\El$ from the notation in what follows.

\begin{proposition} \label{propUomega}
Let
\begin{equation}  \label{UtsUst}
{\U}(t,s) (A_\omega) =U_\omega(t,s) \odot_LA_\omega \odot_R U_\omega(s,t)
\;\;\;\mbox{for \ $A_\omega \in \K_{\odot}\, $.}
\end{equation}
Then $\, {\U}(t,s)$ is a linear operator on $\K_{\odot}$, leaving $
\K_{\odot}$, $\K_\infty$, $\K_1$, and $\K_2$ invariant, with
\begin{align}
{\U}(t,r)\, {\U}(r,s) &= {\U}(t,s) \, ,\\
{\U}(t,t) &=  I \,,\\
\left\{\,{\U}(t,s) (A_\omega)\right\}^\ddagger&=
 {\U}(t,s) (A_\omega^\ddagger)\, \label{UAUdag} .
\end{align}
Moreover, ${\U}(t,s)$ is unitary on $\K_2$ and an isometry in
 $\K_1$ and $\K_\infty$; it
 extends to an isometry
on $\overline{\K}_1$ with the same properties. In addition,
 ${\U}(t,s)$ is jointly strongly continuous
in $t$ and $s$ on  $\overline{\K}_1$ and $\K_2$.
\end{proposition}

 \begin{proof}  The first part of the proposition follows from Propositions~\ref{propK2588},
\ref{propK25}, and \ref{lemRM3}.   $\, {\U}(t,s)$ is clearly  an isometry on $\K_\infty$.
To see that $\, {\U}(t,s)$ is an isometry on
 $\K_1$ and $\K_2$, note that  from Propositions~\ref{propK25} and \ref{lemRM3}
we have
\begin{equation}
\tnorm{\,{\U}(t,s) (A_\omega)}_i \le \tnorm{ A_\omega}_i
 \le \tnorm{\, {\U}(t,s) (A_\omega)}_i
\end{equation}
 for $i=1,2$, where we used  $A_\omega =
{\U}(s,t) \left (\,{\U}(t,s) (A_\omega)\right) $.  As for \eqref{UAUdag},  it follows
from \eqref{BACdag}.

The joint strong continuity of $\, {\U}(t,s)$ on  $\overline{\K}_1$ and
$\K_2$ follows from the  joint strong continuity of $U_\omega(t,s)$ on $\H$ and
Lemmas~\ref{stK2} and \ref{stK1}.
\end{proof}

\begin{lemma}\label{allr}
  Let  $A_\omega\in \K_i$ be such that
$H_\omega(r_0)A_\omega \in \K_i$ for some $r_0 \in [-\infty, \infty)$, where
$i\in \{\odot,1,2,\infty\}$. Then
$H_\omega(r)A_\omega \in \K_i$ for all $r \in [-\infty, \infty)$.
\end{lemma}

\begin{proof} In view of \eqref{HtH2} it suffices to show $\Db_{j,\omega } A_\omega\in \K_i$
if  $H_\omega(r_0)A_\omega \in \K_i$ for some $r_0 \in [-\infty, \infty)$.  But this follows
 immediately from \eqref{DbH}.
\end{proof}

\begin{proposition}\label{derivU1}   Let
 $A_\omega\in \K_i$ be such that
  $H_\omega(r_0)A_\omega$ and
$H_\omega(r_0)A_\omega^\ddagger$ are  in $\K_i$ for some
 $r_0 \in [-\infty, \infty)$
Then the map $r\to {\U}(t,r)(A_\omega)\in \K_i$ is differentiable in
 $\K_i$, and
\begin{equation}\label{eqderivtau}
i\partial_r\,  {\U}(t,r)(A_\omega) =
-{\U}(t,r)([H_\omega(r),A_\omega]_\ddagger) \, ,
\end{equation}
with $[H_\omega(r),A_\omega]_\ddagger$ defined in \eqref{defcomHA}.
\end{proposition}

\begin{proof}  Fix  $i=1$ or $i=2$.
All the expressions make sense as elements of $\K_i$.
 Write
\begin{align}\notag
&\frac ih\left(  \mathcal{U}(t,r+h)(A_\omega)-\mathcal{U}(t,r)(A_\omega)\right) \\
&\quad =
 \frac ih\left( U_\omega(t,r+h)-U_\omega(t,r)\right)
 \odot_L A_\omega \odot_R U_\omega(r+h,t) \label{difftau1}\\
& \qquad \quad  + \
 U_\omega(t,r)\odot_L A_\omega \odot_R
 \frac ih\left( U_\omega(r+h,t)-U_\omega(r,t)\right) .\label{difftau2}
\end{align}
We first focus on \eqref{difftau1}.
Since $H_\omega(r) A_\omega\in K_i$ by Lemma~\ref{allr}, one has
\begin{align}
B_\omega\odot_L A_\omega & =  B_\omega A_\omega
= B_\omega(H_\omega(r)+\gamma)^{-1}
 (H_\omega(r)+\gamma) A_\omega \\ & =
 B_\omega(H_\omega(r)+\gamma)^{-1} \odot_L (H_\omega(r)+\gamma) A_\omega \,
.\nonumber
\end{align}
Theorem~\ref{tpropagator} asserts that $$\frac1h\left(
U_\omega(t,r+h)-U_\omega(t,r)\right)(H_\omega(r)+\gamma)^{-1} \to
iU_\omega(t,r)H_\omega(r)(H_\omega(r)+\gamma)^{-1}$$ strongly with uniformly
bounded norm, as $h \to 0$. Using either Lemma~\ref{stK1} or Lemma~\ref{stK2},
 and the strong continuity of $U_\omega (r,t)$ in $r$, we get
\begin{align}
&\lim_{h\to 0} \frac ih\left( U_\omega(t,r+h)-U_\omega(t,r)\right)
 \odot_L A_\omega \odot_R U_\omega(r+h,t)  \label{diffleft}\\
 & \quad=
 -U_\omega(t,r)H_\omega(r)(H_\omega(r)+\gamma)^{-1}
 \odot_L (H_\omega(r)+\gamma) A_\omega \odot_R U_\omega(r,t) \nonumber \\
& \quad=
 -U_\omega(t,r) \odot_L H_\omega(r) A_\omega \odot_R U_\omega(r,t). \nonumber
\end{align}

We now turn to \eqref{difftau2}. Note that if $B_\omega\in \K_\infty$ then
\begin{equation}
A_\omega \odot_R B_\omega
= (B_\omega^\ast \odot_L A_\omega^\ddagger)^\ddagger
=\left(((H_\omega(r)+\gamma)^{-1}B_\omega)^\ast
\odot_L (H_\omega(r)+\gamma)A_\omega^\ddagger\right)^\ddagger  \, .
\end{equation}
Since the map $A_\omega\to A_\omega^\ddagger$ is an isometry on $\K_i$,
the same argument as above implies that
\begin{align}
&\lim_{h\to 0} U_\omega(t,r)\odot_L A_\omega \odot_R
 \frac ih\left( U_\omega(t,r+h)-U_\omega(t,r) \right)  \label{diffright}\\
 &\quad =
 U_\omega(t,r) \odot_L \left(((H_\omega(r)+\gamma)^{-1}H_\omega(r)U_\omega(r,t))^\ast \odot_L
(H_\omega(r)+\gamma)A_\omega^\ddagger\right)^\ddagger  \nonumber \\
 &\quad = \nonumber
 U_\omega(t,r) \odot_L (H_\omega(r)A_\omega^\ddagger)^\ddagger
 \odot_R  U_\omega(r,t) .
\end{align}
\end{proof}

\begin{proposition}\label{derivU2}   Let
 $A_\omega\in \K_i$ be such that
  $H_\omega(r_0)A_\omega$ and
$H_\omega(r_0)A_\omega^\ddagger$ are  in $\K_i$ for
 some $r_0 \in [-\infty, \infty)$, , where $i\in \{1,2,\infty\}$.
Then $H_\omega(t)U_\omega(t,r)A_\omega$,
 $H_\omega(t)U_\omega(t,r)A_\omega^\ddagger$,
 $H_\omega(t){\, \U}(t,r)(A_\omega)$, and
 $H_\omega(t)\, \U(t,r) (A_\omega^\ddagger) $ are in $\K_i$,
and the map $t\to {\U}(t,r)(A_\omega)\in \K_i$ is differentiable, with
\begin{equation}\label{eqderivU3}
 i\partial_t\,  {\U}(t,r)(A_\omega)
 =
[H_\omega(t),{\U}(t,r)(A_\omega)]_\ddagger  \, ,
\end{equation}
with the proviso that in $\K_\infty$ the meaning of the derivative is as
a bounded and $\P$-a.e.-weak limit.

Moreover,
  we have
\begin{align}\label{HUAW}
&\tnorm{\left(H_\omega(t)+ \gamma\right){\U}(t,r)(A_\omega)}_i \le
\tnorm{W_\omega(t,r)}_\infty
\tnorm{ (H_\omega(r)+\gamma) A_\omega}_i \, ,\\[.05in]
\label{HUAWbound}
&  \tnorm{[H_\omega(t),{\U}(t,r)(A_\omega)]_\ddagger}_i   \le \\
 &\qquad \qquad \tnorm{W_\omega(t,r)}_\infty\left( \tnorm{ (H_\omega(r)+\gamma) A_\omega}_i
+
 \tnorm{ (H_\omega(r)+\gamma) A_\omega^\ddagger}_i
\right) \, ,\notag
\end{align}
and, for all $\varphi\in\H_c\cap \D$,
\begin{align}\label{eqderivU3phi}
& [H_\omega(t),{\U}(t,r)(A_\omega)]_\ddagger \varphi =\\& \qquad \qquad
H_\omega(t)U_\omega(t,r)A_\omega^{\ddagger\ast}U_\omega(r,t)\varphi
-
U_\omega(t,r)A_\omega^{\ddagger\ast}U_\omega(r,t)H_\omega(t)\varphi.\notag
\end{align}
\end{proposition}

We  need the following lemma. (Recall that $\overline{A_\omega} =
A_\omega^{\ddagger\ast}$ for $A_\omega\in \K_{mc,lb}$.)

\begin{lemma}\label{lemadjHA}    Let $A_\omega\in\K_i$ with
$H_\omega(t)A_\omega \in\K_i$ {\rm (}$i\in \{\odot,1,2,\infty\}${\rm )}. If
 $\varphi\in\D(A_\omega^{\ddagger\ast})\cap \D((H_\omega(t)
A_\omega)^{\ddagger\ast})$, it follows that  $A_\omega^{\ddagger\ast}\varphi\in \D$
 and
\begin{equation} \label{adjHAii}
 (H_\omega(t)A_\omega)^{\ddagger\ast}\varphi =
 H_\omega(t)A_\omega^{\ddagger\ast}\varphi .
\end{equation}
As a consequence, $H_\omega(t)( A_\omega \odot_R C_\omega) \in\K_i$ for
  any $C_\omega\in\K_\infty$, and
\begin{equation} \label{adjHAii99}
(H_\omega(t)A_\omega) \odot_R
C_\omega = H_\omega(t)A_\omega^{\ddagger\ast}C_\omega
=H_\omega(t)( A_\omega \odot_R C_\omega) \, .
\end{equation}
\end{lemma}

Lemma~\ref{lemadjHA} can be seen as a generalization of \eqref{BACassoc}, where
$B_\omega\in\K_\infty$ is replaced by the unbounded operator $H_\omega(t)$
whose domain does not contain  $\H_c$.

\begin{proof}[Proof of Lemma~\ref{lemadjHA}.]  Let
 $\varphi\in\D(A_\omega^{\ddagger\ast})\cap \D((H_\omega(t)
A_\omega)^{\ddagger\ast})$  and $\psi\in\H_c\cap \D$, we have, using
Lemma~\ref{lemcomHA},
\begin{equation}\label{adjii1}
\la ({H_\omega(t)} A_\omega)^{\ddagger\ast}\varphi, \psi\ra = \la \varphi, ({H_\omega(t)}
A_\omega)^\ddagger\psi\ra =   \la \varphi,  A_\omega^\ddagger
{H_\omega(t)}\psi \ra  = \la A_\omega^{\ddagger\ast}\varphi,
{H_\omega(t)}\psi \ra\, .
\end{equation}
Since $\H_c\cap \D$ is a core for ${H_\omega(t)}$, it follows that
 $A_\omega^{\ddagger\ast}\varphi \in \D$ and
\begin{equation}
\la ({H_\omega(t)} A_\omega)^{\ddagger\ast}\varphi, \psi\ra = \la {H_\omega(t)}
A_\omega^{\ddagger\ast}\varphi, \psi \ra  .
\end{equation}
Since $\D\cap\H_c$ is dense in $\H$ (it contains
$\mathcal{C}_c^\infty(\R^d)$), \eqref{adjHAii} follows.
\end{proof}

\begin{proof}[Proof of Proposition~\ref{derivU2}.]
Since $H_\omega(r_0)A_\omega\in K_i$, $A_\omega\H_c \subset \D$.
 Since $U_\omega(t,r)\D\subset \D$, the operator
 $H_\omega(t)U_\omega(t,r)A_\omega$ is well-defined on $\H_c$ and (use
Lemma~\ref{allr})
\begin{equation}
H_\omega(t)U_\omega(t,r)A_\omega =
H_\omega(t)U_\omega(t,r)(H_\omega(r)+\gamma)^{-1}  \odot_L
 (H_\omega(r)+\gamma) A_\omega \in K_i \, ,
\end{equation}
as $H_\omega(t)U_\omega(t,r)(H_\omega(r)+\gamma)^{-1}= W_\omega(t,r)
-\gamma U_\omega(t,r)(H_\omega(r)+\gamma)^{-1} $
 is an element of $\K_\infty$.  The estimate \eqref{HUAW} follows.

Furthermore, as in \eqref{diffleft}, on account of Theorem~\ref{tpropagator} we
have
\begin{align}
&\lim_{h\to 0} \frac ih\left( U_\omega(t+h,r)-U_\omega(t,r)\right)
 \odot_L A_\omega \odot_R U_\omega(r,t+h)   \label{deriv2}\\
 &\qquad \quad =
H_\omega(t) U_\omega(t,r)(H_\omega(r)+\gamma)^{-1}
 \odot_L (H_\omega(r)+\gamma) A_\omega \odot_R U_\omega(r,t) \nonumber\\
 &\qquad \quad =
(H_\omega(t)U_\omega(t,r)A_\omega) \odot_R U_\omega(r,t) , \nonumber
\end{align}
where we used associativity of left and right multiplication in $\K_i$
 according to Proposition~\ref{propK2588}, and in $\K_\infty$
we took a bounded and $\P$-a.e.-weak limit.

By the same reasoning as above $H_\omega(t)U_\omega(t,r)A_\omega^\ddagger \in
\K_i$, and we have an estimate similar to \eqref{HUAW}.  Thus we can
differentiate the second term as in \eqref{deriv2} simply by using the
conjugates:
\begin{align}\label{deriv29}
&\lim_{h\to 0}  A_\omega \odot_R
 \frac ih\left( U_\omega(r,t+h)-U_\omega(r,t) \right) \\
 &\quad =
 \left(\lim_{h\to 0}\frac ih\left( U_\omega(t+h,r)-U_\omega(t,r) \right)
 \odot_L A_\omega^\ddagger\right)^\ddagger
=
 (H_\omega(t)U_\omega(t,r)A_\omega^\ddagger)^\ddagger . \nonumber
\end{align}

Combining \eqref{deriv2} and  \eqref{deriv29} we get
\begin{align}\label{eqderivU2}
&i\partial_t\,  \U (t,r)(A_\omega)=\\& \qquad \quad
(H_\omega(t)U_\omega(t,r)A_\omega) \odot_R U_\omega(r,t)
- U_\omega(t,r)\odot_L (H_\omega(t)U_\omega(t,r)A_\omega^\ddagger)^\ddagger
\, .\notag
\end{align}
Recalling that $H_\omega(t)U_\omega(t,r)A_\omega\in\K_i$, it follows from Lemma
\ref{lemadjHA} that
\begin{align}
(H_\omega(t)U_\omega(t,r)A_\omega) \odot_R U_\omega(r,t) &= H_\omega(t)
U_\omega(t,r) A_\omega^{\ddagger \ast} U_\omega(r,t) \nonumber \\ &=
H_\omega(t)\, \U_\omega(t,r) (A_\omega) \, .\label{HUA1}
\end{align}
Likewise, since $H_\omega(t)U_\omega(t,r)A_\omega^\ddagger\in\K_i$, we conclude
that
\begin{align}\nonumber
U_\omega(t,r)\odot_L (H_\omega(t)U_\omega(t,r)A_\omega^\ddagger)^\ddagger
&=
\left((H_\omega(t)U_\omega(t,r)A_\omega^\ddagger)\odot_R
U_\omega(r,t) \right)^\ddagger \\
&=
\left(H_\omega(t)\, \U(t,r) (A_\omega^\ddagger)\right)^\ddagger\, .
\end{align}
Eq.~\eqref{eqderivU3} follows. Furthermore, by Lemma~\ref{lemcomHA} we have
\begin{equation}
(H_\omega U_\omega(t,r)A_\omega^\ddagger)^\ddagger \varphi
=
(U_\omega(t,r)A_\omega^\ddagger)^\ddagger H_\omega\varphi
=
A_\omega^{\ddagger\ast} U_\omega(r,t)H_\omega\varphi
\end{equation}
for any $\varphi\in\D\cap\H_c$, so  \eqref{eqderivU3phi} holds.

The bound \eqref{HUAWbound} follows from \eqref{HUAW} and its counterpart for
$A_\omega^\ddagger$.
\end{proof}

In the special case when  $\El=0$ we have the following corollary,
with
\begin{align}  \label{U00}
\U^{(0)}(t) (A_\omega) = U^{(0)}_\omega(t) \odot_L A_\omega \odot_R
 U^{(0)}_\omega(-t)
\;\;\;\mbox{for \ $A_\omega \in \K_{\odot}\, $,}
\end{align}
where
$U^{(0)}_\omega(t)=\e^{-it H_\omega}$  as in \eqref{U0}.  The operator $\mathcal{L}_i$ introduced in the
following lemma
is usually called the \emph{Liouvillian}.

\begin{corollary}\label{Liouvillian}
 ${\, \U}^{(0)}(t)$ is a one-parameter group of operators
on $\K_{\odot}$, leaving   $\K_i$   invariant for $i=1,2,\infty$.
${\, \U}^{(0)}(t)$ is unitary on $\K_2$ and an isometry on
$\K_1$ and $\K_\infty$, so it extends to an isometry in $\overline{\K_1}$.
It is strongly continuous on   $\overline{\K_1}$ and $\K_2$; we denote by
$\Ll_{i}$, $i=1,2$, the corresponding infinitesimal generators :
\begin{equation}
\U^{(0)}(t) = \e^{-it \Ll_{i}} \quad \mbox{for all $t \in \R$}\, .
\end{equation}
Let
\begin{equation}\label{domainL0}
\D^{(0)}_{i}=
\left\{A_\omega\in \K_i; \;\;
H_\omega A_\omega, H_\omega A_\omega^\ddagger\in \K_i\right \},
\quad i=1,2,\infty\, .
\end{equation}
Then $\D^{(0)}_{i}$ is an operator core for $\Ll_{i}$,
$i=1,2$  (note that
$\Ll_{2}$ is essentially self-adjoint on
$\D^{(0)}_{2}$), and
\begin{equation}
\Ll_{i} (A_\omega) = [ H_\omega, A_\omega]_\ddagger  \quad
\mbox{for all $A_\omega \in \D^{(0)}_{i}$}, \quad i=1,2\, .
\end{equation}
Moreover, for every $B_\omega \in \K_\infty $ there exists a sequence
$B_{n,\omega} \in\D^{(0)}_{\infty} $ such that $B_{n,\omega} \to B_\omega$
as a  bounded and $\P$-a.e.-strong limit.
\end{corollary}

\begin{proof} Most of the Corollary follows immediately from
Propositions~\ref{propUomega}, \ref{derivU1}, \ref{derivU2}, and Stone's
Theorem for the Hilbert space  $\K_2$, the Hille-Yosida Theorem for the Banach
space $\overline{\K_1}$. Since
  $f(H_\omega) A_\omega g(H_\omega) \in \D^{(0)}_i$
for all $f,g \in C_c^\infty(\R)$ and $A_\omega \in \K_i$, $i=1,2,\infty$, we
 conclude that elements in $\K_\infty$ can approximated by sequences in
 $\D^{(0)}_\infty$
as a  bounded and $\P$-a.e.-strong limit, and also that
 $\D^{(0)}_i$ is a core
for $\mathcal{L}_{i}$ for $i=1,2$,
as in the usual proofs of Stone's Theorem  and  the Hille-Yosida Theorem,
 \end{proof}

\subsection{Gauge transformations in spaces of measurable operators}
The map
\begin{align}\label{Gtextends}
\G(t) (A_\omega) =  G(t) A_\omega G(t)^* \; ,
\end{align}
with $G(t)=\e^{i \int_{-\infty}^t \mathbf{E}(s)\cdot \x}$ as in \eqref{gaugedef}, is an
isometry on $\K_\infty$, $\K_1^{(0)}$, and $\K_2^{(0)}$, and hence extends to
an isometry on $\overline{\K}_1$ and on $\K_2$. Moreover, since $G(t)$ and
$\chi_x$ commute, \eqref{Gtextends} holds for $A_\omega$ either in ${\K}_1$ or
$\K_2$.

\begin{lemma} \label{lemGt}
The map $\G(t)$ is strongly continuous on both  $\overline{\K}_1$ and on
$\K_2$, and
\begin{equation}\label{liminfty}
\lim_{t \to -\infty} \G(t) = I \;\;\mbox{strongly}
\end{equation}
on both $\overline{\K}_1$ and on $\K_2$. Moreover, if $A_\omega \in \K_i$,
$i=1$ or $2$, with $[x_j, A_\omega] \in \K_i$ for $j=1,...,d$, then $\G(t)
(A_\omega)$ is continuously differentiable in $\K_i$ with
\begin{equation}\label{eqderivK}
\partial_t \G(t)(A_\omega) = i\left [{\El}(t)\cdot \x,\G(t)(A_\omega)\right]
= i\G(t)\left ( \left [{\El}(t)\cdot \x,A_\omega \right ] \right ) .
\end{equation}
\end{lemma}

\begin{proof}  We start by proving the lemma on $\K_2$.
For $A_\omega \in \K_2$, we have
\begin{align}\label{GGG1A}
\G(t+h)(A_\omega)  - \G(t )(A_\omega)= \G(t) (\G(t+h)\G(-t)-1)(A_\omega) \, .
\end{align}
Since $\G(t) $ is an isometry, continuity follows if we show that
\begin{equation}
\lim_{h \to 0}\tnorm{ (\G_t(h)-1)(A_\omega) }_2 =0 \, , \label{Gth}
\end{equation}
where $ \G_t(h)(A_\omega)= G_t(h)(A_\omega) G_t(h)^*$, with
 $G_t(h) = G(t+h)G(-t)$ being the unitary operator given by multiplication
by the function $\e^{-i  \int_{t}^{t+h} \mathbf{E}(s)\cdot \x\, \di s}$.
Thus
\begin{align}\label{Gh1A}
 (\G_t(h)-1)(A_\omega)   = G_t (h) \left [
(1 -G_t(h)^*)A_\omega +  A_\omega(G_t(h)^*-1) \right ]
\end{align}
Since $G_t(h)$ is unitary, we have
\begin{align}
&\tnorm{ (\G_t(h)-1)(A_\omega)   }_2^2
 \le 2 \left\{ \E\,  \left\| (1 -
G_t(h)^*)A_\omega\chi_0\right\|_2^2 + \E\,  \left\| A_\omega
(G_t(h)^*-1)\chi_0\right\|_2^2
 \right\} \nonumber\\
 & \qquad \qquad  = 2 \left\{ \E\,  \left\| (1 - G_t(h)^*)A_\omega\chi_0\right\|_2^2 +
\E\,  \left\|A_\omega\chi_0  (G_t(h)^*-1)\right\|_2^2\right\}\label{2terms}\, .
\end{align}
Although $G_t(h)^* \notin \K_\infty$  because it is not covariant, we can use the argument in the proof
of Lemma~\ref{stK2} to conclude that both terms in \eqref{2terms} go to $0$ as
 $h \to 0$, obtaining \eqref{Gth}. The limit in \eqref{liminfty} is just continuity at $t=-\infty$ and is proven
in the same way.

The result in $\overline{\K}_1$ now follows from the result in $K_2$ using the
$\Diamond$ map, since for $B_\omega, C_\omega \in \K_2^{(0)}$, we have on
$\K_1$ that
\begin{align}
\G(t )(B_\omega C_\omega)  = \G(t )(B_\omega) \G(t) (C_\omega)= (\G(t
)(B_\omega)) \diamond(\G(t) ( C_\omega))\, ,
\end{align}
and, as $\G(t)$ are isometries, it suffices to prove strong continuity on a
dense subset.

It only remains to prove differentiability and \eqref{eqderivK} assuming $[x_j,
A_\omega] \in \K_i$, since continuity of the derivative follows from
\eqref{Gth} and the strong continuity just obtained for $\G(t)$. We see by
\eqref{GGG1A} that it suffices to show
\begin{align}\label{GGG}
\lim_{h \to 0} \textstyle{\frac 1 h} (\G_t(h)-1) \left (A_\omega \right ) =
i\left [{\El}(t)\cdot \x,A_\omega\right]\, ,
\end{align}
with convergence in $\K_i$. Since $[\x, A_\omega] \in \K_i$, the (Bochner)
integral
\begin{align} \Phi(h) \ =\ i \frac{1}{h} \int_0^h \!\!\di u \,
\G_t(u) \left ( \left [\mathbf{E}(t + u)\cdot \x,  A_\omega \right ] \right )
\end{align}
is, for each $h>0$, a well defined element of $\K_1$. Furthermore, as
$\G_t(\cdot)$ is strongly continuous, the integrand is continuous and
\begin{align}
  \lim_{h \rightarrow 0} \Phi(h) \ = \ i\left [{\El}(t)\cdot \x,A_\omega\right]
  \; .
\end{align}
We claim that $\Phi(h) = h^{-1}(\G_t(h)-1) \left (A_\omega \right )$.  Indeed
it suffices to verify
\begin{align}
  h \chi_x\Phi(h) \chi_y \ = \ (\G_t(h)-1) \left (\chi_x A_\omega \chi_y)
  \right )
\end{align}
for each $x,y$ (since $\chi_x,\chi_y$ commute with $G(t)$).  But this identity
follows since the derivatives of the two sides are equal, and both expressions
vanish at $h=0$. (Derivation is permitted here because of the cut-off induced
by $\chi_x,\chi_y$.)
\end{proof}

\section{Linear response theory and Kubo formula}\label{sec:LinResponse}

In this section we prove our main results. \emph{We assume throughout this section
 that
Assumptions \ref{RandH} and \ref{assumptiona} (stated below) hold.}

\subsection{Adiabatic switching of the electric field}
\setcounter{equation}{0}

We now fix an initial equilibrium state of the system, i.e., we specify a
density matrix ${\zeta}_\omega$ which is in equilibrium, so $[H_\omega,
{\zeta}_\omega ] =0$. For physical applications, we would generally take
${\zeta}_\omega=f(H_\omega)$ with $f$
the Fermi-Dirac distribution at inverse temperature $\beta \in (0,\infty]$ and
\emph{Fermi energy} $E_F \in \R$,  i.e.,
$f(E) = \frac{1}{1+\e^{\beta (E - E_F)}}$ if $\beta < \infty$ and
$f(E)= \chi_{(-\infty,E_F]}(E)$ if $\beta =\infty$; explicitly
\begin{equation}
{\zeta}_\omega \ = \ \begin{cases}  F^{(\beta,E_F)}_\omega \ := \
\frac{1}{1+\e^{\beta (H_\omega - E_F)}} \, , &
\beta < \infty \, , \\
P^{(E_F)}_\omega \ := \ \chi_{(-\infty,E_F]}(H_\omega) \, , &\beta = \infty \, .
\end{cases}
\end{equation}
The fact that we have a Fermi-Dirac distribution is not so important at first,
although when we compute the Hall conductivity we will restrict our attention
to the zero temperature case with the \emph{Fermi projection} $P^{(E_F)}$.

The key property we need is that the hypothesis of either
Proposition~ \ref{fHK}(ii) or Prop. \ref{fHK}(iii) holds:

\begin{assumption}\label{assumptiona}
The initial equilibrium state ${\zeta}_\omega$ is non-negative, i.e.,
${\zeta}_\omega \ge 0$, and, either
\begin{description}
\item[(a)] ${\zeta}_\omega=g(H_\omega)$
 with $g \in \S(\R)$,
\end{description}
 or
\begin{description}
\item[(b)]  ${\zeta}_\omega$
 decomposes as ${\zeta}_\omega=g(H_\omega) h(H_\omega)$
with $g \in \S(\R)$ and $h$ a Borel measurable function which
satisfies $\|h^2
\Phi_{d,\alpha,\beta}\|_\infty < \infty$ and
\begin{align}  \label{assumption}
\E\left\{ \left\| \x \,h(H_\omega) \chi_0\right \|_2^2\right\} < \infty \, .
\end{align}
(Condition \eqref{assumption} is equivalent to
 $\left[x_j, h(H_\omega)\right]\in \K_2$ for all
$j=1,2,\ldots,d$.)
\end{description}
\end{assumption}

\begin{remark}\label{remassump}  We make the following observations
 about Assumption~\ref{assumptiona}:
\begin{description}
  \item[(i)] By   Proposition~\ref{fHK}, either (ii) or (iii), we have
  $\left[x_j,{\zeta}_\omega\right]\in \K_1 \cap\K_2$ for all $j=1,2,\dots,d$.
  \item[(ii)] The equivalence between \eqref{assumption} and $\left[x_j,
  h(H_\omega)\right]\in \K_2$ for $j=1,\dots,d$ follows from the facts
  that $h(H_\omega)\in \K_2$ by Prop. \ref{fHK}(i)  and
  \begin{equation}
    \left\| \x \,h(H_\omega) \chi_0\right \|_2 \ \le  \
    \left\| [\x , \,h(H_\omega)] \chi_0\right \|_2 \ + \
    \left\| h(H_\omega) \chi_0\right \|_2 \; .
  \end{equation}
    Although $|\x|^2=\x\cdot\x$ is not covariant, it follows from
    \eqref{assumption} that for any $a\in\Z^d$ we have
    \begin{equation}  \label{xPchia}
    \E\left\{ \left\| \x \,h(H_\omega) \chi_a \right \|_2^2\right\} < \infty
    \, ,
    \end{equation}
    and hence  the operators  $\left[x_j, h(H_\omega)\right]$ are well defined
    on $\H_c$ for $j=1,\dots,d$.
  \item[(iii)] The Fermi-Dirac distributions
$ f^{(\beta,E_F)}(E)\ := \ (1+\e^{\beta (E - E_F)})^{-1}$ with finite $\beta$
satisfy
 Assumption~\ref{assumptiona}(a).   Just take
$g(E)= k(E) f^{(\beta,E_F)}(E) $, where $k(E)$ is any $C^\infty$ function
which is equal to one for $ E\ge -\gamma$ (defined in \eqref{gamma})
and equal to $0$ for $E\le -\gamma_1$
for some $\gamma_1 >\gamma$.
  \item[(iv)] For a Fermi projection $P^{(E_F)}_\omega$ ($\beta =\infty$), it
  is natural to take $h(H_\omega) = P^{(E_F)}_\omega$ and for $g$ any Schwartz
  function identically $1$ on $[-\gamma, E_F]$.
 Condition  \eqref{assumption} does not hold automatically in this case;
  rather it holds only for $E_F$ in the ``localization regime,'' as discussed
  in the introduction.  The existence of a region of localization
been established for random Landau Hamiltonians
with Anderson-type potentials \cite{CH,Wa,GK4}.
\end{description}
\end{remark}

Let us now switch on, adiabatically, a spatially homogeneous electric field
$\El$, i.e., we take (with $t_- = \min \, \{t, 0\}$, $t_+ = \max \, \{t, 0\}$)
\begin{align} \label{defE}
\mathbf{E}(t)=
\mathrm{e}^{\eta t_-}\mathbf{E}\, ,
\end{align}
and hence
\begin{equation} \label{F(t)}
\F(t) =  \int_{-\infty}^t \mathbf{E}(s) \di s =
\left(\textstyle{\frac{\mathrm{e}^{\eta t_-}} \eta } + t_+  \right) \El \, .
\end{equation}
The system is now described by the ergodic time dependent Hamiltonian
$H_\omega(t)$, as in \eqref{eq:Htilde}.   We write
\begin{equation} \label{P(t)}
{\zeta}_\omega(t) = G(t) {\zeta}_\omega G(t)^* =\G(t) ({\zeta}_\omega),
\quad \text{i.e.,}\quad {\zeta}_\omega(t)=f(H_\omega(t)).
\end{equation}

Assuming the system was in equilibrium at $t=-\infty$ with the density matrix
$\varrho_\omega(-\infty) = {\zeta}_\omega$, the time dependent density matrix
$\varrho_\omega(t)$ would be the solution of the following Cauchy problem
for the Liouville equation:
\begin{align}
\left\{ \begin{array}{l}i\partial_t \varrho_\omega(t) =
[H_\omega(t),\varrho_\omega(t)]_\ddagger\\
\lim_{t \to  -\infty} \varrho_\omega(t)= {\zeta}_\omega
\end{array}\right.  \label{dynamics}  \, ,
\end{align}
where we have written the commutator $[\cdot,\cdot]_\ddagger$ in anticipation
of the fact that this is to be understood as an evolution in $\K_{i}$, $i=1,2$.
The main result of this subsection is the following theorem on solutions to
\eqref{dynamics}, which relies on the ingredients introduced in
Sections~\ref{sectoperator} and \ref{sectrandom}.  In view of
Corollary~\ref{Liouvillian}, we replace the commutator in \eqref{dynamics} by
the Liouvillian at time $t$:
 \begin{equation}
\Ll_i(t)  = \G(t) \Ll_i\G(-t), \quad  i=1,2 \, .
\end{equation}
Note that $\Ll_i(t)  $ has $\D^{(0)}_i$ as an operator core for all $t$,
since it follows from Lemma~\ref{allr} that
$\D^{(0)}_i= \G(t)\D^{(0)}_i$  for $i=1,2,\infty$.

We have the following  generalization of Theorem~\ref{thmrhointro}.

\begin{theorem}\label{thmrho} The Cauchy problem
\begin{align}
\left\{ \begin{array}{l}i\partial_t \varrho_\omega(t) =
\Ll_i(t)(\varrho_\omega(t))\\
\lim_{t \to  -\infty} \varrho_\omega(t)= {\zeta}_\omega
\end{array}\right.  \label{dynamicsL}  \, ,
\end{align}
 has a unique  solution in both $\overline{\K_1}$ and  $\K_2$,
with $\Ll_i(t)  $, $i=1,2$, being the corresponding Liouvillian.
The unique solution
$\varrho_\omega(t)$ is in  $\D^{(0)}_1(t)\cap \D^{(0)}_2(t) \subset \K_1 \cap \K_2$
for all $t$, solves the stronger Cauchy problem \eqref{dynamics}
in both $\K_1$ and $\K_2$, and is
 given by
\begin{align}
\varrho_\omega(t)&=\lim_{s \to -\infty} { \U}(t,s)\left( {\zeta}_\omega
\right)
\label{defrho1}\\
&= \lim_{s \to -\infty}{ \U}(t,s)\left( {\zeta}_\omega(s) \right)
\label{defrho2}\\
&= {\zeta}_\omega(t) - i
 \int_{-\infty}^t \mathrm{d} r \,\mathrm{e}^{\eta r_{\!-}}{ \,\U}(t,r) \left([ \mathbf{E}
\cdot \x, {\zeta}_\omega(r) ]\right) \, .\label{defrho3}
\end{align}
We  also have
\begin{align}
\varrho_\omega(t) ={ \U} (t,s) (\varrho_\omega(s))\, ,
\;\;\tnorm{\varrho_\omega(t)}_i=\tnorm{{\zeta}_\omega}_i \, ,
\end{align}
for all $t,s$ and $i=1,2, \infty$. Furthermore, $\varrho_\omega(t)$ is
non-negative, and if ${\zeta}_\omega = P^{E_F}_\omega$, then $\varrho_\omega(t)$
is an orthogonal projection for all $t$.
\end{theorem}

Before proving the theorem we need a technical but crucial lemma. We write
$\Db_{j,\omega} = \Db_j (\A_\omega)$.
\begin{lemma}\label{lemxHP}
Let $j=1,\cdots, d$.
\begin{description}
\item[(i)] For all $\varphi\in\H_c$ we have  $x_j {\zeta}_\omega\varphi\in \D$ and
\begin{equation}\label{comxHP}
2\Db_{j,\omega} {\zeta}_\omega\varphi =iH_\omega x_j {\zeta}_\omega\varphi - ix_j
H_\omega {\zeta}_\omega\varphi =i[H_\omega,x_j]{\zeta}_\omega \varphi
 \, .
\end{equation}

\item[(ii)]  $H_\omega [x_j,{\zeta}_\omega ]\in \K_1\cap\K_2$.  In fact,
the operators $H_\omega [x_j,{\zeta}_\omega]$ and $[x_j,H_\omega {\zeta}_\omega]$ are
well defined (as commutators) on $\H_c$, we have
\begin{equation}\label{Hcom}
H_\omega [x_j,{\zeta}_\omega] = [x_j,H_\omega {\zeta}_\omega]
 - 2i\Db_{j,\omega}  {\zeta}_\omega \mbox{ on } \H_c \, ,
\end{equation}
and the two operators in the right hand side of \eqref{Hcom} are in
$\K_1\cap\K_2$.

\item[(iii)] $H_\omega [\El\cdot \x,{\zeta}_\omega]\in \K_1\cap\K_2$.
\end{description}
\end{lemma}

\begin{proof}
It follows from \eqref{expression0}  that
\begin{align}\label{chiH99}
H_\omega  x_j\phi = x_j H_\omega \phi -2 i\Db_{j,\omega}\phi
\quad \mbox{for all $\phi \in C_c^\infty (\R^d)$}\, .
\end{align}
Thus if   $\phi\in \D\cap\D(x_j)$ with $H_\omega\phi\in \D(x_j)$, we conclude
by an approximation argument that $x_j \phi \in \D$ and \eqref{chiH99} holds
for $\phi$.

That $[x_j,H_\omega {\zeta}_\omega] \in  \K_1\cap\K_2$ follows from
Assumption~\ref{assumptiona} and
Proposition~\ref{fHK}(ii)-(iii) since the function $Eg(E) \in \S(\R)$.  In particular, this tells us
that $H_\omega {\zeta}_\omega \H_c \subset \D(x_j)$. Thus, given $\varphi\in\H_c$,
we set $\phi=   {\zeta}_\omega\varphi \in \D(x_j)$, so we have $H_\omega\phi\in
\D(x_j)$ and $\phi \in \D(x_j)$ (because $[x_j, {\zeta}_\omega] \in \K_2$). We
conclude that \eqref{comxHP} follows from \eqref{chiH99}.  This proves (i).

Since  $x_j {\zeta}_\omega\varphi\in \D$ for all $\varphi\in\H_c$, the operator
$H_\omega [x_j,{\zeta}_\omega]$ is well defined  on $\H_c$, and \eqref{Hcom}
follows from \eqref{comxHP}. That $\Db_{j,\omega} {\zeta}_\omega \in \K_1\cap\K_2$
follows from Proposition~\ref{propDA}(i). Thus (ii) is proven, and (iii)
follows immediately.
\end{proof}

We now turn to the proof of Theorem~\ref{thmrho}.

\begin{proof}[Proof of Theorem~\ref{thmrho}.]
Let us first apply Proposition~\ref{derivU1} and Lemma~\ref{lemGt} to
\begin{equation}\label{defrhots}
\varrho_\omega(t,s) :={ \U}(t,s)({\zeta}_\omega(s)).
\end{equation}
We get
\begin{align}
i\partial_s \varrho_\omega(t,s) & =
-\U(t,s)\left(\left[H_\omega(s),{\zeta}_\omega(s)\right]_\ddagger\right)
+\U(t,s)\left(-\left[{\El}(s)\cdot \x,{\zeta}_\omega(s)\right]\right)
\nonumber
\\
& = -\U(t,s)\left( \left[{\El}(s)\cdot \x,{\zeta}_\omega(s)\right]\right)
\, ,
\end{align}
where we used \eqref{P(t)}.
As a consequence,
with $\mathbf{E}(r)= \mathrm{e}^{\eta r_{\!-}}\mathbf{E}$,
\begin{equation}
\varrho_\omega(t,t) - \varrho_\omega(t,s)=  i\int_s^t \mathrm{d} r\,
 \mathrm{e}^{\eta r_{\!-}}{ \,\U}(t,r)
\left(\left[{\El}\cdot \x,{\zeta}_\omega(r)\right] \right)
  \, .
\end{equation}
Since
\begin{equation}
\tnorm{{ \, \U}(t,r) \left(\left[{\El}\cdot \x,{\zeta}_\omega(s)\right]
\right)}_i
 =
\tnorm{[{\El}\cdot \x,{\zeta}_\omega]}_i \, ,
\end{equation}
the integral is absolutely convergent and the limit as $s\to-\infty$ can be performed. It
yields the equality between \eqref{defrho2} and \eqref{defrho3}. Equality of \eqref{defrho1}
and \eqref{defrho2} follows from Lemma~\ref{lemGt} which gives
\begin{equation} {\zeta}_\omega = \lim_{s \to -\infty} {\zeta}_\omega(s) \;\;
\mbox{in  both $ {\K}_1$ and $ \K_2$.}
\end{equation}

 Since the  $\, \U(t,s)$ are isometries on $\K_i$, $i=1,2,\infty$
 (Proposition~\ref{propUomega}),
it follows from \eqref{defrho1} that
$\tnorm{\varrho_\omega(t)}_i=|\!|\!|{\zeta}_\omega|\!|\!|_i$. We also get
$\varrho_\omega(t) = \varrho_\omega(t)^\ddagger$, and hence $\varrho_\omega(t)
= \varrho_\omega(t)^*$ as $\varrho_\omega(t) \in \K_\infty$. Moreover,
\eqref{defrho1} with the limit in both $ {\K}_1$ and $ \K_2$  implies that
$\varrho_\omega(t)$ is nonnegative. Furthermore, if ${\zeta}_\omega=
P^{(E_F)}_\omega$ then $\varrho_\omega(t)$ is a projection, since denoting by
$\lim^{(i)}$ the limit in $\K_i$, $i=1,2$, we have
\begin{multline}\varrho_\omega(t) = \sideset{}{^{(1)}}\lim_{ s \to -\infty}{ \U}(t,s)\left( P^{(E_F)}_\omega \right) =\sideset{}{^{(1)}}\lim_{ s \to
-\infty}{ \U}(t,s)\left( P^{(E_F)}_\omega\right)\diamond{ \U}(t,s)\left( P^{(E_F)}_\omega\right) \\
=\left\{ \sideset{}{^{(2)}}\lim_{ s \to -\infty}{ \U}(t,s)\left( P^{(E_F)}_\omega\right)\right\}\diamond
 \left\{ \sideset{}{^{(2)}}\lim_{ s \to -\infty}{ \U}(t,s)\left( P^{(E_F)}_\omega\right)\right\}\ = \
 \varrho_\omega(t)^2\, .
\end{multline}

 To see that $\varrho_\omega(t)$ is a solution of  \eqref{dynamics}
 in $\K_i$, we differentiate the expression \eqref{defrho3} using
Proposition~\ref{derivU2} and  Lemma~\ref{lemGt}; the hypotheses of
Proposition~\ref{derivU2}  are satisfied in view of Lemma~\ref{lemxHP}(iii) and
the fact that $i[{\El}\cdot \x, {\zeta}_\omega(r)] $ is a symmetric operator.
Moreover, it follows from \eqref{HUAWbound} that
\begin{align} \nonumber
&\tnorm{\left[H_\omega(t),{ \U}(t,r) \left(\left[ \mathbf{E}
\cdot \x, {\zeta}_\omega(r) \right]\right)\right]}_i \le
  2\| W_\omega(t,r) \|  \tnorm{ (H_\omega(r)+\gamma)\left[ \mathbf{E}
\cdot \x, {\zeta}_\omega(r) \right]}_i  \\
& \qquad \qquad  \qquad
  = 2 \| W_\omega(t,r) \| \tnorm{ (H_\omega+\gamma)\left[ \mathbf{E} \cdot \x,
{\zeta}_\omega \right]}_i  \label{intderivative}  ,
\end{align}
where
\begin{equation}
\sup_{r; \; r\le t}\| W_\omega(t,r)\| \le C_t < \infty
\end{equation}
by \eqref{HUHest} and \eqref{HUHest45}.  Recalling  \eqref{defrho3}, we
therefore get
\begin{align}\label{partiavarho}
i\partial_t  \varrho_\omega(t)& = -i
 \int_{-\infty}^t \mathrm{d} r \,\mathrm{e}^{\eta r_-}
\left[H_\omega(t),{ \U}(t,r) \left(\left[ \mathbf{E}
\cdot \x, {\zeta}_\omega(r) \right]\right)\right]_\ddagger\\
& =-\left[H_\omega(t), \left\{i
 \int_{-\infty}^t \mathrm{d} r \,\mathrm{e}^{\eta r_{\!-}}{ \,\U}(t,r) \left(\left[ \mathbf{E}
\cdot \x, {\zeta}_\omega(r) \right]\right)\right\}\right]_\ddagger \label{partiavarho2}\\
& =\left[H_\omega(t),  \left\{{\zeta}_\omega(t) - i
 \int_{-\infty}^t \mathrm{d} r \,\mathrm{e}^{\eta r_{\!-}}{\,  \U}(t,r) \left(\left[ \mathbf{E}
\cdot \x, {\zeta}_\omega(r) \right]\right)\right\}\right]_\ddagger \nonumber\\
& =\left[H_\omega(t), \varrho_\omega(t)\right]_\ddagger \label{partiavarho3}\,
,
\end{align}
the integrals being Bochner integrals in $\K_i$.  We justify going from
\eqref{partiavarho} to \eqref{partiavarho2}
as follows:  Since
$H_\omega(t) (H_\omega(t) +\gamma)^{-1} \in \K_\infty$ and
 $(H_\omega(t) +\gamma)^{-1} \in \K_\infty$, we have, as operators on $\H_c$,
\begin{align}\label{interchange}
&\int_{-\infty}^t \mathrm{d} r \,\mathrm{e}^{\eta r_{\!-}} H_\omega(t){ \, \U}(t,r) \left(\left[ \mathbf{E}
\cdot \x, {\zeta}_\omega(r) \right]\right)\\
&\quad = \left(H_\omega(t) (H_\omega(t) +\gamma)^{-1}\right)\odot_L
\int_{-\infty}^t \mathrm{d} r \,\mathrm{e}^{\eta r_{\!-}} (H_\omega(t)
+\gamma)\, \U(t,r) \left(\left[ \mathbf{E} \cdot \x, {\zeta}_\omega(r)
\right]\right)
\nonumber\\
&\quad = H_\omega(t)\left( (H_\omega(t) +\gamma)^{-1}\odot_L \int_{-\infty}^t
\mathrm{d} r \,\mathrm{e}^{\eta r_{\!-}} (H_\omega(t) +\gamma)\, \U(t,r)
\left(\left[ \mathbf{E} \cdot \x, {\zeta}_\omega(r)\right]\right)\right)
\nonumber\\
&\quad = H_\omega(t) \int_{-\infty}^t \mathrm{d} r \,\mathrm{e}^{\eta r_{\!-}}
{\, \U}(t,r) \left(\left[ \mathbf{E} \cdot \x, {\zeta}_\omega(r)
\right]\right)\, . \nonumber
\end{align}
Since the map $A_\omega\to A_\omega^\ddagger$ is an antilinear isometry, we
also have the identity conjugate to \eqref{interchange}. We thus have
\eqref{partiavarho3}.

It remains to show that the solution of \eqref{dynamicsL} is unique in
both $\overline{\K_1}$ and  $\K_2$. It suffices to show that if
$\nu_\omega(t)$ is a solution of \eqref{dynamicsL} with $\zeta_\omega =0$ then
$\nu_\omega(t)= 0$ for all $t$. We give the proof for $\overline{\K_1}$,
the proof  for $\K_2$ being similar and slightly easier. For any $s\in \R$, set
$\tilde{\nu}_\omega^{(s)}(t)= \U(s,t)(\nu_\omega (t))$.
 If $A_\omega \in \D^{(0)}_\infty$,
 we have, using Lemma~\ref{derivU2}  in $\K_\infty$ and \eqref{dynamicsL}, that
\begin{align}
&i\partial_t \T \left\{ A_\omega \odot_L\tilde{\nu}_\omega^{(s)}(t)\right\}=
i\partial_t \T \left\{{ \U}(t,s)(A_\omega)\odot_L {\nu}_\omega(t)\right\}\\
& \   = \T \left\{[H_\omega(t),\U(t,s)(A_\omega)]_\ddagger \odot_L
{\nu}_\omega(t)\right\} + \T \left\{{ \U}(t,s)(A_\omega)\odot_L
\Ll_{1}(t)(\nu_\omega(t))\right\} \nonumber
 \\
 & \   = -\T \left\{\U(t,s)(A_\omega) \odot_L
\Ll_{1}(t)({\nu}_\omega(t))\right\} + \T \left\{{ \U}(t,s)(A_\omega)\odot_L
\Ll_{1}(t)(\nu_\omega(t))\right\} = 0 . \nonumber
\end{align}
In the final step we have used the fact that for $A_\omega \in \D^{(0)}_\infty$
and $B_\omega \in \D_1$ we have
\begin{equation}\label{LLcycle}
\T \left\{[H_\omega(t),A_\omega]_\ddagger \odot_L {B}_\omega\right\} =
 - \T \left\{A_\omega\odot_L \Ll_{1}(t)({B}_\omega)\right\}.
\end{equation}
Indeed, since $\D_1^{(0)}$ is a core for $\Ll_{1}(t)$ it suffices to consider
$B_\omega \in \D_1^{(0)}$. For such $B$, \eqref{LLcycle} follows by cyclicity
of the trace, with some care needed since $H_\omega(t)$ is unbounded:
\begin{align}
    & \T \left\{[H_\omega(t),A_\omega]_\ddagger \odot_L {B}_\omega\right\} \\
    \nonumber
    & \quad = \ \T \left \{ H_\omega(t) A_\omega \odot_L
    B_\omega \right \} \ - \ \T \left \{ (H_\omega(t) A_\omega^\ddagger)
    ^\ddagger \odot_L B_\omega \right \} \\ \nonumber
    & \quad = \ \T \left \{ (H_\omega(t) + \gamma) A_\omega \odot_L
    ((H_\omega(t) + \gamma) B_\omega^{\ddagger})^{\ddagger} \odot_R
    (H_\omega(t) + \gamma)^{-1} \right \} \\
    \nonumber & \qquad \qquad \ -
    \T \left \{ ((H_\omega(t) + \gamma) A_\omega^\ddagger)^\ddagger
    \odot_L (H_\omega(t)+\gamma)^{-1}
     (H_\omega(t) + \gamma) B_\omega \right \} \\
    & \nonumber \quad = \ - \T
    \left\{A_\omega \odot_L [H_\omega(t),{B}_\omega]_\ddagger\right\}
    \ = \ - \T \left\{A_\omega\odot_L \Ll_{1}(t)({B}_\omega)\right\}.
\end{align}
We conclude that for all $t$ and $A_\omega \in \D^{(0)}_\infty$ we have
  \begin{equation}\label{Tanu}
\T \left\{ A_\omega \odot_L\tilde{\nu}_\omega^{(s)}(t)\right\}=
\T \left\{ A_\omega \odot_L\tilde{\nu}_\omega^{(s)}(s)\right\}=
\T \left\{ A_\omega \odot_L{\nu}_\omega(s)\right\},
\end{equation}
and hence \eqref{Tanu} holds for all $A_\omega \in \K_\infty$ by
Corollary~\ref{Liouvillian} and Lemma~\ref{stK1} (or Lemma~\ref{lemmaduality2}) .  Thus
 $\tilde{\nu}_\omega^{(s)}(t)  = \nu_\omega(s)$ by Lemma~\ref{lemmaduality},
that is,  $\nu_\omega(t)=\U(t,s)(\nu_\omega (s))$.
 Since $\lim_{s \rightarrow -\infty} \nu_\omega(s) =0$
by hypothesis, we get $\nu_\omega(t)=0$ for all $t$.
\end{proof}


\subsection{The current and the conductivity}
From now on $\varrho_\omega(t)$ will denote the unique solution to
\eqref{dynamicsL}, given explicitly in \eqref{defrho3}.  We set
\begin{equation}
\Db_{\omega}(t)= \Db(\A_\omega+\F(t))=
G(t)\Db(\A_{\omega}) G(t)^*= G(t)\Db_{\omega}G(t)^* .
\end{equation}
 Since $H_\omega(t) \varrho_\omega(t) \in \K_{1,2}$ we
have
 $\varrho_\omega(t)\H_c\subset\D$,  hence the operators
 $\Db_{j,\omega}(t)\varrho_\omega(t)$ are
well-defined on $\H_c$, $j=1,2,\ldots,d$,
and we have
\begin{align}\label{Dvarrho}
\Db_{j,\omega}(t)\varrho_\omega(t) = \left(\Db_{j,\omega}(t)
 (H_\omega(t)+\gamma)^{-1}\right) \odot_L \left( (H_\omega(t)+\gamma)
\varrho_\omega(t)\right) \in \K_{1,2}\, .
\end{align}

\begin{definition}\label{defcurrent}
Starting with a system in equilibrium in state ${\zeta}_\omega$,
the net current (per
unit volume), $\J(\eta,\El;{\zeta}_\omega)\in\R^d$, generated by switching on an electric
field $\El$ adiabatically  at rate $\eta>0$  between time $-\infty$ and time
$0$,  is defined as
\begin{equation}\label{defJ}
\J(\eta,\El;{\zeta}_\omega) =  \T \left( \v_{\omega}(0) \varrho_\omega(0)\right)
 - \T \left( \v_\omega {\zeta}_\omega\right)  ,
\end{equation}
where the velocity operator  $\v_{\omega}(t)$ at time $t$ is as in \eqref{velocity}, i.e.,
\begin{equation}\label{velocityt}
\v_{\omega}(t)=  2 \Db_{\omega}(t) =
\left\{2 \Db_{j,\omega}(t)\right)\}_{j=1,\cdots,d}\, ,
\end{equation}
a vector of essentially self-adjoint operators on $\D$ (or $C_c^\infty(\R)$).
\end{definition}

\begin{remark}
\begin{description}
\item[(a)]  The term $  \T \left( \v_\omega {\zeta}_\omega\right)=
\left\{\T \left( \v_{j,\omega} {\zeta}_\omega\right)\right\}_{j=1,\cdots,d}$ is the
current at time $t=-\infty$.   Since the system is then at equilibrium one
expects this term to be zero, a fact which we prove in Lemma~\ref{lemequilib}.
It follows that the net current is equal to the first term of \eqref{defJ},
which is the current at time $0$. We will simply call this the current.

\item[(b)] The   current  $\J(\eta,\E;{\zeta})$  is a real vector.
This follows from the fact that $0\le\varrho_\omega(t) \in \K_{1} $, and hence
$\sqrt{\varrho_\omega(t)} \in \K_{2} $, the fact that
$\Db_{j,\omega}(t)\sqrt{\varrho_\omega(t)} \in \K_{2}$ by the same argument
as in \eqref{Dvarrho},
 the centrality of $\T$, and the essential self-adjointness of the
 components of $\v_{\omega}(t)$.
\end{description}
\end{remark}

\begin{lemma}\label{lemequilib}
Let $f$ be a  Borel measurable function on the real line, such that $\|
  f \widetilde \Phi_{d,\alpha,\beta}  \|_\infty$ is finite. Then
\begin{equation}
\T(\Db_{j,\omega} f(H_\omega))=0 \, .
\end{equation}
As a consequence, we have  $  \T \left( \v_\omega P^{(E_F)}_\omega\right)=0$.
\end{lemma}

This result appears in \cite{BESB}, with a detailed proof in the discrete case
and some remarks for the continuous case. The latter is treated in \cite{KSB}.
Their proof relies on a Duhamel formula and the Fourier transform.  We give an
alternative proof based on the Helffer-Sj\"ostrand formula.

\begin{proof}[Proof of Lemma~\ref{lemequilib}.]
First note that by a limiting argument it suffices to consider $f \in \S(\R)$.
In fact, we may find a sequence $g_n\in \S(\R)$ such that $\sup_n \| g_n
\widetilde \Phi_{d,\alpha,\beta} \|_\infty < \infty $ and $
  g_n(H_\omega) \to f(H_\omega) $ strongly.
Then
\begin{align}
  &\Db_{j,\omega} (f(H_\omega)-g_n(H_\omega)) =\\
& \quad \Db_{j,\omega} \frac{1}{\sqrt{H_\omega+\gamma}}
  \odot_L \frac{1}{(H_\omega+\gamma)^{2[[\frac d 4]]}} \odot_R
  (H_\omega+\gamma)^{2[[\frac d 4]]+\frac 1 2} (f(H_\omega)-g_n(H_\omega))
\nonumber \; ,
\end{align}
where the left hand factor is in $\K_\infty$ by Proposition \ref{propDA}(i),
the middle factor is in $\K_1$ by Proposition \ref{trSGEE}, and the right hand
factor is a uniformly bound sequence in $\K_\infty$ converging strongly to
zero.  By dominated convergence, we conclude that the $\K_1$ norm, and thus the
trace per unit volume, converges to zero.

Therefore, suppose $f \in \S(\R)$. Let $G(t) = \int_t^\infty \mathrm{d} t\, f
(t)$, and set $ F(t) = b(t) G(t)$, where $b(t) \in C^\infty(\R)$ is such that
$b(t)= 1$ for $ t > -\gamma $ and $b(t)=0$ for $t <- \gamma -1$ (so $b(t) =1$
in  a neighborhood of the spectrum of $H_\omega$).  We have $F \in \S(\R)$,
$G(H_\omega)= F(H_\omega)$,  and $f(H_\omega)= F^\prime (H_\omega)$.

We now recall the generalization of the Helffer-Sj\"ostrand formula given in
\cite[Lemma B.2]{HuS}: given a self-adjoint operator $A$ and $f \in \S(\R)$ we have
\begin{align}\label{HSp}
\textstyle{\frac 1 {p!}} f^{(p)} (A) = \int d\tilde{f}(z) (z-A)^{-p-1}
\;\;\;\mbox{for $p=0,1,\ldots$}\, ,
\end{align}
where the integral converges absolutely in operator norm by
\eqref{HShigherorder}. (See \cite[Appendix B]{HuS} for details.)

By \eqref{x,R=RDR} from the proof of Proposition~\ref{x,f},   we have
\begin{equation}
[x_j,R_\omega(z)]= 2 i R_\omega(z)\Db_{j,\omega}R_\omega(z) \in \K_\infty\, ,
\end{equation}
for  $R_\omega(z) = (H_\omega -z)^{-1}$ with $\mathrm{Im} \, z \ne 0$.
By the usual
Helffer-Sj\"ostrand formula \eqref{HS} we have
\begin{equation}
  \left[x_j, F(H_\omega)\right] \ = \ -
 \int d\tilde{F}(z) [x_j, R_\omega(z)]
\ = \  - 2 i \int d\tilde{F}(z)R_\omega(z)\Db_{j,\omega} R_\omega(z) \, ,
\end{equation}
which in particular gives another proof to the fact that
$\left[x_j, F(H_\omega)\right] \in \K_\infty$, which we already knew by
Proposition~ \ref{fHK}(ii).

There is a slight technical difficulty due to the fact that
$R_\omega(z)\Db_{j,\omega} R_\omega(z)$ may not be in $\K_1$ (although $
\left[x_j, F(H_\omega)\right]$ is). Thus we introduce a cutoff by picking a
sequence $h_n \in C^\infty_c(\R)$, $|h_n| \le 1$, $h_n = 1$ on $[-n,n]$, and
apply \eqref{HSp} with $p=0$ and $p=1$ to obtain
\begin{align}\nonumber
&\T \left \{ \left[x_j, F(H_\omega)\right] \odot_L h_n(H_\omega) \right \}  =
- 2 i \int d\tilde{F}(z)\T \left \{ R_\omega(z)\Db_{j,\omega} R_\omega(z) \odot_L
h_n(H_\omega) \right \} \\
& \, =  -2 i\int d\tilde{F}(z)\T\left\{\Db_{j,\omega} R_\omega(z)^2
\odot_L h_n(H_\omega)
\right\}
=-2i\T\left\{ \Db_{j,\omega}  f(H_\omega) \odot_L h_n(H_\omega) \right\}\! .
\end{align}
In the limit $n \to \infty$, we get
\begin{align}
\T\left\{ \Db_{j,\omega}  f(H_\omega)\right\}= \tfrac i 2
 \T\left\{ \left[ F(H_\omega), x_j\right]\right\}=0
\end{align}
by Proposition~\ref{fHK}(v).
\end{proof}

It is useful to rewrite the current \eqref{defJ}, using \eqref{defrho3} and the
argument in \eqref{interchange}, as
\begin{align} \label{formuleJ}
\J(\eta,\El;{\zeta}_\omega)&  = \T \left\{ 2\Db_{\omega}(0)
\left( \varrho_\omega(0) - {\zeta}_\omega(0)\right)\right\} \\
 & = -\T \left\{2
 \int_{-\infty}^0  \mathrm{d} r\,  \mathrm{e}^{\eta r}\Db_{\omega}(0) \,{ \U}(0,r) \left(
i[ \mathbf{E} \cdot \x, {\zeta}_\omega(r) ] \right) \right\}  \, , \nonumber
\end{align}
which is justified, since
\begin{equation}
\T \left( \Db_\omega(0) {\zeta}_\omega(0)\right)= \T \left(G(0) \Db_\omega
{\zeta}_\omega G(0)^*\right)= \T \left( \Db_\omega {\zeta}_\omega\right)
\end{equation}
by cyclicity of the trace, and all three terms are equal to zero.

The conductivity tensor $\sigma(\eta;{\zeta}_\omega)$ is  defined as the derivative (or
differential)  of  the function $\J(\eta, \cdot;{\zeta}_\omega)\colon\R^d \to \R^d$ at
$\El=0$.  Note that
 $\sigma(\eta;{\zeta}_\omega)$ is a $d \times d$ matrix
 $\left\{\sigma_{jk}(\eta;{\zeta}_\omega)\right\}$:
\begin{definition}
For $\eta >0$  the conductivity tensor $\sigma(\eta;{\zeta}_\omega)$ is defined as
\begin{equation}\label{defsigmajketa}
\sigma(\eta;{\zeta}_\omega) = \partial_{\El} \J(\eta,0;{\zeta}_\omega) \, ,
\end{equation}
if it exists. The conductivity tensor $\sigma({\zeta}_\omega)$
is
defined by
\begin{equation}\label{defsigmajk}
\sigma({{\zeta}_\omega}) :=  \lim_{\eta\downarrow 0} \sigma(\eta;{{\zeta}_\omega})\, ,
\end{equation}
whenever the limit exists.
\end{definition}


\subsection{Computing the linear response:  a Kubo formula for the conductivity}

The next theorem gives  a ``Kubo formula" for the conductivity.
\begin{theorem}\label{thmsgmjk}  Let $\eta > 0$. The current $\J(\eta,\El;{{\zeta}_\omega})$ is differentiable with
respect to $\El$ at $\El=0$ and the derivative $\sigma(\eta;{{\zeta}_\omega})$ is given by
\begin{equation}\label{sigmajk}
\sigma_{jk}(\eta;{{\zeta}_\omega}) =\, -\T \left\{ 2  \int_{-\infty}^0  \mathrm{d} r\,
\mathrm{e}^{\eta r}
  \Db_{j,\omega} \, \U^{(0)}(-r) \left(i [  x_k, {\zeta}_\omega ] \right)  \right\} ,
\end{equation}
where $\U^{(0)}(r)(A_\omega)=\mathrm{e}^{-irH_\omega} \odot_L A_\omega \odot_R \mathrm{e}^{irH_\omega}$.
\end{theorem}

We also have  the analogue of \cite[Eq.~(41)]{BESB} and
\cite[Theorem~1]{SBB2};   $\mathcal{L}_1$ is the  Liouvillian on $\overline{\K_1}$
(see Corollary~\ref{Liouvillian}).

\begin{corollary} The conductivity  $\sigma_{jk}(\eta;{{\zeta}_\omega}) $ is given by
\begin{align}\label{sigmajkbis}
\sigma_{jk}(\eta;{{\zeta}_\omega}) =\, -\T \left\{ 2 \Db_{j,\omega} \,
(i\mathcal{L}_1 +\eta)^{-1} \left(i [  x_k, {\zeta}_\omega ] \right)
 \right\} ,
\end{align}
\end{corollary}

\begin{proof}  Since  $H_\omega [  x_k, {\zeta}_\omega ] \in \K_1 \cap \K_2$ by
Lemma~\ref{lemxHP}(ii), we have
\begin{align}\nonumber
& \Db_{j,\omega} \, \U^{(0)}(-r) \left(i [  x_k, {\zeta}_\omega ] \right)=
\Db_{j,\omega}(H_\omega + \gamma)^{-1}
\odot_L (H_\omega + \gamma)\, \U^{(0)}(-r) \left(i [  x_k, {\zeta}_\omega ] \right)\\
& \qquad \qquad = \Db_{j,\omega}(H_\omega + \gamma)^{-1}
\odot_L \U^{(0)}(-r) \left( (H_\omega + \gamma)i [  x_k, {\zeta}_\omega ] \right),
\end{align}
and it follows from \eqref{sigmajk} that
\begin{align}\label{sigmajk555}\nonumber
\sigma_{jk}(\eta;{{\zeta}_\omega})& =\, -2\,\T \left\{
 \Db_{j,\omega}(H_\omega + \gamma)^{-1}\odot_L
(i\mathcal{L}_1 +\eta)^{-1} \left((H_\omega + \gamma) i [  x_k, {\zeta}_\omega ]
\right)  \right\}\\
& =\, -2\,\T \left\{
 \Db_{j,\omega}
(i\mathcal{L}_1 +\eta)^{-1} \left( i [  x_k, {\zeta}_\omega ]
\right)  \right\} ,
\end{align}
since
 $(i\mathcal{L}_1 +\eta)^{-1} \left((H_\omega + \gamma) i [  x_k, {\zeta}_\omega ]
\right)$ and $(i\mathcal{L}_1 +\eta)^{-1} \left( i [  x_k, {\zeta}_\omega ]
\right) $ are in $\K_1 \cap \K_2$ and hence in $\K_1$ (not just in
$\overline{\K_1}$), where
\begin{align}
(H_\omega + \gamma)^{-1}\odot_L
(i\mathcal{L}_1 +\eta)^{-1} \left((H_\omega + \gamma) i [  x_k, {\zeta}_\omega ]
\right) =
(i\mathcal{L}_1 +\eta)^{-1} \left( i [  x_k, {\zeta}_\omega ]
\right) .
\end{align}
\end{proof}

\begin{proof}[Proof of Theorem~\ref{thmsgmjk}.]
From \eqref{formuleJ} and $\J_j(\eta,0;{\zeta_\omega})= 0$ (Lemma \ref{lemequilib}), we
have
\begin{equation}\label{defsigmajk3}
    \sigma_{jk}(\eta;{{\zeta}_\omega})=-\lim_{E
\to 0} 2\T \left\{
 \int_{-\infty}^0  \mathrm{d} r\,  \mathrm{e}^{\eta r}
\Db_{j,\omega}(0)
\,  \U(0,r) \left(i
[ {x_k}, {\zeta}_\omega(r) ] \right) \right\}\, ,
\end{equation}
where $\Db_{j,\omega}(0)=\Db_{j,\omega}(\El,0)$ and
${\zeta}_\omega(r)={\zeta}_\omega(\El,r)$ depend on $\El$ through the gauge
transformation $\G$ and $U_\omega(0,r)=U_\omega(\El,0,r)$ also depends on
$\El$. (For clarity, in this proof we display the argument $\El$ in all
functions which depend on $\El$.)

Let us first understand that we can interchange integration and the limit $\El
\rightarrow 0$, i.e., that
\begin{equation}
\label{defsigmajk31}
    \sigma_{jk}(\eta;{{\zeta}_\omega}) \ = \ -2 \int_{-\infty}^0  \mathrm{d} r\,  \mathrm{e}^{\eta r}
 \lim_{{E} \to 0}  \T \left\{  \Db_{j,\omega}({\El}, 0)
{ \,\U}({\El},0,r) \left(i[ {x_k}, {\zeta}_\omega({\El}, r) ] \right)
\right\}\, .
\end{equation}
Note that
\begin{align}\nonumber
& \Db_{j,\omega}({\El}, 0){ \, \U}({\El},0,r) \left(i[ {x_k}, {\zeta}_\omega({\El}, r) ] \right) \\
 &\ =  \left\{ \Db_{j,\omega}({\El}, 0) (H_\omega(\El,0) +\gamma)^{-1}
(H_\omega(\El,0) +\gamma) U_\omega({\El},0,r) (H_\omega({\El}, r)
+\gamma)^{-1}\right\}\nonumber
 \\
& \qquad \qquad \qquad\odot_L \left\{(H_\omega({\El},r)+\gamma) \left(i[ {x_k},
{\zeta}_\omega({\El}, r) ] \right)\right\}\odot_R U_\omega({\El},r,0)
\label{limsigmajk256} \\
& \ = \left\{ \G({\El}, 0) \left(\Db_{j,\omega} (H_\omega
+\gamma)^{-1}\right)\right\} \odot_L W_\omega({\El}, 0,r)
\nonumber
 \\
&\qquad \qquad\qquad \odot_L \left\{ \G({\El}, r) \left((H_\omega +\gamma)[ i{x_k},
{\zeta}_\omega] \right) \right\}\odot_R U_\omega({\El},r,0)
\nonumber \, .
\end{align}
Using
\eqref{DbH}, \eqref{HUAW},
gauge invariance of the norms, \eqref{HUHest}, \eqref{HUHest45},
and Lemma~\ref{lemxHP}(ii), we get
\begin{align} \label{uniformbd}
&\sup_{|{E}| \le 1,r\le0 }\tnorm{\Db_{j,\omega}({\El}, 0){\, \U}({\El},0,r) \left(i
[ {x_k}, {\zeta}_\omega({\El}, r) ] \right)}_1 \\
&\quad \le  \tnorm{\Db_{j,\omega} (H_\omega +\gamma)^{-1}}_\infty
\left\{\sup_{|{E}| \le 1,r\le0 }\tnorm{W_\omega({\El}, 0,r)}_\infty \right\}
\tnorm{ (H_\omega+\gamma)[ {x_k}, {\zeta}_\omega ] }_1   < \infty . \nonumber
\end{align}
Eq.~\eqref{defsigmajk31} follows from  \eqref{defsigmajk3},
\eqref{uniformbd}, \eqref{Testimate},
and dominated convergence.

Next, we note that for any $s$ we have
\begin{align}\label{GEk}
\lim_{{E} \to 0}\G({\El}, s)= I \;\;\mbox{strongly in $\K_1$}\, ,
\end{align}
which can be proven  by a argument similar to the one used to prove
Lemma~\ref{lemGt}. Along the same lines, for $B_\omega \in \K_\infty$   we have
\begin{align}\label{GEk2}
\lim_{{E} \to 0}\G({\El}, s)(B_\omega)= B_\omega  \;\;\mbox{strongly in $\H$,
with  $\tnorm{\G({\El}, s)(B_\omega)}_\infty=
\tnorm{B_\omega}_\infty $}\, .
\end{align}
It therefore follows from  \eqref{limsigmajk256} that
\begin{align}\label{limit101}
  & \lim_{{E} \to 0}\T \left\{  \Db_{j,\omega}({\El}, 0){\,  \U}({\El},0,r)
\left(i[ {x_k}, {\zeta}_\omega({\El}, r) ]
  \right)\right\} \\
  &\  = \lim_{{\El} \to 0}  \T \left\{
    \left(\Db_{j,\omega} - \F_j(0) \right )
    U_\omega({\El}, 0,r) (H_\omega(\El,r) + \gamma)^{-1}
    \odot_L \right . \nonumber  \\
  & \quad \qquad \qquad  \qquad \qquad  \qquad \qquad
  \odot_L \left .  (H_\omega
    +\gamma)[ i{x_k}, {\zeta}_\omega ] \odot_R U_\omega({\El},r,0)
    \right\}
     \nonumber \\
  &\  = \lim_{{\El} \to 0}  \T \left\{
    \Db_{j,\omega} U_\omega({\El}, 0,r) (H_\omega(\El,r) + \gamma)^{-1}\odot_L
    (H_\omega +\gamma)[ i{x_k}, {\zeta}_\omega ] \odot_R
    U_\omega^{(0)}(r) \right\} \nonumber\\
 & \ = \lim_{{\El} \to 0}  \T \left\{
    \Db_{j,\omega} U_\omega({\El}, 0,r)
(H_\omega +\gamma)^{-1}
 \left\{ (H_\omega +\gamma)
(H_\omega({\El}, 0) +\gamma)^{-1}\right\} \odot_L \nonumber\right.\\
& \quad \qquad  \qquad  \qquad   \qquad  \qquad  \qquad  \left.
\odot_L
    (H_\omega +\gamma)[ i{x_k}, {\zeta}_\omega ] \odot_R
    U_\omega^{(0)}(r) \right\} \nonumber\\
 &\  = \lim_{{\El} \to 0}  \T \left\{
    \Db_{j,\omega} U_\omega({\El}, 0,r) (H_\omega + \gamma)^{-1}\odot_L
    (H_\omega +\gamma)[ i{x_k}, {\zeta}_\omega ] \odot_R
    U_\omega^{(0)}(r) \right\} \nonumber,
\end{align}
where we used \eqref{GEk}, \eqref{DuhamelU2}, the fact that $
\Db_{j,\omega}({\El}, 0) = \Db_{j,\omega} - \F_j
(0)$, \eqref{C(t,s)}-\eqref{DbH},  and Lemma~\ref{stK1}.
(Technically, we have not shown convergence yet. This equation should be read
as saying that if  any of these  limits exists, then they all exist and agree.)

To proceed it is convenient to introduce a cutoff so that we can deal with
$\Db_{j,\omega}$ as if it were in $\K_\infty$.  Thus we pick $f_n\in
C^\infty_c(\R)$, real valued, $|f_n|\le 1$, $f_n=1$ on $[-n,n]$. Using
Proposition~\ref{propDA}(i) and Lemma~\ref{stK1} we have
\begin{align}\label{DUQ}
&\T \left\{
    \Db_{j,\omega} U_\omega({\El}, 0,r) (H_\omega +\gamma)^{-1} \odot_L
    (H_\omega +\gamma)[ i{x_k}, {\zeta}_\omega ] \odot_R
    U_\omega^{(0)}(r) \right\} \\
& \quad =  \label{DUQ2} \lim_{n\to\infty} \T \left\{ \Db_{j,\omega}f_n(H_\omega)
    U_\omega({\El},0,r) \odot_L [ i{x_k}, {\zeta}_\omega ] \odot_R U_\omega^{(0)}(r)
    \right \} \\
&\quad = \label{DUQ4} \lim_{n\to\infty} \T \left\{  U_\omega({\El},0,r) \odot_L
    i[ {x_k}, {\zeta}_\omega ] \odot_R
    \left(U_\omega^{(0)}(r) \Db_{j,\omega}f_n(H_\omega)\right)
    \right\}  \\
&\quad = \label{DUQ5} \lim_{n\to\infty} \T \left\{  U_\omega({\El},0,r) \odot_L
    \left( (H_\omega+\gamma) i[  x_k, {\zeta}_\omega ]\right)^\ddagger\odot_R
 \right . \\
& \qquad \qquad \qquad \qquad \qquad \qquad\left . \odot_R
U_\omega^{(0)}(r) ( H_\omega+\gamma)^{-1} \Db_{j,\omega}f_n(H_\omega) \right \} \nonumber \\
&\quad = \label{DUQ6} \T \left\{  U_\omega({\El},0,r) \odot_L \left(
(H_\omega+\gamma) i[  x_k, {\zeta}_\omega ]\right)^\ddagger \odot_R
U_\omega^{(0)}(r) (H_\omega+\gamma )^{-1} \Db_{j,\omega}\right \} \; ,
\end{align}
where we used  Lemma~\ref{lemmacentral} to go from \eqref{DUQ2} to
\eqref{DUQ4}. The step from \eqref{DUQ4} to \eqref{DUQ5} is justified because
$ (H_\omega+\gamma )^{-1}$ commutes with $U^{(0)}$.  Finally, since
 $ ( H_\omega+\gamma)^{-1}  \Db_{j,\omega} \in \K_\infty$
 (that is, its bounded closure is
in $\K_\infty$), we can take the limit $n \to \infty$, using Lemma~\ref{stK1}
again. (Note $\left(i[ x_k, {\zeta}_\omega ]\right)^\ddagger = i[ x_k, {\zeta}_\omega
]$.)

Finally, combining \eqref{limit101} and
\eqref{DUQ}-\eqref{DUQ6}, we get
\begin{align}
 &  \lim_{{E} \to 0} \T \left\{  \Db_{j,\omega}({\El}, 0)\,{ \U}({\El},0,r) \left(i[ {x_k}, {\zeta}_\omega({\El}, r) ] \right)\right\}
 \\ &\quad  =  \T \left\{  U_\omega^{(0)}(-r) \odot_L \left(
(H_\omega+\gamma) i[  x_k, {\zeta}_\omega ]\right)^\ddagger \odot_R
U_\omega^{(0)}(r)  \left ( \Db_{j,\omega}(H_\omega+\gamma)^{-1} \right
)^*\right \} \nonumber \\
&\quad = \T \left\{  \Db_{j,\omega}(H_\omega+\gamma)^{-1} U^{(0)}(-r)
\odot_L (H_\omega+\gamma) i[  x_k, {\zeta}_\omega ] \odot_R U_\omega^{(0)}(r)
\right \} \label{finally2}  \\
& \quad = \T \left\{  \Db_{j,\omega}{\,  \U}^{(0)}(-r) \left ( i[  x_k, {\zeta}_\omega
] \right ) \right \} \label{finally} \; ,
\end{align}
where to obtain \eqref{finally2} we used
\eqref{DUQ}-\eqref{DUQ6} in the reverse direction, with $U_\omega^{(0)}(r) $
substituted for  $ U_\omega({\El},0,r)$, and in the last step used again that
$(H_\omega+\gamma)^{-1}$ commutes with $U^{(0)}(r)$.

The Kubo formula \eqref{sigmajk} now follows from
\eqref{defsigmajk31}  and \eqref{finally}.
\end{proof}


\subsection{The  Kubo-St\u{r}eda formula for the Hall conductivity}
\label{subsectHall}

Following \cite{BESB,AG}, we now recover the well-known  Kubo-St\u{r}eda formula
 for the Hall
conductivity at zero temperature. We write
\begin{equation}
    \sigma_{j,k}^{(E_f)} = \sigma_{j,k}(P^{(E_F)}_\omega) \; , \text{ and } \
    \sigma_{j,k}^{(E_f)}(\eta) = \sigma_{j,k}(\eta;P^{(E_F)}_\omega) \; .
\end{equation}

\begin{theorem}\label{thmHall}
If ${\zeta}_\omega = P^{(E_F)}_\omega$ is a
Fermi projection satisfying \eqref{assumption}, we have
\begin{align}\label{expHall}
\sigma_{j,k}^{(E_F)}
=
-i \T \left\{ P^{(E_F)}_\omega
\odot_L \left[ \left[x_j, P^{(E_F)}_\omega \right],
\left[x_k, P^{(E_F)}_\omega\right ]
\right]_\diamond \right\}
\end{align}
for all $j,k=1,2,\ldots,d$.  As a consequence,  the conductivity tensor is antisymmetric;
in particular $\sigma_{j,j}^{(E_F)} =0$ for $j=1,2,\ldots,d$.
\end{theorem}

Clearly the direct conductivity vanishes, $\sigma_{jj}^{(E_F)}=0$. Note that,
if the system is time-reversible the off diagonal elements are zero in the
region of localization, as expected.
\begin{corollary} Under the assumptions of Theorem \ref{thmHall}, if $\A=0$ (no magnetic field),
we have $\sigma_{j,k}^{(E_F)} =0$ for all $j,k=1,2,\ldots,d$.
\end{corollary}

\begin{proof}  Let $J$ denote complex conjugation on $\H$, i.e.,
$J \varphi = \bar{\varphi}$, an  antiunitary  operator on $\H$. The time
reversal operation is given by $\Theta(S) = JSJ$, where $S$ is a self-adjoint
operator (an observable). We have $J \H_c = \H_c$, and hence
$\Theta(A_\omega)\varphi = JA_\omega J\varphi$ gives a complex conjugation on
 $\K_i$, $i=1,2,\infty$.

If $\A=0$, we have $\Theta(H_\omega)=H_\omega$, and thus
$\Theta(f(H_\omega))=f(H_\omega)$ for any real valued Borel measurable function
$f$. Moreover $\Theta(i[x_j,P^{(E_F)}_\omega])=-i[x_j,P^{(E_F)}_\omega]$ and
$\Theta([A_\omega,B_\omega]_\diamond) =
[\Theta(A_\omega),\Theta(B_\omega)]_\diamond$. On the other hand if
$A_\omega\in\K_1$ is symmetric, then $\T (\Theta(A_\omega)) = \T (A_\omega)$.
Since $P^{(E_F)}_\omega \odot_L i\left[ i[x_j, P^{(E_F)}_\omega ], i[x_k,
P^{(E_F)}_\omega ]\right]_\diamond
 \odot_R P^{(E_F)}_\omega$ is symmetric, it follows from Theorem~\ref{thmHall} and
the above remarks that
\begin{align}
\sigma_{j,k}^{(E_F)} &=
 \T \left\{ P^{(E_F)}_\omega
\odot_L i\left[ i[x_j, P^{(E_F)}_\omega ], i[x_k, P^{(E_F)}_\omega
]\right]_\diamond
 \odot_R P^{(E_F)}_\omega \right\} \\
& =   - \T \left\{ P^{(E_F)}_\omega \odot_L i\left[ i[x_j, P^{(E_F)}_\omega ],
i[x_k, P^{(E_F)}_\omega ]\right]_\diamond
 \odot_R P^{(E_F)}_\omega \right\}=-\sigma_{j,k}^{(E_F)} \nonumber ,
\end{align}
and hence $\sigma_{j,k}^{(E_F)}=0$.
\end{proof}

Before proving Theorem~\ref{thmHall}, we recall that
under Assumption~\ref{assumptiona} the operator $[x_k, P^{(E_F)}_\omega ]
\in \K_1 \cap \K_2$ is defined on $\H_c$ as $x_k P^{(E_F)}_\omega-
P^{(E_F)}_\omega x_k$ thanks to  \eqref{assumption}.

\begin{lemma}\label{lemtriplecom} We have (as operators on $\H_c$)
\begin{equation}\label{triplecom}
\left[P^{(E_F)}_\omega, \left[P^{(E_F)}_\omega, [  x_k, P^{(E_F)}_\omega]
\right]_\odot\right]_\odot = [x_k, P^{(E_F)}_\omega] .
\end{equation}
\end{lemma}

\begin{proof}
Since $P^{(E_F)}_\omega \in \K_\infty$ and $[ x_k, P^{(E_F)}_\omega ] \in \K_1
\cap \K_2$, the left hand side of \eqref{triplecom} makes sense in $\K_1$ and
$\K_2$, and thus as an operator on $\H_c$.

Note that the orthogonal projection $1-P^{(E_F)}_\omega$ is in $\K_\infty$,
although it
is {\em not} in $\K_1$ or $\K_2$. Furthermore $(1- P^{(E_F)}_\omega) \H_c
\subset \H_c + P^{(E_F)}_\omega \H_c \subset \D(\x)$. Thus $P^{(E_F)}_\omega
x_k (1-P^{(E_F)}_\omega)$ and $(1-P^{(E_F)}_\omega) x_k P^{(E_F)}_\omega$ make
sense as operators on $\H_c$ (almost surely), and we have
\begin{equation}
  \left [ x_k ,
P^{(E_F)}_\omega \right ] \ = \ (1-P^{(E_F)}_\omega) x_k P^{(E_F)}_\omega
 - P^{(E_F)}_\omega x_k (1-P^{(E_F)}_\omega) \quad \text{on $ \H_c$}.
\end{equation}
Since $P^{(E_F)}_\omega (1-P^{(E_F)}_\omega)=0$, the right hand side
 of this expression
is unchanged if we replace $x_k$ by $[x_k,P^{(E_F)}_\omega]$ in the first term
and by $-[x_k,P^{(E_F)}_\omega]$ in the second. As technically
$[x_k,P^{(E_F)}_\omega]$ is defined on $\H_c$, we should introduce the products
$\odot_{L,R}$ here. Thus,
\begin{align}\label{triplecommid}
&  \left [ x_k ,
 P^{(E_F)}_\omega \right ] \ = \ (1-P^{(E_F)}_\omega) \odot_L
[x_k,P^{(E_F)}_\omega] \odot_R
 P^{(E_F)}_\omega \\ &\qquad \qquad  \qquad \qquad \quad  + P^{(E_F)}_\omega \odot_L
[x_k,P^{(E_F)}_\omega]  \odot_R (1-P^{(E_F)}_\omega)  \; . \notag
\end{align}

Now, given any $A_\omega \in \K_{\odot}$ we have
\begin{equation}
  \left[P^{(E_F)}_\omega, A_\omega
\right]_\odot  \ = \ -\left[1-P^{(E_F)}_\omega, A_\omega \right]_\odot \;
,
\end{equation}
and thus
\begin{align}\label{offdiag}
 & \left[P^{(E_F)}_\omega, \left[P^{(E_F)}_\omega, A_\omega
\right]_\odot\right]_\odot = \\
&\qquad \qquad \ P^{(E_F)}_\omega \odot_L A_\omega
\odot_R (1-P^{(E_F)}_\omega) \ + \  (1-P^{(E_F)}_\omega)\odot_L A_\omega
\odot_R P^{(E_F)}_\omega \; ,\nonumber
\end{align}
using that $P^{(E_F)}_\omega \odot (1-P^{(E_F)}_\omega) = 0$. Finally,
\eqref{triplecom} follows from \eqref{triplecommid} and \eqref{offdiag}.
\end{proof}

\begin{remark}
(i)Eq. \eqref{triplecommid} appears in \cite{BESB} (and then in \cite{AG}) as a
key step in the derivation of the expression of the Hall conductivity.
\\ (ii) In \eqref{triplecom} we use crucially the fact that we work at temperature zero, i.e. that
the initial density matrix is the  orthogonal projection $P^{(E_F)}_\omega$. The argument
does not go through at positive temperature.
\end{remark}

\begin{proof}[Proof of Theorem~\ref{thmHall}.] We first regularize the velocity
$\Db_{j,\omega}$ with a smooth function $f_n\in\mathcal{C}^\infty_c(\R)$,
$|f_n|\le 1$, $f_n=1$ on $[-n,n]$, so that
 $\Db_{j,\omega}
f_n(H_\omega)\in \K_1\cap \K_2 \in \K_\infty$. We have, using the centrality of the trace
 $\T$ (see Lemma~\ref{lemmacentral}), that
\begin{align}\label{eqhall00a}
\widetilde \sigma_{jk}^{(E_F)}(r)& :=   -\T\left\{ 2\Db_{j,\omega}
\, \U^{(0)}(-r) (i[  x_k, P^{(E_F)}_\omega ])  \right\}
\\
& =  \nonumber - \lim_{n\to\infty} \T\left\{
(2\Db_{j,\omega}f_n(H_\omega))\odot_L \U^{(0)}(-r) (i[  x_k,
P^{(E_F)}_\omega ])  \right\} \\
& =    \label{eqhall00} -\lim_{n\to\infty} \T\left\{
\U^{(0)}(r)(2\Db_{j,\omega}f_n(H_\omega)) \odot_L  i[  x_k,
P^{(E_F)}_\omega ]  \right\} .
\end{align}
 Next, it follows from Lemma~\ref{lemmacentral} that, for
 $A_\omega, B_\omega\in
\K_\infty$ and $ C_\omega\in\K_1$, we have
\begin{equation}
\T \left\{ A_\omega \odot_L [B_\omega, C_\omega]_\odot \right\} =
\T \left\{ [A_\omega, B_\omega] \odot_L  C_\omega \right\}\, .
\end{equation}
 It follows, on the account of Lemma~\ref{lemtriplecom}, that
\begin{align}\label{eqhall009}
&\T\left\{ {\U^{(0)}}(r)(2\Db_{j,\omega}f_n(H_\omega)) \odot_L  i[
x_k, P^{(E_F)}_\omega ]  \right\}
\\ &\quad =
\T\left\{ {\U^{(0)}}(r)(2\Db_{j,\omega}f_n(H_\omega)) \odot_L
\left[P^{(E_F)}_\omega, \left[P^{(E_F)}_\omega, i[  x_k, P^{(E_F)}_\omega]
\right]_\odot\right]_\odot  \right\} \nonumber
\\ &\quad =
\T\left\{{\U^{(0)}}(r)\left( \left[P^{(E_F)}_\omega,
\left[P^{(E_F)}_\omega, 2\Db_{j,\omega}f_n(H_\omega)
\right]\right] \right) \odot_L  i[  x_k,
P^{(E_F)}_\omega]\right\} ,\nonumber
\end{align}
where we used that $P^{(E_F)}_\omega$ commutes with $U^{(0)}_\omega$.

We now claim that
\begin{equation}\label{equalcom}
\left[ P_\omega^{(E_F)}, 2\Db_{j,\omega}f_n(H_\omega)\right]
=\left[H_\omega , i[x_j,P_\omega^{(E_F)}] \right]_\ddagger \odot_R f_n(H_\omega) \, .
\end{equation}
To see this, we use \eqref{Hcom} to conclude that
\begin{align}
&\left[H_\omega , i[x_j,P_\omega^{(E_F)}] \right]_\ddagger \odot_R f_n(H_\omega)\\
& \qquad \qquad \quad \notag
= 2 \left( \Db_{j,\omega}P_\omega^{(E_F)}\right)^\ddagger
 \odot_R f_n(H_\omega) -
 2 \Db_{j,\omega} P_\omega^{(E_F)}f_n(H_\omega) \nonumber
\\ & \qquad \qquad \quad  = 2 \left( P_\omega^{(E_F)} \Db_{j,\omega} f_n(H_\omega) -
 \Db_{j,\omega}P_\omega^{(E_F)} f_n(H_\omega)\right) \notag\\
& \qquad \qquad\quad  =
2 \left( P_\omega^{(E_F) }\Db_{j,\omega}f_n(H_\omega) -
 \Db_{j,\omega}f_n(H_\omega)P_\omega^{(E_F)}\right) \, ,\nonumber
\end{align}
which is just  \eqref{equalcom}. Combining \eqref{eqhall00}, \eqref{eqhall009},
and  \eqref{equalcom}, we get after taking $n \rightarrow \infty$,
\begin{equation}\label{eqhall0090} \widetilde \sigma_{jk}^{(E_F)}(r) \
 = \
-\T\left\{{\U^{(0)}}(r)\left( \left[P^{(E_F)}_\omega, \left[H_\omega ,
i[x_j,P_\omega^{(E_F)}] \right]_\ddagger \right]_\odot \right) \diamond i[
x_k, P^{(E_F)}_\omega]\right\} \, .
\end{equation}
Here it is useful to note that, by Proposition \ref{propDA}(i),
the restriction to $\H_c$ of
$ \overline{\left[
P_\omega^{(E_F)}, 2\Db_{j,\omega}\right]}$ is in $
\K_\infty\cap\K_1 \cap \K_2$, and
 \begin{equation}\label{equalcoma}
\left[H_\omega , i[x_j,P_\omega^{(E_F)}] \right]_\ddagger = \overline{\left[
P_\omega^{(E_F)}, 2\Db_{j,\omega}\right]} \in \K_1 \cap \K_2 \, .
\end{equation}

In addition, on $\K_i$, $i=1,2$, we have
\begin{equation}\label{notecom1}
P_\omega^{(E_F)}\odot_L (H_\omega i[x_j,P_\omega^{(E_F)}])
=
H_\omega (P_\omega^{(E_F)}\odot_L i[x_j,P_\omega^{(E_F)}])\, ,
\end{equation}
and, on the account of Lemma~\ref{lemadjHA},
\begin{equation}\label{notecom2}
(H_\omega i[x_j,P_\omega^{(E_F)}])\odot_R P_\omega^{(E_F)} = H_\omega
(i[x_j,P_\omega^{(E_F)}]\odot_R P_\omega^{(E_F)}) \, .
\end{equation}
It also follows from \eqref{notecom1} and \eqref{notecom2}  that
\begin{align}\label{notecom25}
H_\omega \left[P_\omega^{(E_F)}, i[x_j,P_\omega^{(E_F)}] \right]_\odot=
 \left[P_\omega^{(E_F)},  H_\omega i[x_j,P_\omega^{(E_F)}]
\right]_\odot \, ,
\end{align}
all terms being well defined in $\K_i$. Therefore,
\begin{align}
\left[ P_\omega^{(E_F)},\left[H_\omega , i[x_j,P_\omega^{(E_F)}]
\right]_\ddagger \right]_\odot  =  \left[H_\omega, \left[ P_\omega^{(E_F)}
, i[x_j,P_\omega^{(E_F)}] \right]_\odot \right]_\ddagger\, .
\end{align}

We thus get
\begin{align}\nonumber
\widetilde \sigma_{jk}^{(E_F)}(r)& =  -\T\left\{\U^{(0)}_\omega(r)\left(
\left[H_\omega, \left[ P_\omega^{(E_F)} ,
 i[x_j,P_\omega^{(E_F)}] \right]_\odot
\right]_\ddagger \right)
\diamond  i[  x_k, P^{(E_F)}_\omega]\right\}\\
&= -\left \la \left \la   \e^{-ir\Ll_2} { \Ll_2} \left ( \left[
P_\omega^{(E_F)} , i[x_j,P_\omega^{(E_F)}] \right]_\odot \right ),
        i[  x_k, P^{(E_F)}_\omega] \right\ra \right\ra \label{eqhall01}
 \, ,
\end{align}
where we used \eqref{centralK2} and Corollary~\ref{Liouvillian}. Recall that $
\la \la \cdot, \cdot \ra \ra$ is the inner product on $\H_2$ and $\Ll_2$
is the Liouvillian in $\K_2$ -- the self-adjoint generator of the unitary group
$\, \U^{(0)}(t)$. Combining \eqref{sigmajk}, \eqref{eqhall00a}, and \eqref{eqhall01},
we get
\begin{equation} \label{sigmaL}
\sigma_{jk}^{(E_F)}(\eta)  = - \left \la \left \la    i  \left(\Ll_2+
i\eta \right)^{-1}{ \Ll_2} \left ( \left[ P_\omega^{(E_F)} ,
 i[x_j,P_\omega^{(E_F)}] \right]_\odot \right ),
        i[  x_k, P^{(E_F)}_\omega] \right\ra \right\ra \, .
\end{equation}

It follows from the spectral theorem  (applied to
$\mathcal{L}_2$) that
\begin{equation}\label{stL}
\lim_{\eta\to 0} \left(\Ll_2+ i\eta \right)^{-1}{ \Ll_2} =
P_{(\mathrm{Ker}\, \mathcal{L}_2)^\perp}  \;\;\mbox{strongly in $\K_2$} \, ,
\end{equation}
where
 $P_{(\mathrm{Ker}\, \mathcal{L}_2)^\perp}$
is the orthogonal projection onto ${(\mathrm{Ker} \,\mathcal{L}_2)^\perp}$.  Moreover,
 we have
\begin{equation}\label{Pperp}
\left[ P_\omega^{(E_F)} , i[x_j,P_\omega^{(E_F)}] \right]_\odot
\in{(\mathrm{Ker}\, \mathcal{L}_2)^\perp}\, .
 \end{equation}
To see this, note that if $A_\omega \in \mathrm{Ker}\, \mathcal{L}_2$,
then for all $t$
we have
$\, \U^{(0)}(r)(A_\omega) = A_\omega$, and hence
$ \e^{-itH_\omega } \odot_L  A_\omega  =   A_\omega \odot_R \e^{-itH_\omega } $,
so it follows that $f(H_\omega) \odot_L  A_\omega =  A_\omega \odot_R f(H_\omega)$
for all $f \in \S(\R)$, i.e., $[A_\omega, f(H_\omega)]_\odot = 0$.  An approximation
 argument using Lemma~\ref{stK2} gives  $[A_\omega, P^{(E_F)}_\omega]_\odot = 0$.
Thus
\begin{align}
\left \la \left \la A_\omega,
 \left[ P_\omega^{(E_F)} , i[x_j,P_\omega^{(E_F)}] \right]_\odot  \right\ra \right\ra=
\left \la \left \la[A_\omega, P^{(E_F)}_\omega]_\odot,
 i[x_j,P_\omega^{(E_F)}]   \right\ra \right\ra = 0 \, ,
\end{align}
and  \eqref{Pperp} follows.

Combining \eqref{sigmaL}, \eqref{stL},  \eqref{Pperp}, and
 Lemma~\ref{formulaCommutator}, we get
\begin{align}\nonumber
 \sigma_{j,k}^{(E_F)}& = i  \left \la \left \la
\left[ P_\omega^{(E_F)} , i[x_j,P_\omega^{(E_F)}] \right]_\odot,
        i[  x_k, P^{(E_F)}_\omega] \right\ra \right\ra \\
&= - i \T\left\{    \left[ P_\omega^{(E_F)} , i[x_j,P_\omega^{(E_F)}] \right]_\odot
   \diamond     i[  x_k, P^{(E_F)}_\omega]  \right\} \\
&=  -i \T\left\{    P_\omega^{(E_F)} \odot_L \left[ i[x_j,P_\omega^{(E_F)}] ,
        i[  x_k, P^{(E_F)}_\omega] \right]_\diamond\right\}\,, \nonumber
\end{align}
which is just  \eqref{expHall}.  The theorem is proved.
\end{proof}


\begin{acknowledgement}  We thank Jean Bellissard for many discussions on the
 Kubo formula. We are grateful to Daniel Lenz for the proof that
 $\K_\infty$ is a von Neumann algebra.
We thank Vladimir Georgescu for bringing the inequality
 \eqref{diamnabla} to our attention.
\end{acknowledgement}



\begin{thebibliography}{ThKNN}


\bibitem[AENSS]{AENSS} Aizenman, M., Elgart, A., Naboko, S., Schenker, J.H.,
Stolz, G.: Moment Analysis for Localization in Random Schr\"odinger Operators.
2003 Preprint, math-ph/0308023.

\bibitem[AES]{AES} Aizenman, M., Elgart, A., Schenker, J.H.:
Adiabatic charge transport, localization, and the Kubo formula for $2D$
Hall conductance in a gapless state. In preparation.

\bibitem[AG]{AG} {Aizenman, M.,   Graf, G.M.:}
{ Localization bounds for an electron gas}, J. Phys. A: Math.
Gen. {\bf 31}, 6783-6806, (1998).


\bibitem[AvSS]{ASS} Avron, J., Seiler, R., Simon, B.: Charge deficiency,
charge transport and comparison of dimensions.  Comm. Math. Phys.~{\bf 159},
399-422 (1994).

\bibitem[B]{Be}  Bellissard, J.: Ordinary quantum {H}all effect and
noncommutative cohomology. In {\em Localization in disordered systems (Bad
Schandau, 1986)}, pp. 61-74. Teubner-Texte Phys. {\bf 16}, Teubner, 1988.

\bibitem[BES]{BESB} Bellissard, J., van Elst, A., Schulz-Baldes, H.:
The non commutative geometry of the quantum Hall effect.
{J. Math. Phys.}~{\bf 35}, 5373-5451 (1994).

\bibitem[BeG]{BeG}  Berthier, A., Georgescu, V.:
On the point spectrum of Dirac operators.  J. Funct. Anal.  {\bf 71},  309-338 (1987).


\bibitem[BoGK]{BGK}  Bouclet, J.M.,   Germinet, F., Klein, A.:
Sub-exponential decay of operator kernels
 for functions of generalized Schr\"odinger operators.
Proc. Amer. Math. Soc.    \textbf{132} ,  2703-2712  (2004).

\bibitem[BrR]{BR} Bratteli, O., Robinson, D.W.: \emph{Operator Algebras and Quantum
Statitical Mechanics I}.  Springer-Verlag, 1979.


\bibitem[CH]{CH} Combes,  J.M.,   Hislop, P.D.:{ Landau Hamiltonians with
random potentials: localization and the density of states}. Commun. Math.
Phys. {\bf 177}, 603-629 (1996).



\bibitem[CT]{CT}
Combes,  J.M.,  Thomas, L.: Asymptotic behaviour of eigenfunctions for
multiparticle {S}chr\"odinger operators. {Comm. Math. Phys.}~{\bf 34}, 251--270
(1973).

\bibitem[CoJM]{CJM} Cornean, H.D., Jensen, A., Moldoveanu, V.:
A rigorous proof for the Landauer-B\"uttiker formula. Preprint  on mp\_arc 04-71.

\bibitem[CyFKS]{CFKS}  Cycon, H.L.,   Froese, R.G.,  Kirsch, W., Simon, B.:
 {\em Schr\"odinger operators}. Heidelberg: Springer-Verlag, 1987.


\bibitem[D]{D} Davies, E.B.: Kernel estimates for functions of second order
 elliptic operators.  Quart. J. Math. Oxford (2) {\bf 39}, 37-46 (1988).




\bibitem[ES]{ES} Elgart, A. , Schlein, B.: Adiabatic charge transport
 and the Kubo formula for Landau Type Hamiltonians.
 Comm. Pure Appl. Math.  57,  590-615 (2004).


\bibitem[F]{F} Faris, W.G.: \emph{Self-Adjoint Operators}.
Lecture Notes in Mathematics {\bf 433}. Springer-Verlag, 1975.

\bibitem[GK1]{GK1}  Germinet, F., Klein, A.: {Bootstrap Multiscale Analysis
and Localization in Random Media}. Commun. Math. Phys.~{\bf  222}, 415-448
(2001).

\bibitem[GK2]{GK2}  Germinet, F.,  Klein, A.: Operator kernel estimates for  functions of generalized Schr\"odinger operators.
Proc. Amer. Math. Soc. 131,  911-920  (2003).

\bibitem[GK3]{GK3} Germinet, F.,  Klein, A.: {A characterization of the
Anderson metal-insulator transport transition}.
Duke Math. J. \textbf{124}, 309-351 (2004).

\bibitem[GK4]{GK4} Germinet, F,  Klein, A.:
Explicit finite volume criteria for localization in continuous
 random media and applications. Geom. Funct. Anal. 13, 1201-1238 (2003).

\bibitem[GK5]{GK5} Germinet, F,  Klein, A.: New characterizations of
the region of dynamical localization for random Schr\"odinger operators.
In preparation.


\bibitem[HS]{HS} Helffer, B.,  Sj\"ostrand, J.:  Equation de Schr\"odinger
avec champ magn\'etique et \'equation de Harper. In
\emph{Schr\"odinger Operators}, H. Holden and A. Jensen, eds.,
pp. 118-197. Lectures Notes in Physics \textbf{345},
Springer-Verlag, 1989.

\bibitem[HuS]{HuS}  Hunziker W.,  Sigal, I.M.:  Time-dependent scattering theory for
$N$-body quantum systems.  Rev. Math. Phys. \textbf{12}, 1033-1084  (2000).

\bibitem[K]{Ka}   Kato, T.: \emph{Perturbation Theory for Linear Operators}.
Springer-Verlag, 1976.

\bibitem[KeS]{KSB} Kellendonk, J., Schulz-Baldes, H.,
Quantization of edge currents for continuous magnetic operators.
 J. Funct. Anal.~{\bf 209},  388-413 (2004).


\bibitem[Ku]{Ku} Kunz, H.:
{The Quantum Hall Effect for Electrons in a Random Potential}.
Commun. Math. Phys.~{\bf 112}, 121-145 (1987).

\bibitem[LS]{LS} { H. Leinfelder, C.G.  Simader},
{\it Schr\"odinger operators with singular magnetic potentials},
Math. Z. 176, 1-19 (1981).

\bibitem[NB]{NB} Nakamura, S., Bellissard, J.:
{Low Energy Bands do not Contribute to Quantum Hall Effect}.
Commun. Math. Phys.~{\bf 131}, 283-305 (1990).

\bibitem[Na]{Na} Nakano, F.: Absence of transport in Anderson localization.
 Rev. Math. Phys.~{\bf 14},  375-407 (2002).

\bibitem[P]{Pa}  Pastur, L., Spectral properties of disordered systems in the one-body approximation.  Comm. Math. Phys.~{\bf  75}, 179-196  (1980).

\bibitem[PF]{PF}  Pastur, L.,  Figotin, A.: {\em Spectra of Random and
Almost-Periodic Operators}.   Springer-Verlag, 1992.

\bibitem[RS1]{RS1}  Reed, M.,  Simon, B.:  \emph{Methods of Modern
Mathematical Physics I: Functional Analysis}, revised and enlarged edition.
 Academic Press, 1980.

\bibitem[RS2]{RS2}  Reed, M.,  Simon, B.:  \emph{Methods of Modern
Mathematical Physics II: Fourier Analysis, Self-Adjointness}.
Academic Press, 1975.

\bibitem[RS4]{RS4}  Reed, M.,  Simon, B.:  \emph{Methods of Modern
Mathematical Physics IV: Analysis of Operators}.
Academic Press, 1978.



\bibitem[SB1]{SBB1} Schulz-Baldes, H., Bellissard, J.:
Anomalous transport: a mathematical framework.
Rev. Math. Phys.~{\bf 10}, 1-46 (1998).

\bibitem[SB2]{SBB2} Schulz-Baldes, H., Bellissard, J.: A Kinetic Theory
 for Quantum Transport in Aperiodic Media. J. Statist. Phys.~{\bf 91},
 991-1026 (1998).

\bibitem[Si1]{Si1}  Simon, B.: Maximal and minimal Schr\"odinger forms.
J. Operator Theory \textbf{1}, 37-47 (1979).


\bibitem[Si2]{Si2}  Simon, B.: {Schr\"odinger semi-groups}.  Bull. Amer.
Math. Soc. {\bf 7},  447-526 (1982).

\bibitem[St]{St} St\u{r}eda, P.: {Theory of quantised Hall conductivity
    in two dimensions}. J. Phys. C. {\bf 15}, L717-L721 (1982).

\bibitem[T]{T}  Takesaki, M.:  \emph{ Theory of Operator Algebras I.}  Springer-Verlag, 1979.


\bibitem[ThKNN]{TKNN}  Thouless, D. J., Kohmoto, K., Nightingale, M. P.,
den Nijs, M.: {Quantized Hall conductance in a two-dimensional periodic
potential}. Phys. Rev. Lett. {\bf 49}, 405-408 (1982).

\bibitem[W]{Wa} Wang, W.-M.: {Microlocalization, percolation, and
Anderson localization for the magnetic Schr\"odinger operator with a
random potential}.  J. Funct. Anal.~\textbf{146}, 1-26  (1997).

\bibitem[Y]{Y} Yosida, K.:  \emph{Functional Analysis, 6th edition.}
Springer-Verlag, 1980.

\end{thebibliography}
\end{document}